\documentclass[traditabstract]{aa} % for the abstract without structuration 
\usepackage{graphicx}
\usepackage{txfonts}
\usepackage{ulem}
\usepackage{natbib}
\def\ms{\hbox{\,m\,s$^{-1}$}}         %m.s -1
\def\m2s2{\hbox{\,m$^{2}$\,s$^{-2}$}} %m2.s -2
\def\kms{\hbox{\,km\,s$^{-1}$}}       %km.s -1
\def\vsini{\hbox{$v$\,sin\,$i_{\star}$}}      %vsini
\def\Msun{\hbox{$M_{\odot}$}}             %Msun
\def\Rsun{\hbox{$R_{\odot}$}}

\def\Kepler{{\it Kepler}}
\def\Ksix{{KOI-206}}
\def\Kfour{{KOI-614}}
\def\Kzero{{KOI-680}}
\def\Ksixb{{KOI-206b}}
\def\Kfourb{{KOI-614b}}
\def\Kzerob{{KOI-680b}}
\newcommand{\e}[1]{{\;\times\; 10^{#1}}}
\begin{document}

   \title{SOPHIE velocimetry of \Kepler\ transit candidates\thanks{Based on observations made with SOPHIE on the 1.93-m telescope at the Observatoire de Haute-Provence (CNRS), France.}.}

   \subtitle{XV. \Kfourb, \Ksixb,\ and \Kzerob: a massive warm Jupiter orbiting a G0 metallic dwarf and two highly inflated planets with a distant companion around evolved F-type stars.}
\author{J.M. Almenara\inst{\ref{lam},\ref{grenoble1},\ref{grenoble2}} 
\and C. Damiani\inst{\ref{lam}}
\and F. Bouchy\inst{\ref{lam}}
\and M. Havel\inst{\ref{nasa}}
\and G. Bruno\inst{\ref{lam}}
\and G. H\'ebrard\inst{\ref{ohp},\ref{iap}}
\and R.F. Diaz\inst{\ref{lam},\ref{geneva}}
\and M. Deleuil\inst{\ref{lam}}
\and S.C.C. Barros\inst{\ref{lam}}
\and I. Boisse\inst{\ref{lam}}
\and A.S. Bonomo \inst{\ref{inaf}}
\and G. Montagnier\inst{\ref{ohp},\ref{iap}}
\and A. Santerne\inst{\ref{porto},\ref{porto2},\ref{lam}}
             }

\institute{
Aix Marseille Universit\'e, CNRS, LAM (Laboratoire d'Astrophysique de Marseille) UMR 7326, 13388, Marseille, France\label{lam}
\and Univ. Grenoble Alpes, IPAG, F-38000 Grenoble, France\label{grenoble1}
\and CNRS, IPAG, F-38000 Grenoble, France\label{grenoble2}
\and NASA Postdoctoral Program Fellow, Ames Research Center, P.O. Box 1, Moffett Field, CA 94035\label{nasa}
\and Observatoire de Haute Provence, 04670 Saint Michel l'Observatoire, France\label{ohp}
\and Institut d'Astrophysique de Paris, UMR7095 CNRS, Universit\'e Pierre \& Marie Curie, 98bis boulevard Arago, 75014 Paris, France\label{iap}
\and Observatoire Astronomique de l'Universit\'e de Gen\`eve, 51 chemin des Maillettes, 1290 Versoix, Switzerland\label{geneva}
\and INAF - Osservatorio Astrofisico di Torino, via Osservatorio 20, 10025 Pino Torinese, Italy \label{inaf}
\and Centro de Astrof\'{i}sica, Universidade do Porto, Rua das Estrelas, 4150-762 Porto, Portugal \label{porto}
\and Instituto de Astrof\'isica e Ci\^{e}ncias do Espa\c co, Universidade do Porto, CAUP, Rua das Estrelas, PT4150-762 Porto, Portugal\label{porto2}
}

   \date{}

% \abstract{}{}{}{}{} 
% 5 {} token are mandatory
 
  \abstract 
  % context heading (optional)
  % {} leave it empty if necessary    
   {We report the validation and characterization of three new transiting exoplanets using SOPHIE radial velocities: \Kfourb, \Ksixb, and \Kzerob.  \Kfourb\ has a mass of $2.86\pm0.35~{\rm M_{Jup}}$ and a radius of $1.13^{+0.26}_{-0.18}~{\rm R_{Jup}}$, and it orbits a G0, metallic ([Fe/H]=$0.35\pm0.15$) dwarf in 12.9~days. Its mass and radius are familiar and compatible with standard planetary evolution models, so it is one of the few known transiting planets in this mass range to have an orbital period over ten~days. With an equilibrium temperature of $T_{eq}=1000 \pm 45$~K, this places \Kfourb\ at the transition between what is usually referred to as \lq\lq{}hot\rq\rq{} and \lq\lq{}warm\rq\rq{} Jupiters. \Ksixb\ has a mass of $2.82\pm 0.52~{\rm M_{Jup}}$ and a radius of $1.45\pm0.16~{\rm R_{Jup}}$, and it orbits a slightly evolved F7-type star in a 5.3-day orbit. It is a massive inflated hot Jupiter that is particularly challenging for planetary models because it requires unusually large amounts of additional dissipated energy in the planet. On the other hand, \Kzerob\ has a much lower mass of $0.84\pm0.15~{\rm M_{Jup}}$ and requires less extra-dissipation to explain its uncommonly large radius of $1.99\pm0.18~{\rm R_{Jup}}$. It is one of the biggest transiting planets characterized so far, and it orbits a subgiant F9-star well on its way to the red giant stage, with an orbital period of 8.6~days. With host stars of masses of $1.46\pm0.17~\Msun$ and $1.54 \pm 0.09~\Msun$, respectively, \Ksixb, and \Kzerob\ are interesting objects for theories of formation and survival of short-period planets around stars more massive than the Sun. For those two targets, we also find signs of a possible distant additional companion in the system.} 
  % aims heading (mandatory)
  % {...}
  % methods heading (mandatory)
  % {...}
  % results heading (mandatory)
  % {...}
  % conclusions heading (optional), leave it empty if necessary 
  % {...}

   \authorrunning{J.M. Almenara et al.}
   \titlerunning{SOPHIE velocimetry of \Kepler\ transit candidates}
   \maketitle
%
%________________________________________________________________

\section{Introduction}

Since 2010, we have been conducting a radial-velocity follow-up of \Kepler\ \citep{2010Sci...327..977B} transiting planetary candidates with the SOPHIE spectrograph \citep{2008SPIE.7014E..17P,2013A&A...549A..49B}. Transits combined with radial velocity measurements enable us to infer the masses and radii of the star and planet up to a one-parameter degeneracy \citep{2013pss3.book..489W}. To derive the absolute planetary parameters, an independent stellar parameter is needed, which can be estimated with the aid of stellar models. With the mass and radius of the planet, we can determine the mean density that gives insight into the composition and formation of the objects. In addition, such a survey allows the rate of false-positive of the \Kepler\ close-in giant exoplanet candidates to measured \citep{2012A&A...545A..76S}. 

  In this paper, we present the validation and characterization of three new transiting exoplanets: \Kfourb, \Ksixb, and \Kzerob. Two of them, \Ksixb\ and \Kzerob,\, have already been announced by our group to be planetary companions in \citet{2012A&A...545A..76S}.  The third one, \Kfourb,\ was not in the sample considered by \citet{2012A&A...545A..76S} owing to its transit depth of 0.39\%, slightly lower than the 0.4\% selection cut-off. 

The structure of the paper is as follows. We present the \Kepler\ and SOPHIE observations (Section~\ref{obs}), the derivation of the spectral parameters (Section~\ref{specan}), the stellar and planetary parameter estimation (Section~\ref{pastis}), the planetary evolution models (Section~\ref{set}), and finally the discussion (Section~\ref{discussion}) and the conclusions (Section~\ref{conclusion}) of our work.

\section{Observations} \label{obs}
\subsection{Photometric detection with \Kepler}\label{keplerobs}
The three targets analyzed in this paper were identified as KOI (\Kepler\ Object of Interest) by \citet{2011ApJ...736...19B}.  Their IDs, coordinates, and magnitudes are reported in Table~\ref{tableparam}. Two of them were found to be hosting a transiting companion  of planetary nature with high probability, namely \Ksix\ and \Kzero\ (vetting flag of 2), whereas \Kfour\ was considered a moderate probability candidate because of the V-shaped transit (vetting flag of 3). There is no significant spatial shift at the time of the transits in the \Kepler\ photometric centroid for the three targets.  No other transits with different periods were detected in any of the light curves, so there are no signs of multiple transiting systems. 
At the time of the beginning of the analysis, the \Kepler\ light curves of the first 15~quarters are publicly available at the MAST archive\footnote{http://archive.stsci.edu/kepler/data\_search/search.php}. Long cadence (29.4~minutes sampling) data only is available for \Kfour, but \Ksix\ and \Kzero\ have short-cadence data (1-minute sampling) available in 3 of the 15~first quarters. We used the light curve of quarters Q1 to Q15 calibrated through the \Kepler\ Science Pipeline and corrected for instrumental artifacts and other known systematics errors \citep{2010ApJ...713L.120J}. The three light curves present transits with depths of about 0.5\% and a typical uncertainty for individual points at the level of 160-280~ppm for the long-cadence data. 

\begin{figure*}
\centering
\includegraphics[width=6cm]{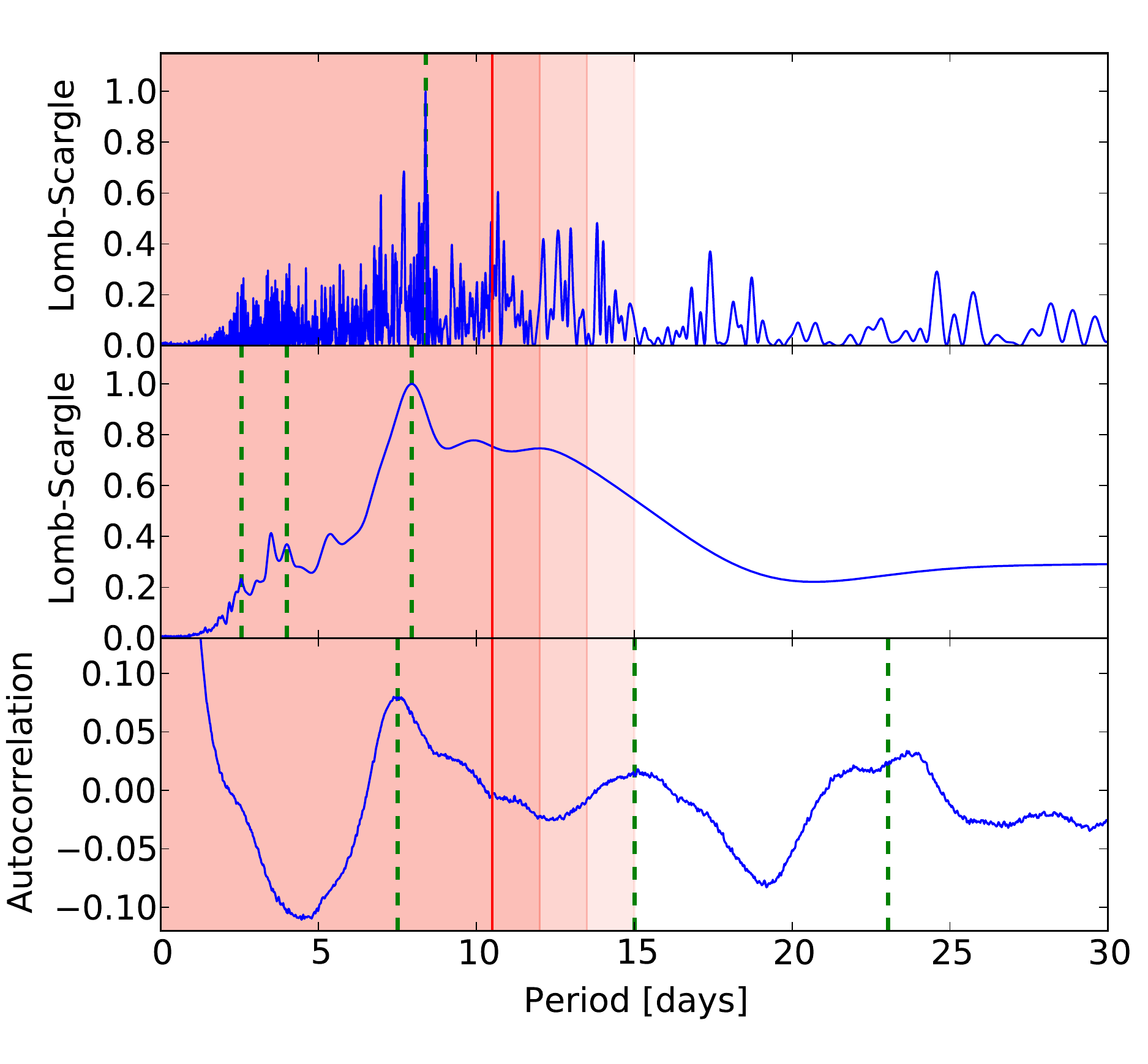} 
\includegraphics[width=6cm]{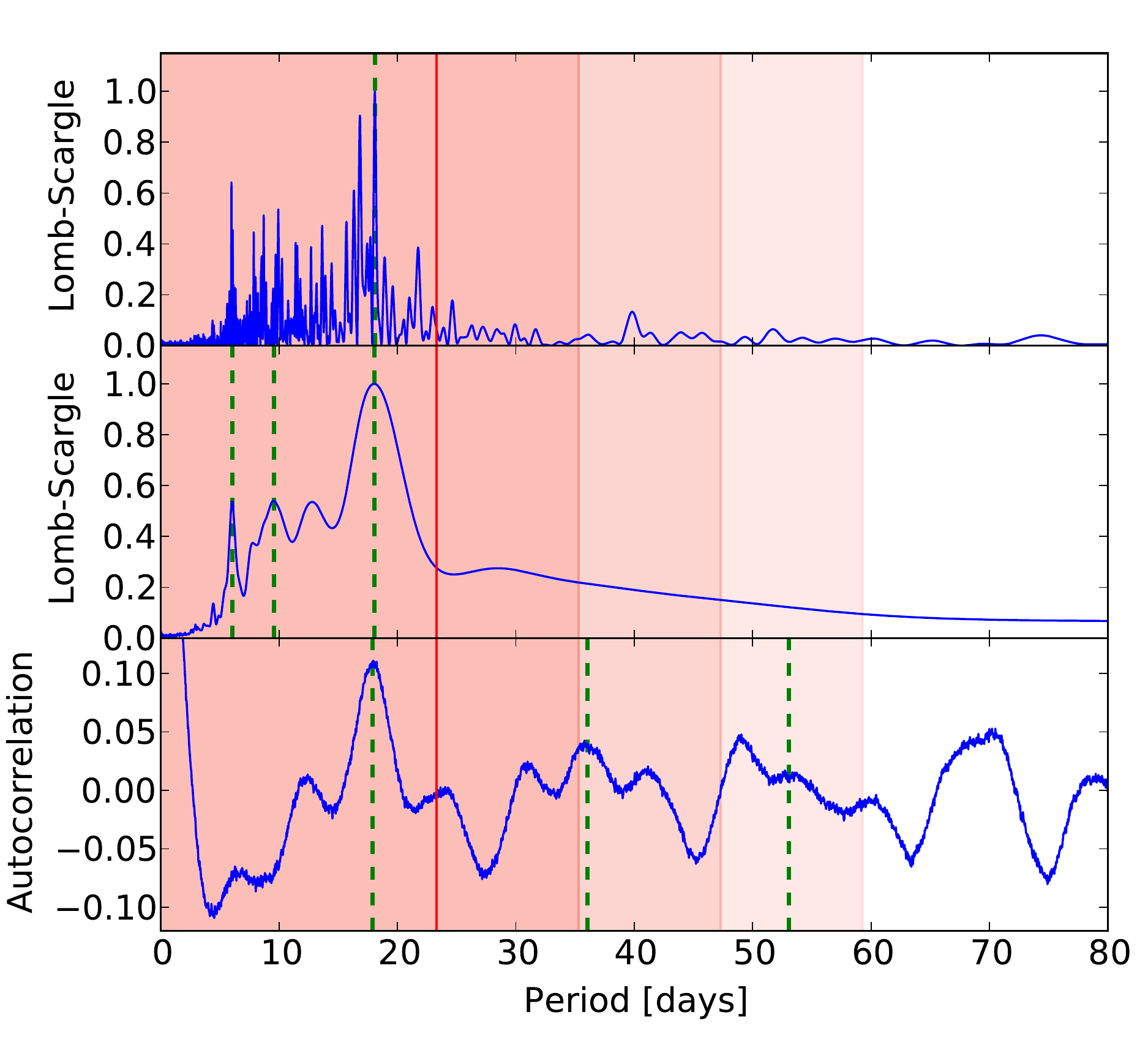}
\includegraphics[width=6cm]{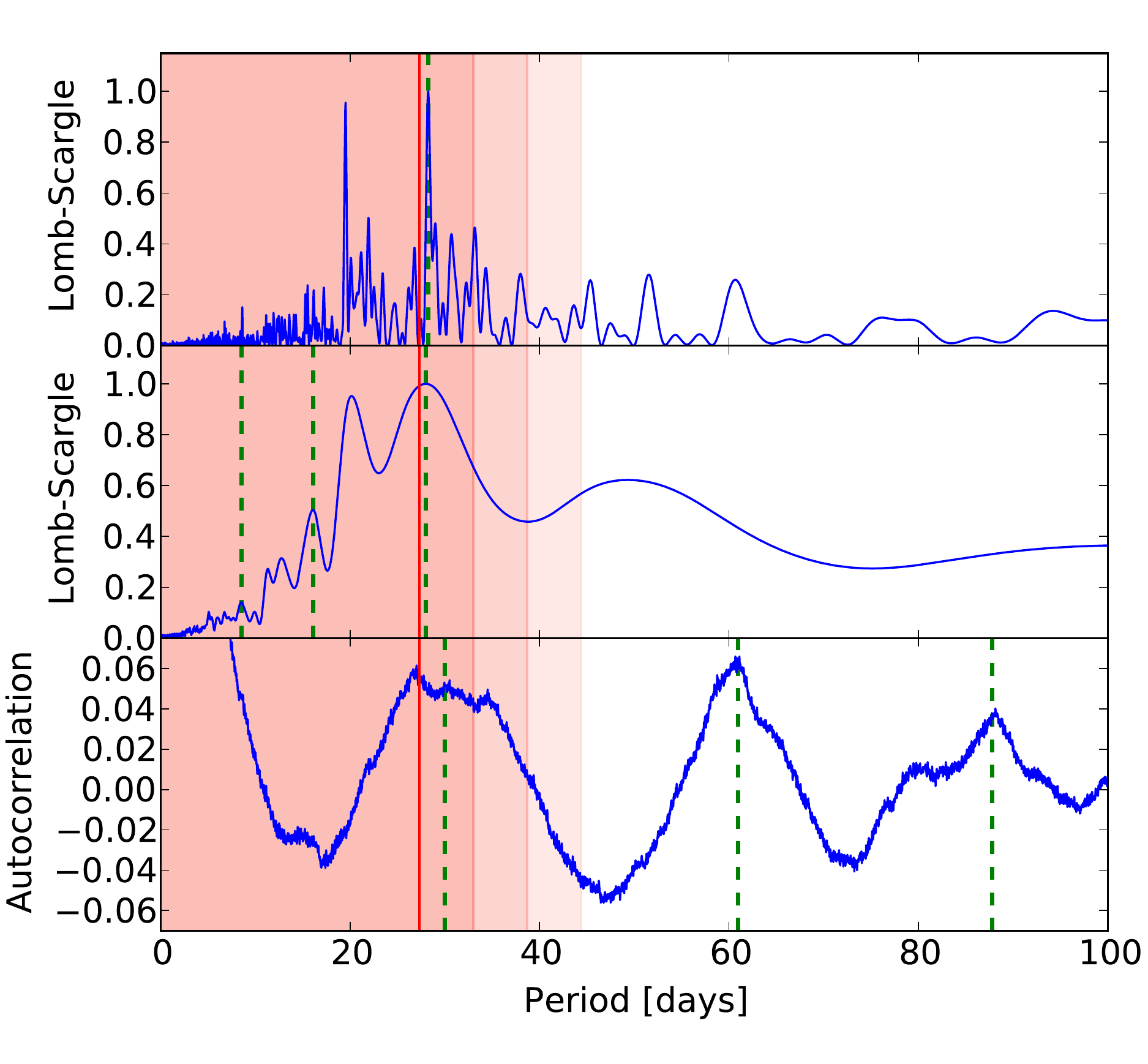}
\caption{From left to right: periodograms of \Ksix, \Kfour, and \Kzero. Top panels: Lomb-Scargle periodogram of the complete light curve. Middle panels: mean Lomb-Scargle periodogram of the light curve divided in chunks. Bottom panel: light curve autocorrelation. The green dashed lines indicate the center of Gaussians fitted to the data to derive the main frequency and its harmonics. The red vertical line corresponds to the maximum 
rotation period deduced from the \vsini\ and the star radius, and the shaded areas are the associated 1, 2, and 3-$\sigma$ upper limit regions from the darkest to the lighter ones, respectively.}
\label{LS}
\end{figure*}

Besides the transits, the three light curves show signs of stellar variability with flux modulations of about 90, 75, and 45~ppm amplitudes for \Ksix, \Kfour, and \Kzero,\ respectively. As a result, we computed their Lomb-Scargle periodograms \citep{1989ApJ...338..277P} using the long-cadence sampling after subtracting the transits (Fig.~\ref{LS}, top panels). They all present peaks compatible with the rotational period of the stars as expected from their measured projected rotational velocities (\vsini, Section~\ref{specan}) and radii (Section~\ref{pastis}), which give upper limits at $10.5 \pm 1.5$, $23.3^{+12}_{-7.1}$, and 27.3 $\pm$ 5.7~days for \Ksix, \Kfour, and \Kzero,\ respectively. Given the amplitude of the variations and their period, we conclude that those peaks are likely due to spots covering the stellar surface. The evolution of the spots, which vary in number and position with time, may scramble the periodicity of the signal, so we divided the light curves in chunks of the same duration, chosen to span over about five expected rotation period, and we again computed the Lomb-Scargle in each one. For each light curve, the resulting mean periodograms were plotted in Fig.~\ref{LS} (middle panels). We also give the autocorrelation functions computed on the whole light curve (Fig.~\ref{LS}, bottom panels).

The main period and its harmonics are clearly detectable for the three targets. All methods are in good agreement, and we estimate stellar rotational periods of 7.95 $\pm$ 0.90, 18.09 $\pm$ 0.10, and 28.24 $\pm$ 0.23~days for \Ksix, \Kfour, and \Kzero,\ respectively. Finally, owing to the high level of noise at high frequency, we cannot detect any evidence of pulsations in the expected range for those stars \citep{2011A&A...530A.142B}.

\subsection{Radial velocities with SOPHIE}\label{rvsec}

\begin{table}
  \caption{SOPHIE radial velocity measurements.}
\begin{tabular}{lccrrr}
\hline
\hline
BJD & RV & $\pm$$1-\sigma$ & bisect.$^\ast$  & exp. & S/N$^\dagger$  \\
-2\,400\,000 & [km\,s$^{-1}$] & [km\,s$^{-1}$] & [km\,s$^{-1}$] & [s] &   \\
\hline
\multicolumn{3}{l}{\textbf{\hspace{0.7cm}\Ksix}}  \\
55998.6935$^\ddagger$&  -1.010  & 0.089&        -0.195& 2700&   16.5\\
56011.6479&     -1.494  & 0.061&        0.062&  2678&   16.4\\
56013.6259&     -1.263  & 0.062&        0.269&  2685&   13.7\\
56037.5652&     -1.486  & 0.060&        0.024&  3600&   13.5\\
56099.5792&     -1.138  & 0.043&        -0.058& 3600&   19.0\\
56100.5893&     -1.180  & 0.061&        0.351&  2146&   12.4\\
56103.4277&     -1.538  & 0.043&        -0.252& 3600&   18.9\\
56111.5394$^\ddagger$&  -1.387  & 0.082&        0.029&  3600&   20.6\\
56125.5699&     -1.420  & 0.047&        0.126&  3600&   16.3\\
56153.5630&     -1.127  & 0.071&        0.301&  2703&   14.7\\
56158.5576&     -1.285  & 0.053&        -0.015& 2599&   17.8\\
56159.4445&     -1.395  & 0.053&        -0.113& 3600&   18.7\\
\hline
\multicolumn{3}{l}{\textbf{\hspace{0.7cm}\Kfour}}  \\
55830.4663&     -51.355 &0.065& -0.008& 900&    7.1\\
55975.7222$^\ddagger$&  -51.096 &0.093& -0.392& 779&    5.1\\
55977.7202$^\ddagger$&  -50.994 &0.083& -0.189& 1126&   5.9\\   
56038.6270&     -51.252 &0.041& -0.023& 1800&   9.4\\
56042.6242&     -50.933 &0.037& -0.076& 1800&   12.5\\
56086.5856&     -51.350 &0.034& 0.078&  1800&   12.9\\
56157.4901&     -50.955 &0.021& 0.073&  1800&   14.7\\
56160.5560&     -50.951 &0.029& 0.097&  1800&   11.4\\
56161.4765&     -51.088 &0.031& 0.081&  1800&   10.7\\
56162.5625&     -51.206 &0.042& -0.013& 1800&   11.8\\
56163.5253&     -51.313 &0.069& 0.074&  1800&   6.7\\
\hline
\multicolumn{3}{l}{\textbf{\hspace{0.7cm}\Kzero}}  \\
55765.5358$^\ddagger$  &  -26.361  &  0.0238  &  -0.0086 &  3600 & 20.99\\
55777.5181  &  -26.436  &  0.0241  &  -0.0087 &  2400 & 20.62\\
55801.5416  &  -26.392  &  0.0257  &  0.0232  &  3600 & 19.54\\
55802.4406  &  -26.424  &  0.0173  &  0.0673  &  3600 & 29.18\\
55804.4901  &  -26.451  &  0.0205  &  0.0398  &  2549 & 22.90\\
55809.4062  &  -26.330  &  0.0305  &  0.1083  &  2587 & 16.79\\
55831.3506  &  -26.426  &  0.0170  &  0.0673  &  3600 & 28.79\\
55832.3516  &  -26.347  &  0.0182  &  0.0489  &  3600 & 26.47\\
55833.3257  &  -26.312  &  0.0215  &  0.0036  &  3600 & 24.20\\
56534.4898  &  -26.358  &  0.0177  &  -0.0426 &  3600 & 26.90\\
56538.4772  &  -26.257  &  0.0176  &  -0.0216 &  3600 & 29.21\\
56552.4366$^\ddagger$  &  -26.380  &  0.0313  &  -0.0718 &  3600 & 20.63\\
\hline
\multicolumn{6}{l}{$\ast$: bisector spans; associated error bars are twice those of RVs.} \\ 
\multicolumn{6}{l}{$\dagger$: signal-to-noise ratio per pixel at 550~nm.} \\
\multicolumn{6}{l}{$\ddagger$: measurements corrected for moonlight pollution.} \\
  \label{table_rv}
\end{tabular}
\end{table}

\begin{figure*}
\begin{center}
\hspace{-0.3cm}
\includegraphics[width=6.1cm]{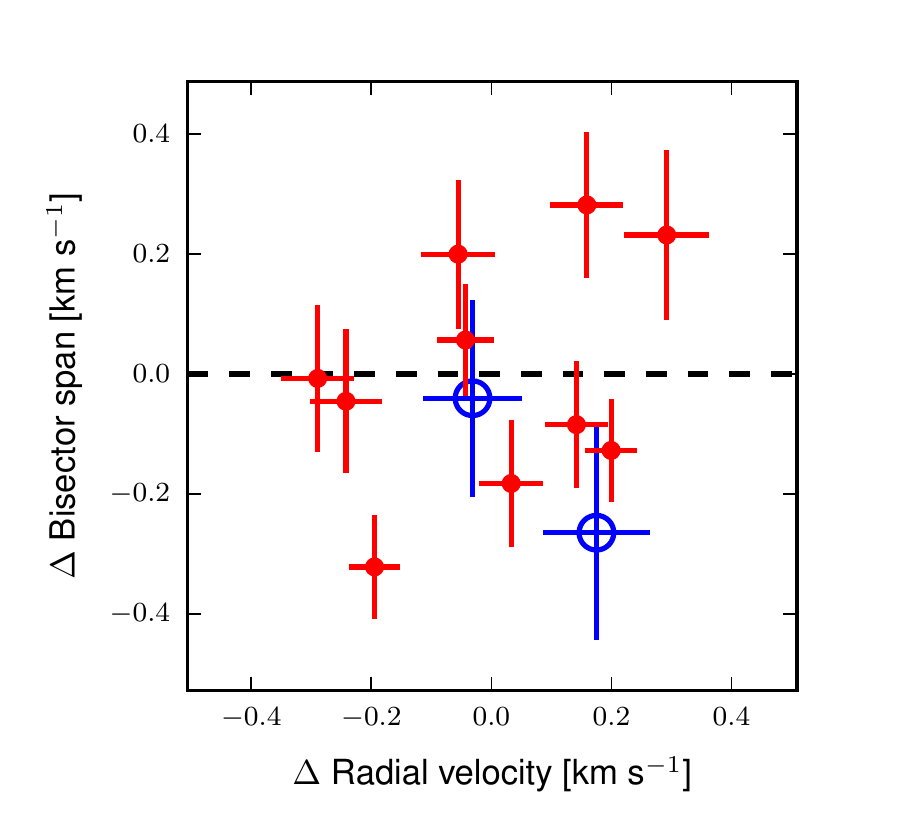}
\includegraphics[width=6.1cm]{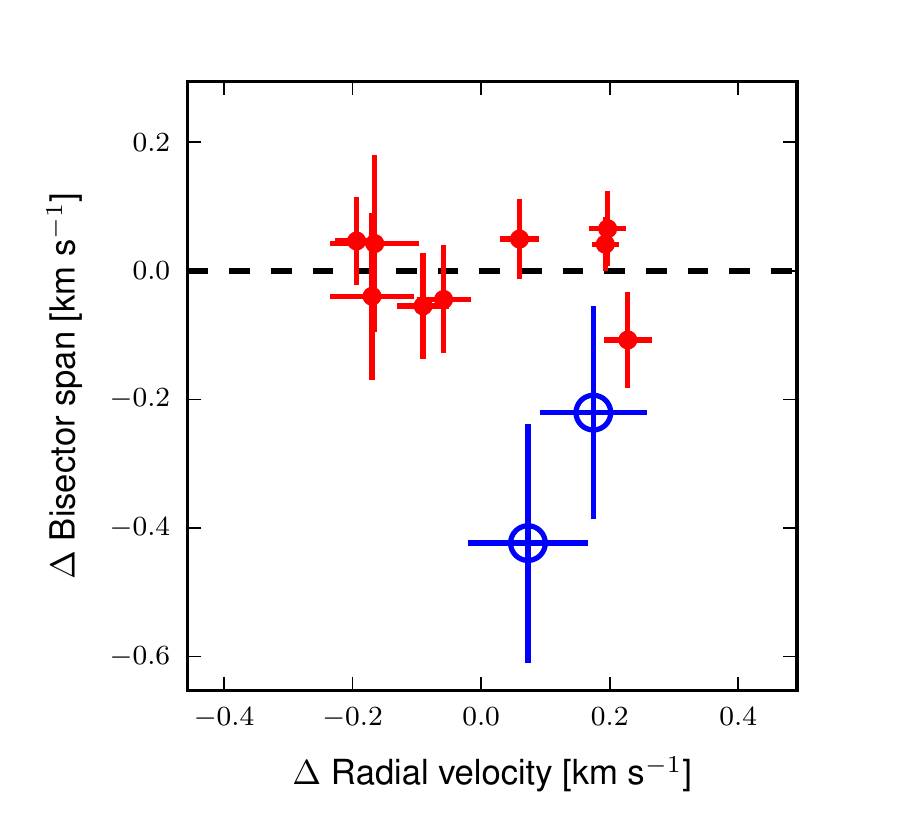}
\includegraphics[width=6.1cm]{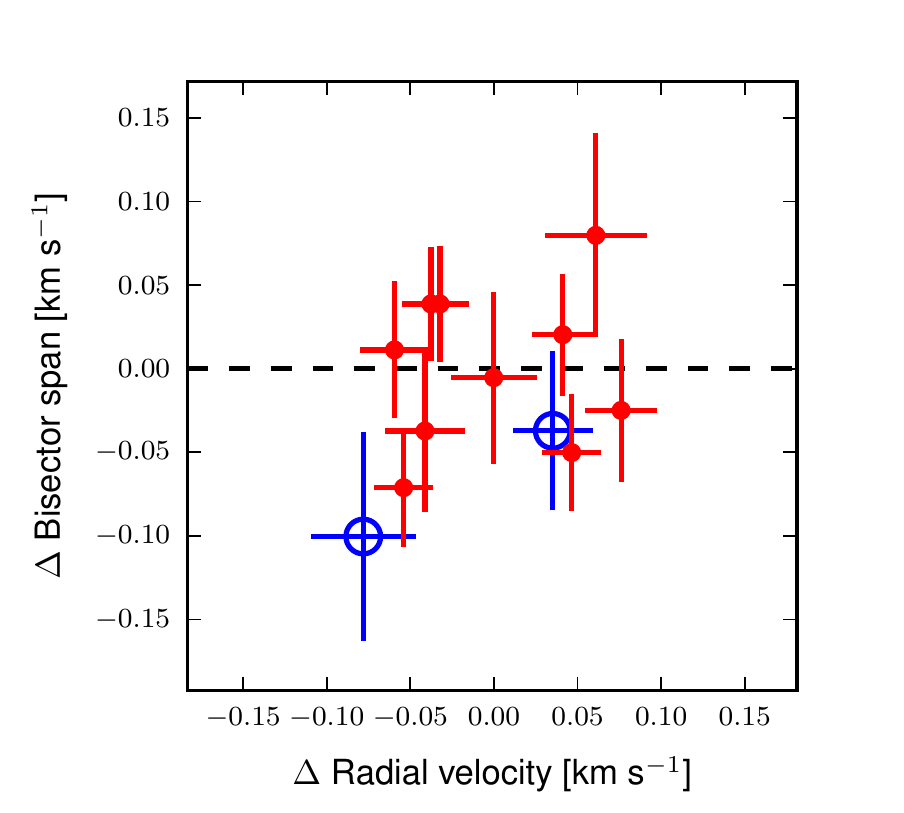}
\caption{Bisector span as a function of the detrended radial velocities with 1-$\sigma$ error bars
for \Ksix, \Kfour, and \Kzero\ (from left to right). Points affected and corrected by Moon contamination are indicated in blue. The ranges here have the same extents in the $x$- and $y$-axes.
}
\label{fig_bisectors}
\end{center}
\end{figure*}

The spectroscopic follow-up of the three targets was performed with the SOPHIE spectrograph \citep{2008SPIE.7014E..17P} after its upgrade made in 2011 \citep{2013A&A...549A..49B}. The three \Kepler\ targets were observed from July 2011 to August 2012 with a few additional measurements made in September 2013 for \Kzero. To increase the throughput for these faint targets, the high-efficiency mode of SOPHIE was used (resolving power $R\sim40\,000$), as well as the slow read-out mode of the detector. The exposure times ranged from 15 to 60\,min depending on the targets and the weather conditions, in order to keep the signal-to-noise ratio as constant as possible for a given target. While the first fiber pointed at the target, the sky background was acquired using the second fiber aperture of SOPHIE pointed at $\sim$2~arcmin from the target.

The spectra were extracted using the SOPHIE pipeline \citep{2009A&A...505..853B}, which includes localization of the spectral orders on the 2D-images, optimal order extraction,  cosmic-ray rejection, wavelength calibration, and corrections of flat field. They were corrected for the spectrograph drifts, measured from thorium calibration exposures every two to three hours during the night. Radial velocities were obtained from a Gaussian fit to the weighted cross-correlation function (CCF) with F0-type numerical masks for the three targets \citep{1996A&AS..119..373B,2002A&A...388..632P} and are reported in Table~\ref{table_rv}. Some spectra were contaminated by moonlight and were corrected following the method described in \citet{2008MNRAS.385.1576P} and \citet{2008A&A...488..763H} using the background sky obtained with the second fiber. The corrected exposures are identified in Table~\ref{table_rv}. Radial velocities were also corrected using the RV reference star HD185144 \citep[][]{2010Sci...330..653H,2013A&A...549A..49B,2014A&A...571A..37S} which was systematically observed the same nights of the three \Kepler\ targets. 

The RV dispersion over the two years of measurements on this constant star is about 10 m/s but with some variations at the level of 20-25 m/s over a few days that seem to be correlated with the dome temperature. The correction by this constant star slightly improve the residual of the RVs, but without any significant change of the parameters. The error bars on the radial velocities were computed from the CCF using the method presented by \citet{2010A&A...523A..88B}. For faint targets such as these, some spectral domains are too noisy and degrade the accuracy of the radial-velocity measurement, therefore the number of spectral orders used in the CCF was adjusted to decrease the dispersion of the measurements.

The radial velocity measurements are displayed in the lower plots of Fig.~\ref{solplot}, together with their best Keplerian fits and the residuals. They show variations in phase with the transit ephemeris derived from \Kepler\ and with semi-amplitudes implying a companion in the planet-mass range. 

For \Kfourb,\ no significant linear drift, mask effect, or correlation between the bissector span (Fig.~\ref{fig_bisectors}, middle panel) were detected, reinforcing the conclusion that the radial-velocity variations are not caused by spectral-line profile changes attributable to blends or stellar activity. On the other hand, the radial velocities of \Ksixb\ and \Kzerob\ show a significant linear drift of $-500\pm200$ and $43.9\pm9.2$~\ms~yr$^{-1}$, suggesting the presence of a distant companion in those two systems of minimum mass of about $4.3\pm1.6$ and $2.9\pm0.6$~M$_{\rm Jup}$ respectively. 

The nature of these companions is difficult to assess with the available data. Using different stellar masks to compute the CCF produced no significant effect in both cases. After removing for each target the two measurements affected by the moonlight, the hypothesis of a null Spearman\rq{}s rank correlation coefficient between the bisector span and radial velocities cannot be rejected at the 5\% level for both targets, supporting the hypothesis that the companion has no detectable effect on the CCF. We find the same result when the long-term drift is subtracted from the radial velocities as can be seen in Fig.~\ref{fig_bisectors} (left and right panels), where the CCF bisector spans show neither variations nor trends as a function of radial velocity. Neither is there a significative correlation between the radial-velocity long-term drift and the CCF bisector spans (Spearman\rq{}s rank correlation p-value of 0.49 for \Ksix\ and 0.32 for \Kzero).

The flux of identified (separations $\gtrsim2\arcsec$) contaminant stars in the \Kepler\ photometric mask is estimated for each quarter, and the light curves are corrected. There is still the possibility that the light curve and the CCF are contaminated by stars closer than $\sim2\arcsec$. This contaminant star can be physically related to the target or by chance close to the line of sight. We estimated the probability of the latter being 1.9, 1.8, and 2.0\% for \Ksix, \Kfour, and \Kzero,\ respectively, using the Besan\c{c}on model \citep{2003A&A...409..523R}. The magnitude of the stars in the simulation is limited according to the observed transit depth, and its distance to the target to less than~$2\arcsec$. We try to constrain the contaminant adding synthetic spectra to the co-added SOPHIE spectrum. The synthetic secondary CCF peak is detectable for early than K0V-K3V stars at the same distance of the target (or equivalent flux ratio for an unrelated contaminating star), if this CCF peak is well detached from the main one in radial velocity. That translated into an upper limit for the possible flux contamination in the light curve and CCF of $\sim$ 5\% for \Ksix, $\sim$ 10\% for \Kfour, and $\sim$ 3\% for \Kzero. 

\begin{figure*}
\includegraphics[width=8.5cm]{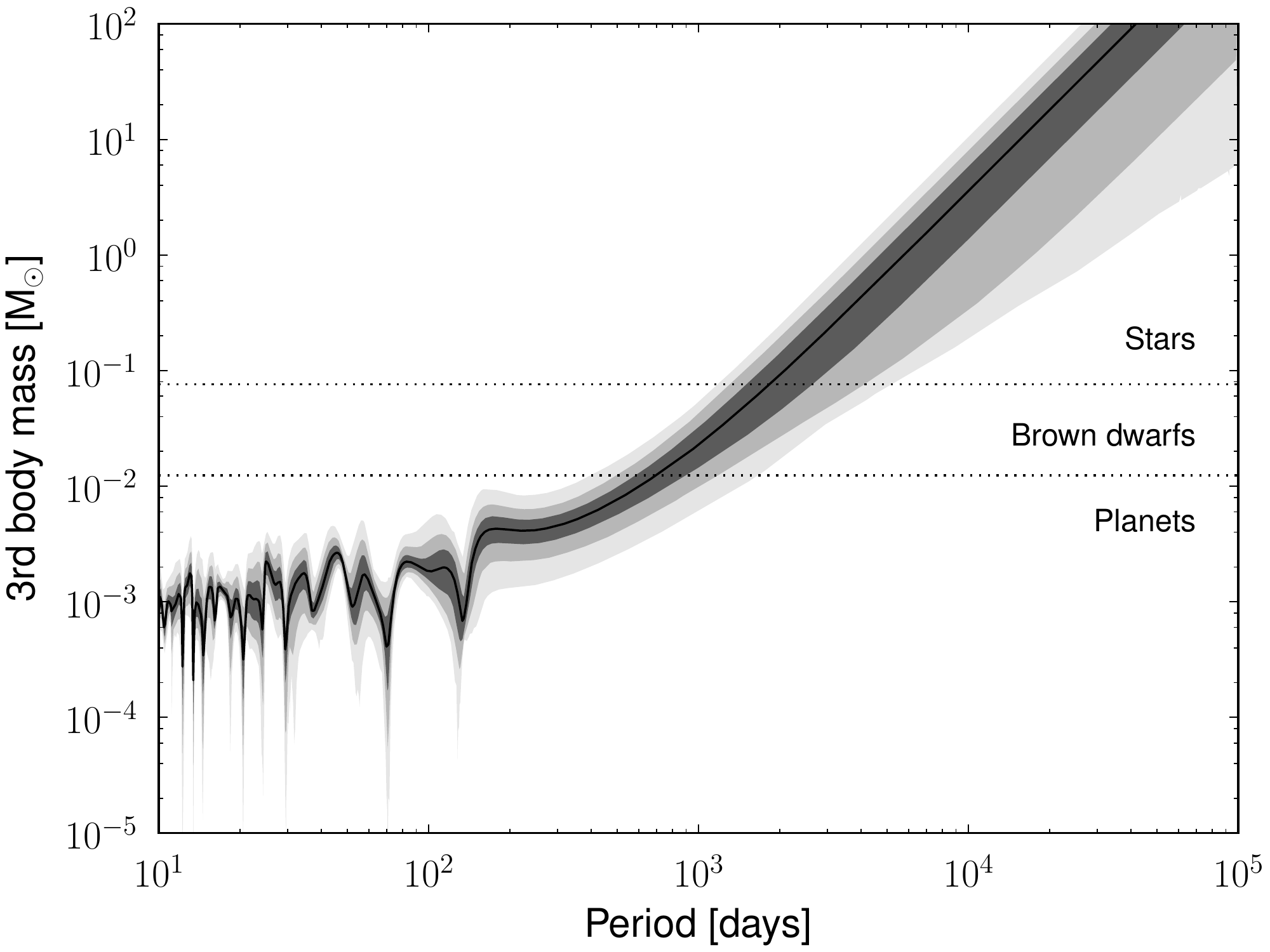}
\hspace{1cm}\includegraphics[width=8.5cm]{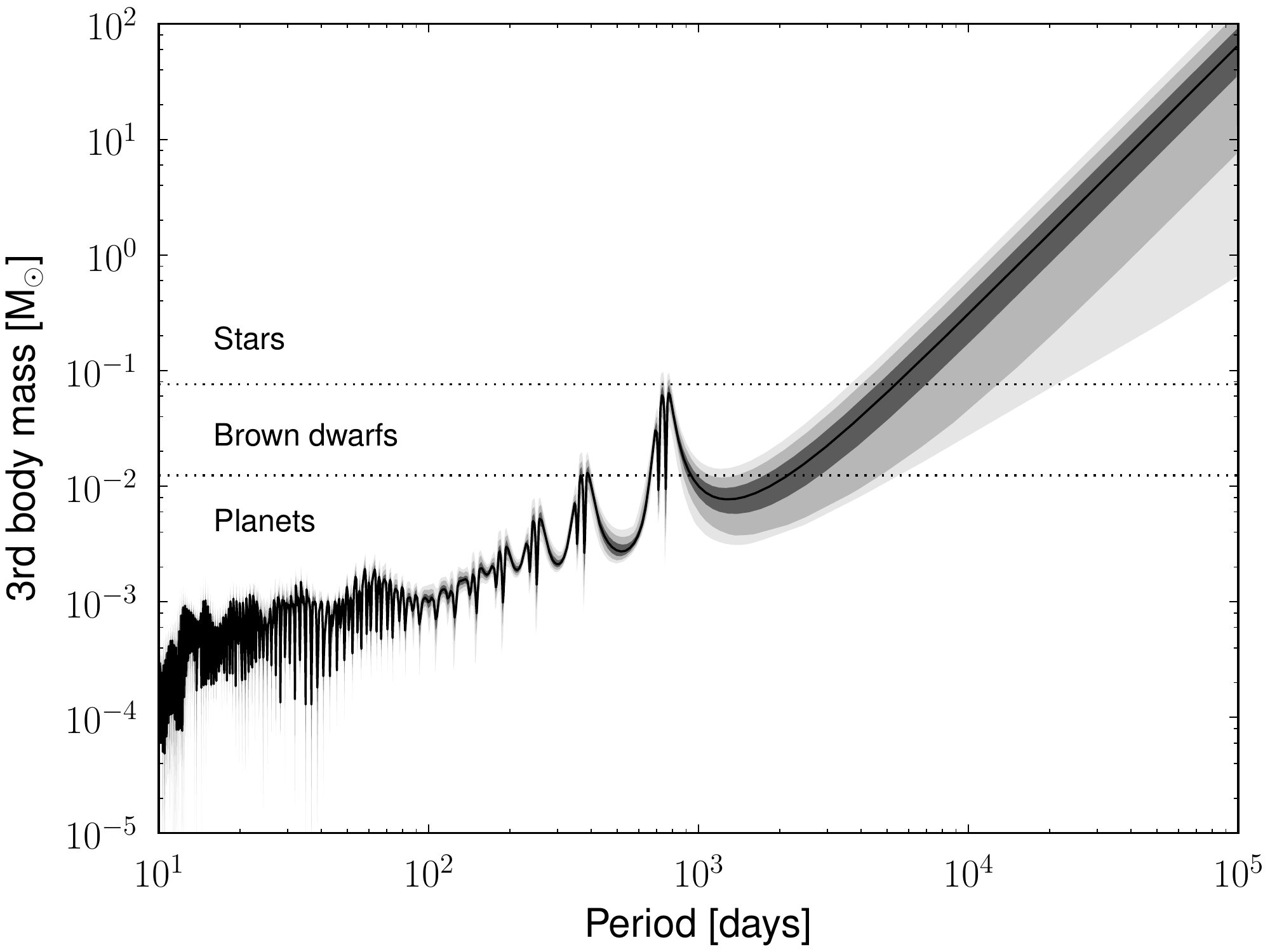}
\caption{Third-body mass allowed by the RV drift for a given period assuming circular orbits for \Ksix\ (left) and \Kzero\ (right). The median of the distribution is represented by a solid line, and the 1-, 2-, and 3-$\sigma$ confidence intervals are represented with three different gray levels. The dotted lines separate the planetary, brown dwarf, and star regimes.}
\label{rvdrift}
\end{figure*}

Assuming the RV drift observed for \Ksix\ and \Kzero\ is caused by a companion in a circular orbit, we estimated the allowed masses at each period fitting the RV residuals after having subtracted the Keplerian (Section~\ref{pastis}). The result is presented in Fig.~\ref{rvdrift}. We estimated the probability that the companion allowed by the drift is stellar in 34\% for \Ksix\ and 31\% for \Kzero\ using the statistics of \citet{2010ApJS..190....1R}.
As a conclusion, for the three targets, the radial-velocity variations detected at the \Kepler\ period are compatible with a companion of a planetary nature, and in the case of for \Ksixb\ and \Kzerob,\ there is no evidence that their estimated parameters are significantly affected by the possible additional companions.

\section{Spectroscopic parameters}\label{specan}

For \Ksix\ and \Kzero\ we used the SOPHIE spectra described in Section~\ref{rvsec}. For \Kfour\ we used spectra obtained with the ESPaDOnS spectrograph. The target was observed in `object+sky' mode, with a spectral resolution $\lambda/\Delta \lambda \sim 65000$ (Program 12AF95). The six observations ran from June 30 to July 5, 2012, with exposure times ranging from 3515~s for the first one to 1715~s for the five others. Each spectrum was corrected for its radial velocity shift, and then the respective spectra were co-added. 

The final spectrum has a signal-to-noise ratio per element of resolution at 5500\AA\ of 50, 90, and 250 for \Kfour, \Ksix, and \Kzero,\ respectively.
We derived the effective temperature T$_{\mathrm{eff}}$, surface gravity $\log g$, metallicity [Fe/H], and microturbulence $v_\mathrm{micro}$ of the stars with the VWA package \citep{2004A&A...425..683B,2008A&A...478..487B,2010A&A...519A..51B}. We used a grid of MARCS models \citep{2008A&A...486..951G} to reach the ionization equilibrium over a large selection of spectral lines. Then, we fitted a rotational profile on a set of isolated metallic lines, solving for the projected rotational velocity \vsini\ and the macroturbulence $v_\mathrm{macro}$. For \Ksix\ and \Kzero, instead, the \vsini\ was measured through the Gaussian width of the cross-correlation function (CCF) \citep[and references therein]{2010A&A...523A..88B}, because of the paucity of isolated lines to be fitted with VWA.\\
The spectrum of \Kfour\ did not present any particular problem, and the relative atmospheric parameters are presented in Table~\ref{tableparam}. For \Ksix, we found a considerably high value for $v_\mathrm{micro}$ ($2.8 \pm 0.8 \, \mathrm{km \, s}^{-1}$). The large uncertainty can be influenced by its degeneracy with $v_\mathrm{macro}$, which is problematic for constraining a spectrum with a low signal-to-noise ratio. However, the measurement favors a value higher than $2 \, \mathrm{km \, s}^{-1}$, indicating an evolved object \citep[e.g.,][]{2005oasp.book.....G}. We imposed different values for $v_\mathrm{macro}$ (between 5.0 and 7.5 $\mathrm{km \, s}^{-1}$) and fitted all the other atmospheric parameters. In this way, we verified that the latter are not significantly affected. Eventually, we measured \vsini$ = 11 \pm 1 \, \mathrm{km \, s}^{-1}$ with the CCF. In conclusion, the ionization balance required $T_{\mathrm{eff}} = 6340 \pm 140$ K, $\log g = 4.0 \pm 0.3$ dex, and [Fe/H]$= 0.06 \pm 0.19$ dex.\\
For \Kzero, the analysis yielded $T_{\mathrm{eff}} = 6090 \pm 110$ K, $\log g = 3.5 \pm 0.1$ dex, $\mathrm{[Fe/H]}= -0.17 \pm 0.10$ dex, and $v_\mathrm{micro} = 1.3 \pm 0.2 \, \mathrm{km \, s}^{-1}$. Because of the loose constraint we obtained on $v_\mathrm{macro}$ with VWA, we chose to fix it through the T$_{\mathrm{eff}}$, according to the relation in \citet{2010MNRAS.405.1907B}, to $3.6 \, \mathrm{km \, s}^{-1}$. Finally, from the CCF we measured \vsini$ = 6 \pm 1 \, \mathrm{km \, s}^{-1}$.\\
We used the SME package from \cite{1996A&AS..118..595V} and \cite{2005ApJS..159..141V} to check the reliability of the results for \Ksix\ and \Kzero. We verified $T_{\mathrm{eff}}$ on H$_\alpha$, the $\log g$ on the pressure-sensitive lines Ca$\lambda$6122, Ca$\lambda$6162, and Ca$\lambda6439,$ and on the MgIb triplet. Fixing the $T_{\mathrm{eff}}, \, \log g$, and [Fe/H] on \Ksix, we fitted $v_\mathrm{macro} \simeq 7 \, \mathrm{km \, s}^{-1}$ and $v_\mathrm{micro} \simeq 2.8 \, \mathrm{km \, s}^{-1}$. 

\section{Modeling with PASTIS}\label{pastis}

\begin{table*}
\centering
\caption{\Ksix, \Kfour, and \Kzero\ system parameters.}            
\begin{minipage}[t]{\textwidth} 
\setlength{\tabcolsep}{3.0mm}
\renewcommand{\footnoterule}{}                          
\begin{tabular}{l c c c}        
\hline\hline                 
Object                       & Kepler-433      & Kepler-434       & Kepler-435      \\
\Kepler\ object of interest & KOI-206          & KOI-614          & KOI-680          \\
\Kepler\ input catalog      & KIC 5728139      & KIC 7368664      & KIC 7529266      \\
USNO-A2 ID                & 1309-0358523     & 1329-0421364     & 1331-0387694     \\
2MASS ID                  & 19502247+4058381 & 19342073+4255440 & 19290895+4311502 \smallskip\\

\multicolumn{4}{l}{\hspace{-0.4cm} Coordinates} \\
\hline    
RA (J2000) [hh:mm:ss.sss] &  19:50:22.476 &  19:34:20.729 &  19:29:08.959 \\
Dec (J2000) [dd:mm:ss.ss] & +40:58:38.17  & +42:55:44.08  & +43:11:50.21  \smallskip\\

\multicolumn{4}{l}{\hspace{-0.4cm} Magnitudes} \\ 
\hline
Kepmag$^\S$ (AB)   & 14.463     & 14.517     & 13.643     \\
d51mag$^\S$ (AB)   & 14.745     & 14.766     & 13.830     \\
g' Sloan-Gunn (AB) & 14.953     & 14.970     & 14.013     \\
r' Sloan-Gunn (AB) & 14.413     & 14.445     & 13.610     \\
i' Sloan-Gunn (AB) & 14.253     & 14.322     & 13.485     \\
z' Sloan-Gunn (AB) & 14.166     & 14.277     & 13.448     \\
J 2MASS (Vega)     & 13.201 $\pm$ 0.020 & 13.400 $\pm$ 0.022 & 12.593 $\pm$ 0.020 \\
H 2MASS (Vega)     & 12.948 $\pm$ 0.024 & 13.035 $\pm$ 0.031 & 12.345 $\pm$ 0.019 \\
Ks 2MASS (Vega)    & 12.826 $\pm$ 0.026 & 13.039 $\pm$ 0.030 & 12.291 $\pm$ 0.022 \\
WISE W1 (Vega)     & 12.716 $\pm$ 0.025 &                   & 12.261 $\pm$ 0.023 \\
WISE W2 (Vega)     & 12.803 $\pm$ 0.032 &                   & 12.299 $\pm$ 0.024 \\
WISE W3 (Vega)     & 12.961 $\pm$ 0.531 &                   &                   \smallskip\\

\multicolumn{4}{l}{\hspace{-0.4cm} Spectroscopic parameters} \\
\hline
Effective temperature, $T_{\mathrm{eff}}$[K]   & 6340 $\pm$ 140  & 5970 $\pm$ 100   & 6090 $\pm$ 110  \\
Surface gravity, log\,$g$ [cgs]              & 4.0 $\pm$ 0.30  & 4.22 $\pm$ 0.10  &  3.5 $\pm$ 0.1 \\
Metallicity, $[\rm{Fe/H}]$ [dex]             & 0.06 $\pm$ 0.19 & 0.35 $\pm$ 0.15  & -0.17 $\pm$ 0.10 \\
Stellar rotational velocity, {\vsini} [\kms] & 11 $\pm$ 1     &  3.0 $\pm$ 1.0   &     6 $\pm$ 1 \\
Microturbulent velocity, $v_{micro}$ [\kms]   & 2.8 $\pm$ 0.8  &  1.0 $\pm$ 0.2   &   1.3 $\pm$ 0.2 \\
Macroturbulent velocity, $v_{macro}$ [\kms]   & 5 $\pm$ 1      &  2.5 $\pm$ 1.0   &    3.6 (fixed) \smallskip\\

\multicolumn{4}{l}{\hspace{-0.4cm} Result from light curve, radial velocity, and SED combined analysis} \\
\hline
Planet orbital period, $P$ [days]$^{\bullet}$       & 5.33408384 $\pm$ 1.1$\e{-6}$  &  12.8747099 $\pm$ 5.0$\e{-6}$ & 8.6001536 $\pm$ 1.8$\e{-6}$ \\
Mid-transit time, $T_{c}$ [BJD]$^{\bullet}$          & 2454964.98152 $\pm$ 1.5$\e{-4}$ &  2455003.02283 $\pm$ 4.2$\e{-4}$ & 2455010.64241 $\pm$ 4.0$\e{-4}$ \\
$cov(P,T_{c})$ [days$^2$]                           & -1.25$\e{-10}$                   &  -1.07$\e{-9}$                & -2.20$\e{-10}$         \\
Orbital eccentricity, $e$$^{\bullet}$               & 0.119 $\pm$ 0.079,$\;$ $\textless$ 0.29$^{\dagger}$ &  0.131 $\pm$ 0.072,$\;$ $\textless$ 0.32$^{\dagger}$ & 0.114 $\pm$ 0.077,$\;$ $\textless$ 0.30$^{\dagger}$  \\
Argument of periastron, $\omega$ [deg]$^{\bullet}$  & 68$^{+67}_{-36}$              &  82 $\pm$ 33              &  104 $\pm$ 36        \\
Orbit inclination, $i$ [deg]$^{\bullet}$            & 89.21$^{+0.52}_{-0.90}$       &  86.46$^{+0.38}_{-0.74}$    &  85.51 $\pm$ 0.52    \\
Orbital semi-major axis, $a$ [AU]                  & 0.0679 $\pm$ 0.0027         &  0.1143 $\pm$ 0.0030      &  0.0948 $\pm$ 0.0018   \\
semi-major axis / radius of the star, $a/R_{\star}$ & 6.44 $\pm$ 0.62             & 17.9 $\pm$ 1.6            & 6.35 $\pm$ 0.51  \smallskip \\

Radius ratio, $k=R_{p}/R_{\star}$$^{\bullet}$            & 0.06590 $\pm$ 0.00015 & 0.0835$^{+0.014}_{-0.0080}$ & 0.06384 $\pm$ 0.00020 \\
Linear limb darkening coefficient, $u_a$$^{\bullet}$    & 0.325 $\pm$ 0.022     & 0.00 $\pm$ 0.43          & 0.374 $\pm$ 0.024 \\
Quadratic limb darkening coefficient, $u_b$$^{\bullet}$ & 0.220 $\pm$ 0.045     & 0.01 $\pm$ 0.49          & 0.180 $\pm$ 0.042 \\
Transit duration [h]                         & 6.178 $\pm$ 0.015     & 2.288 $\pm$ 0.045        & 9.003 $^{+0.059}_{-0.036}$ \\
Impact parameter, $b$                                 & 0.082 $\pm$ 0.081     & 0.979 $\pm$ 0.031        & 0.448 $\pm$ 0.024 \smallskip \\

Systemic velocity, $V_{r}$ [\kms]$^{\bullet\;\ddag}$        & -1.284 $\pm$ 0.052 & -51.156 $\pm$ 0.028 & -26.3665 $\pm$ 0.0079 \\
Radial velocity linear drift [\ms yr$^{-1}$]$^{\bullet}$ & -500 $\pm$ 200     & 83 $\pm$ 69         &  43.9 $\pm$ 9.2 \\
Radial velocity semi-amplitude, $K$ [\ms]$^{\bullet}$     & 259 $\pm$ 44      & 221 $\pm$ 23        &  64 $\pm$ 12 \smallskip\\

Jitter \Kepler\ long cadence, $\sigma_{Kepler, \mathrm{LC}}$ [ppm] $^{\bullet}$  & 65 $\pm$ 16     & 36 $\pm$ 28            &  27.5$^{+9.3}_{-14}$  \\
Jitter \Kepler\ short cadence, $\sigma_{Kepler, \mathrm{SC}}$ [ppm] $^{\bullet}$ & 88 $\pm$ 58     &                        & 48 $\pm$ 32         \\
Jitter radial velocity, $\sigma_{\mathrm{RV}}$ [\ms]$^{\bullet}$              & 34 $\pm$ 32        & 16$^{+19}_{-12}$         & 10 $\pm$ 10 \\
Jitter SED, $\sigma_{\mathrm{SED}}$ [mag]$^{\bullet}$                         & 0.056 $\pm$ 0.024 & 0.017$^{+0.025}_{-0.014}$ & 0.019 $\pm$ 0.016 \smallskip\\

Effective temperature, $T_{\mathrm{eff}}$[K]$^{\bullet}$   & 6360 $\pm$ 140     & 5977 $\pm$ 95     & 6161 $\pm$ 94       \\
Metallicity, $[\rm{Fe/H}]$ [dex]$^{\bullet}$             & -0.01 $\pm$ 0.20   & 0.25 $\pm$ 0.14   & -0.18 $\pm$ 0.11     \\ 
Stellar density, $\rho_{\star}$ [$\rho_\odot$]$^{\bullet}$ & 0.126 $\pm$ 0.033  & 0.46 $\pm$ 0.11   & 0.047 $\pm$ 0.011    \\
Star mass, $M_\star$ [\Msun]                            & 1.46 $\pm$ 0.17     & 1.198 $\pm$ 0.093 & 1.538 $\pm$ 0.088      \\
Star radius, $R_\star$ [\Rsun]                          & 2.26 $\pm$ 0.25     & 1.38 $\pm$ 0.13   & 3.21 $\pm$ 0.30        \\
Deduced stellar surface gravity, $\log$\,$g$ [cgs]    & 3.892 $\pm$ 0.056   & 4.242 $\pm$ 0.055 & 3.613$^{+0.047}_{-0.070}$ \\
Age of the star [$Gyr$]                               & 2.67 $\pm$ 0.91     & 4.0 $\pm$ 1.7     & 2.25 $\pm$ 0.42        \\
Luminosity of the star, $\log(L/L_\odot)$ [dex]        & 0.88 $\pm$ 0.11     & 0.33 $\pm$ 0.11  & 1.129 $\pm$ 0.089       \\
Distance of the system [pc]$^{\bullet}$                 & 1870 $\pm$ 210     & 1240 $\pm$ 120     & 2070 $\pm$ 200        \\
Color excess, $E_{(B-V)}$ [mag]$^{\bullet}$               & 0.210 $\pm$ 0.039 & 0.049 $\pm$ 0.029  & 0.044 $\pm$ 0.025 \smallskip\\

Planet mass, $M_{p}$ [M$_{\rm Jup}$ ]                    & 2.82 $\pm$ 0.52  & 2.86 $\pm$ 0.35      & 0.84 $\pm$ 0.15 \\
Planet radius, $R_{p}$[R$_{\rm Jup}$]                    & 1.45 $\pm$ 0.16  & 1.13$^{+0.26}_{-0.18}$ & 1.99 $\pm$ 0.18 \\
Planet mean density, $\rho_{p}$ [$g\;cm^{-3}$]          & 1.13 $\pm$ 0.32  & 2.4 $\pm$ 1.3         & 0.131 $\pm$ 0.037 \\
Planet surface gravity, $\log$\,$g_{p}$ [cgs]          & 3.518$^{+0.071}_{-0.10}$ & 3.74$^{+0.13}_{-0.21}$ & 2.72 $\pm$ 0.11 \\
Planet equilibrium temperature$^\ast$, $T_{eq}$ [K]     & 1776 $\pm$ 87      &  1000 $\pm$ 45         & 1729 $\pm$ 70 \smallskip\\
\hline
\hline       
\vspace{-0.5cm}
\end{tabular}
\begin{list}{}{}
\item $^\S$ Not used in the SED fitting. $^{\bullet}$ MCMC jump parameter. $^{\dagger}$ upper limit, 99\% confidence. $^{\ddag}$ reference time BJD 2456000. 
 \item $^{\ast}$ $T_{eq}=T_{\mathrm{eff}}\left(1-A\right)^{1/4}\sqrt{\frac{R_\star}{2 a}}$, using an albedo $A=0$.
\item $\Msun = 1.98842\e{30}$~kg, \Rsun = 6.95508$\e{8}$~m, M$_{\rm Jup}$ = 1.89852$\e{27}$~kg, R$_{\rm Jup}$ = 7.1492$\e{7}$~m
\end{list}
\end{minipage}
\label{tableparam}  
\end{table*}

We use the Planet Analysis and Small Transit Investigation Software \citep[PASTIS,][]{2014MNRAS.441..983D}, that allows the light curve, the radial velocity, and the spectral energy distribution (SED) to be fit with a Markov Chain Monte Carlo (MCMC) algorithm. The photometric magnitudes used for the SED fit are listed in Table~\ref{tableparam}. Only the sections of the light curve around the transits are used, after normalization to a linear trend fitted outside the transit. A 3-$\sigma$ clipping is performed to eliminate outliers, and the measured flux is corrected for the contamination factors provided for each quarter at MAST archive. PASTIS uses the EBOP code \citep{1972ApJ...174..617N,1981psbs.conf..111E,1981AJ.....86..102P} extracted from the JKTEBOP package \citep{2011MNRAS.417.2166S} to model the photometric transits. Both long- and short-cadence (when available) are used, with an oversampling factor of 10 for the long-cadence data to account for the long integration time when comparing with the model. Radial velocity curves were simultaneously fitted with eccentric Keplerian orbit and a linear drift. We also account for additional sources of Gaussian noise in the light curves, radial velocities, and SED by fitting a jitter value to each data set. We used the stellar evolutionary tracks from {\sl STAREVOL} (Palacios, {\sl priv. com.}), Dartmouth \citep{2008ApJS..178...89D}, and PARSEC \citep{2012MNRAS.427..127B} to estimate the stellar parameters, and the PHOENIX/BT-Settl synthetic spectral library \citep{Allard} to model the SED. The interpolated spectrum is scaled to a given distance and corrected from interstellar extinction. The distance $d$ and the color excess $E_{(B-V)}$ were included as free parameters in our model. The model has therefore 20 jump parameters for \Kfourb, and two more for \Ksixb\ and \Kzerob\ that have short-cadence data, indicated in Table~\ref{tableparam}. We used non-informative priors (uniform or Jeffreys distributions) except for $T_{\mathrm{eff}}$, [Fe/H], and $\rho_\star$, for which we used spectroscopic estimations as normal and asymmetric normal distribution priors, and $P$ and $T_c$, for which we used the determination by \citet{2011ApJ...736...19B} as normal distribution priors, but the width of the distribution was increased by an order of magnitude to ensure that it would not bias the results. 

\begin{figure*}
\hspace{-0.25cm} \Ksix \hspace{5cm} \Kfour \hspace{5cm} \Kzero\
\centering
\includegraphics[width=6cm]{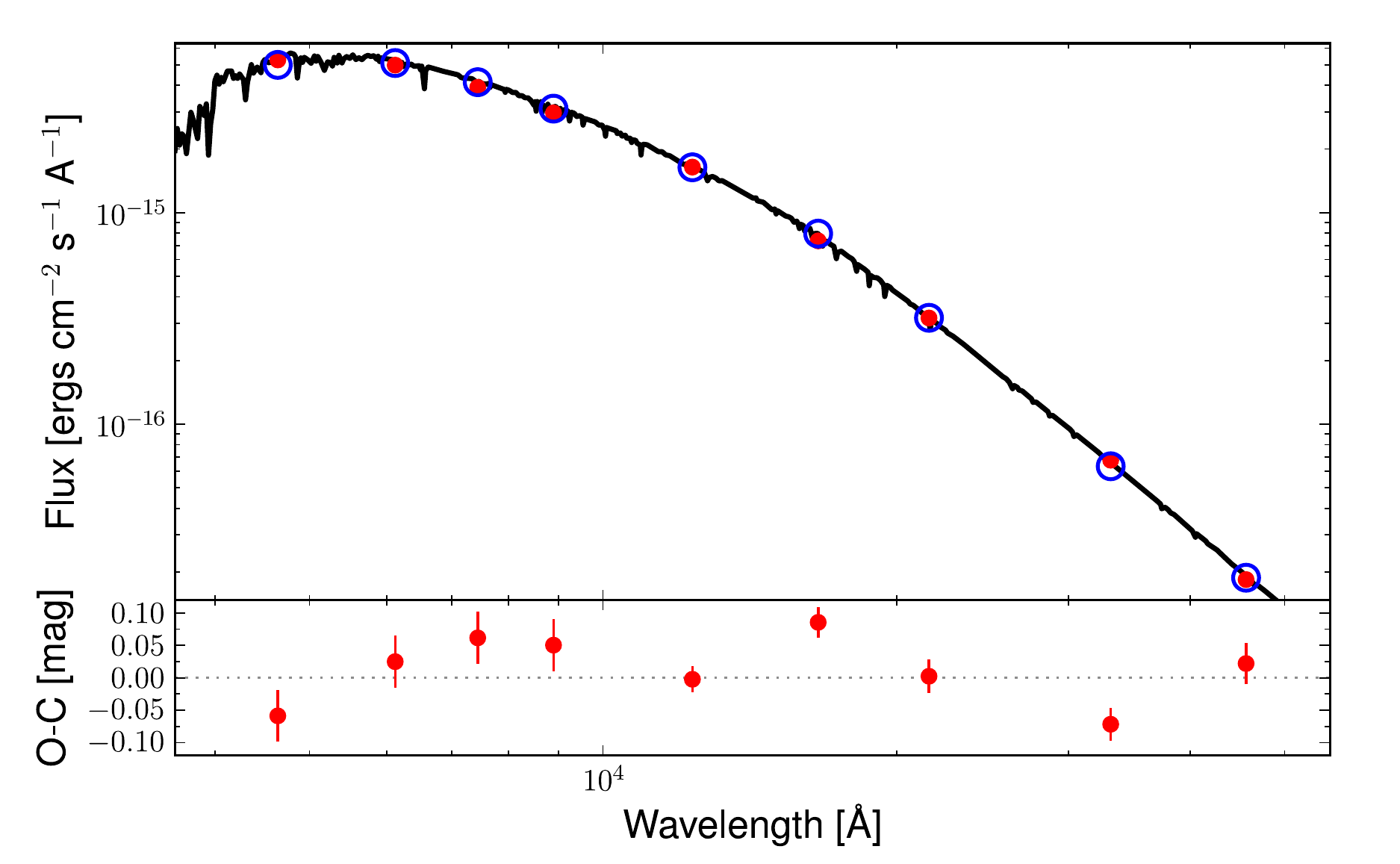}\includegraphics[width=6cm]{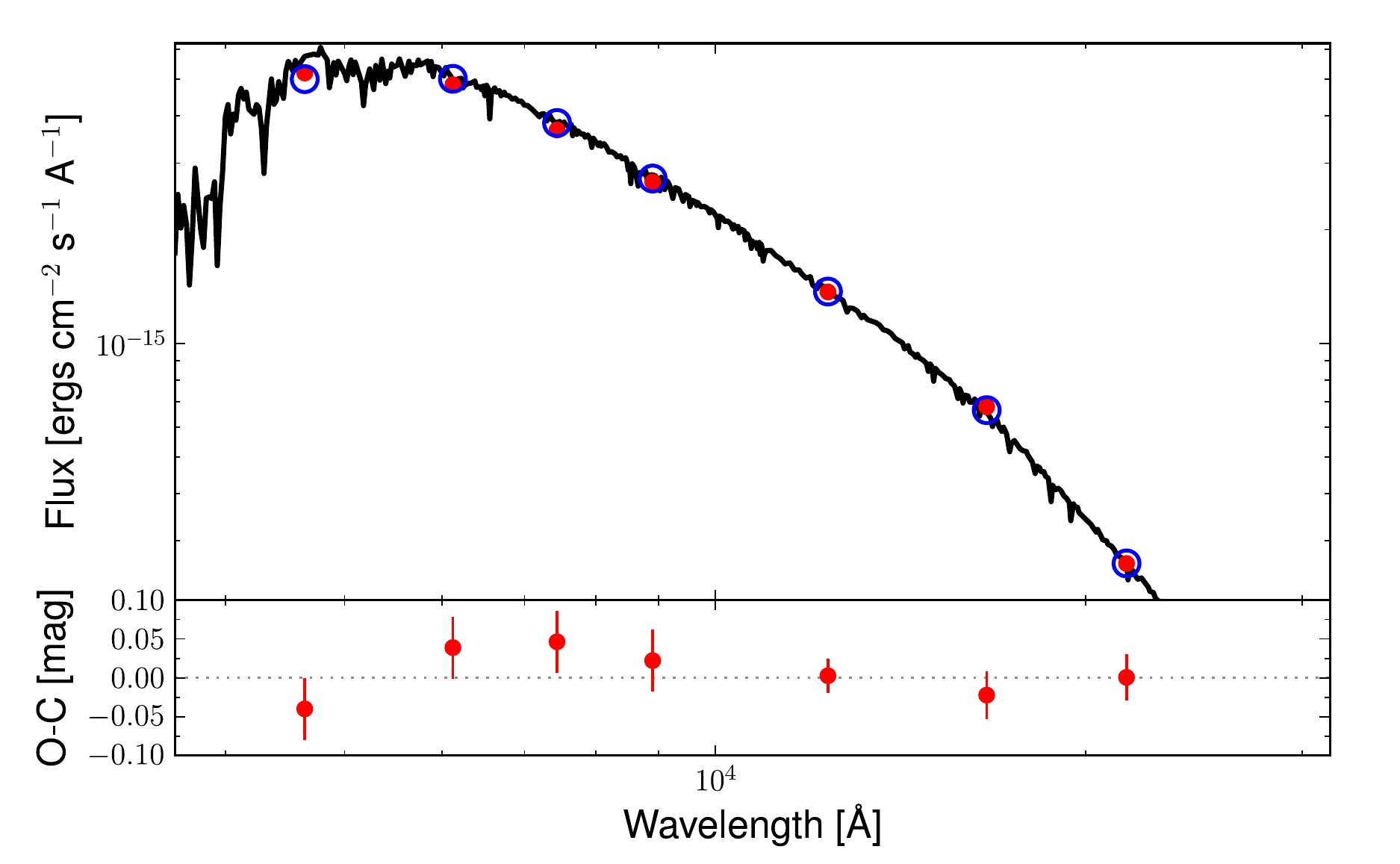}\includegraphics[width=6cm]{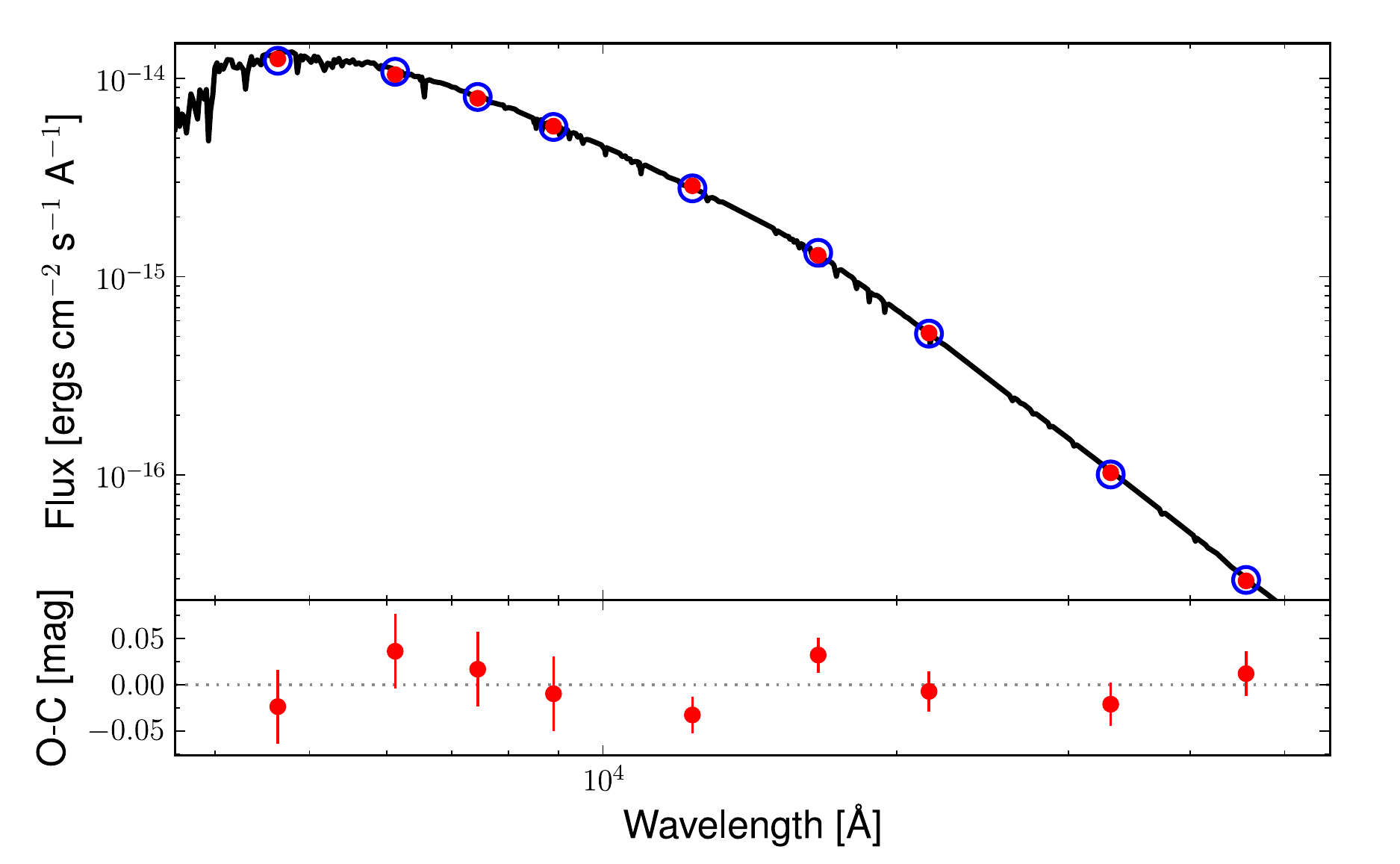}\\
\includegraphics[width=6cm]{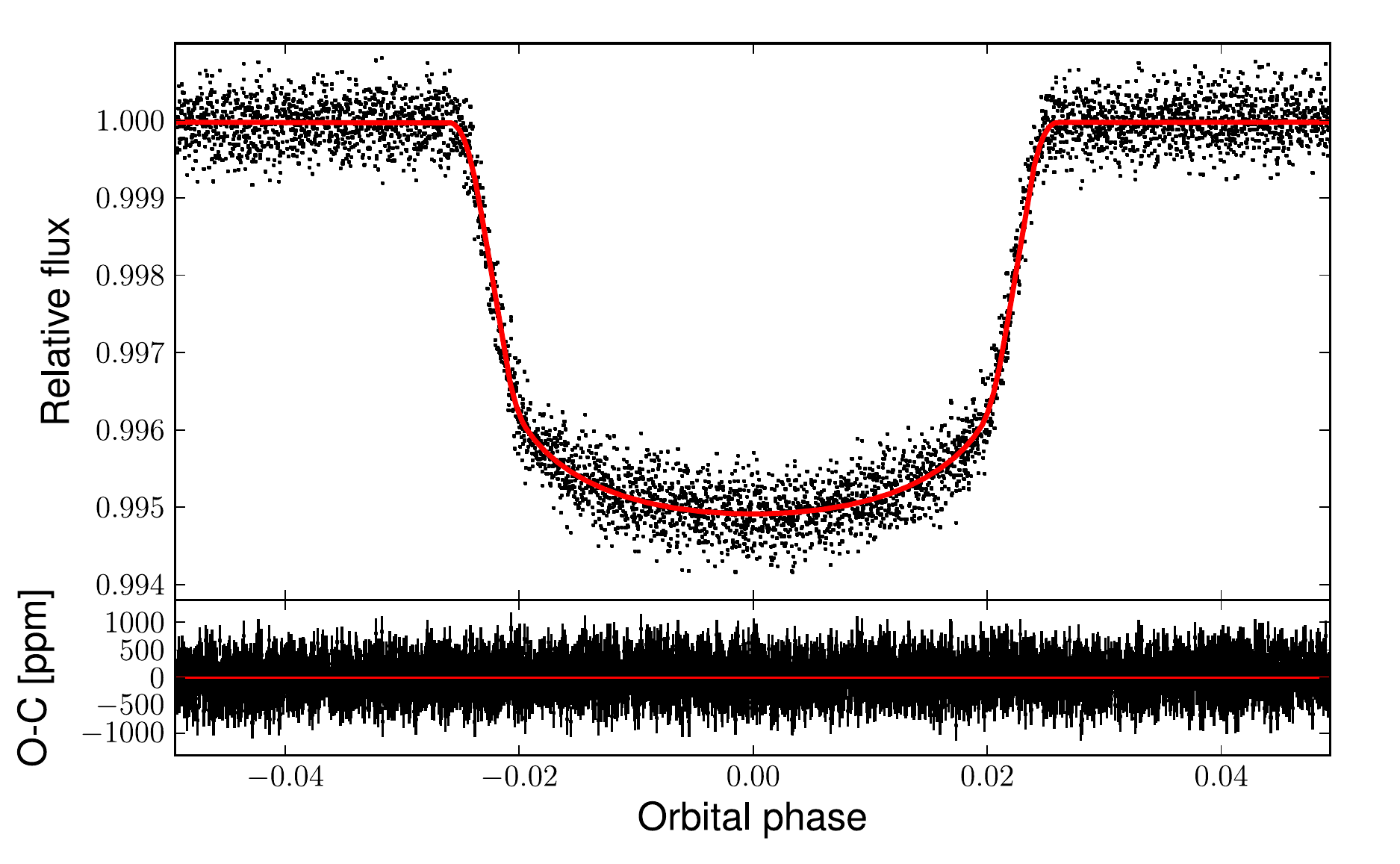}\includegraphics[width=6cm]{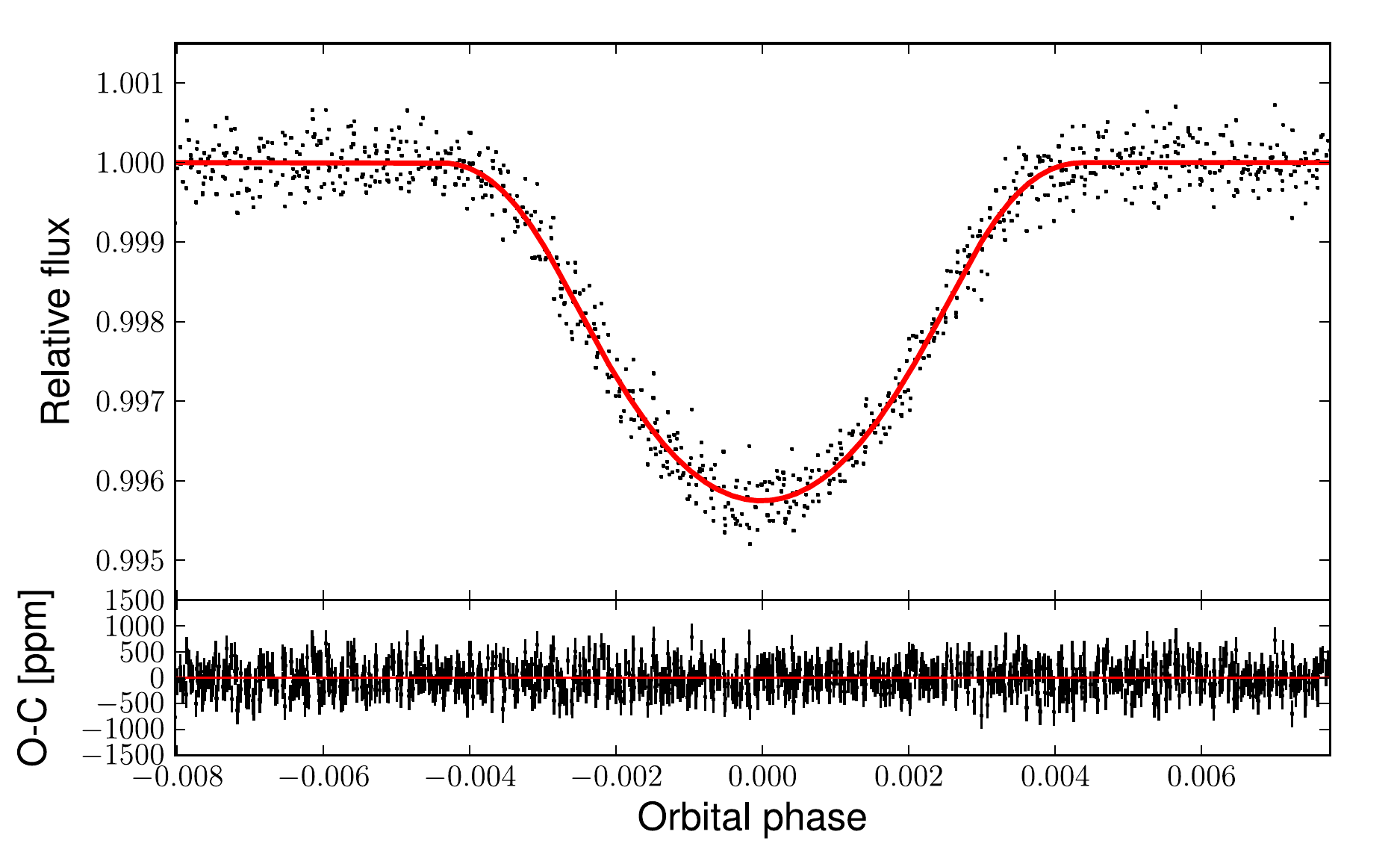}\includegraphics[width=6cm]{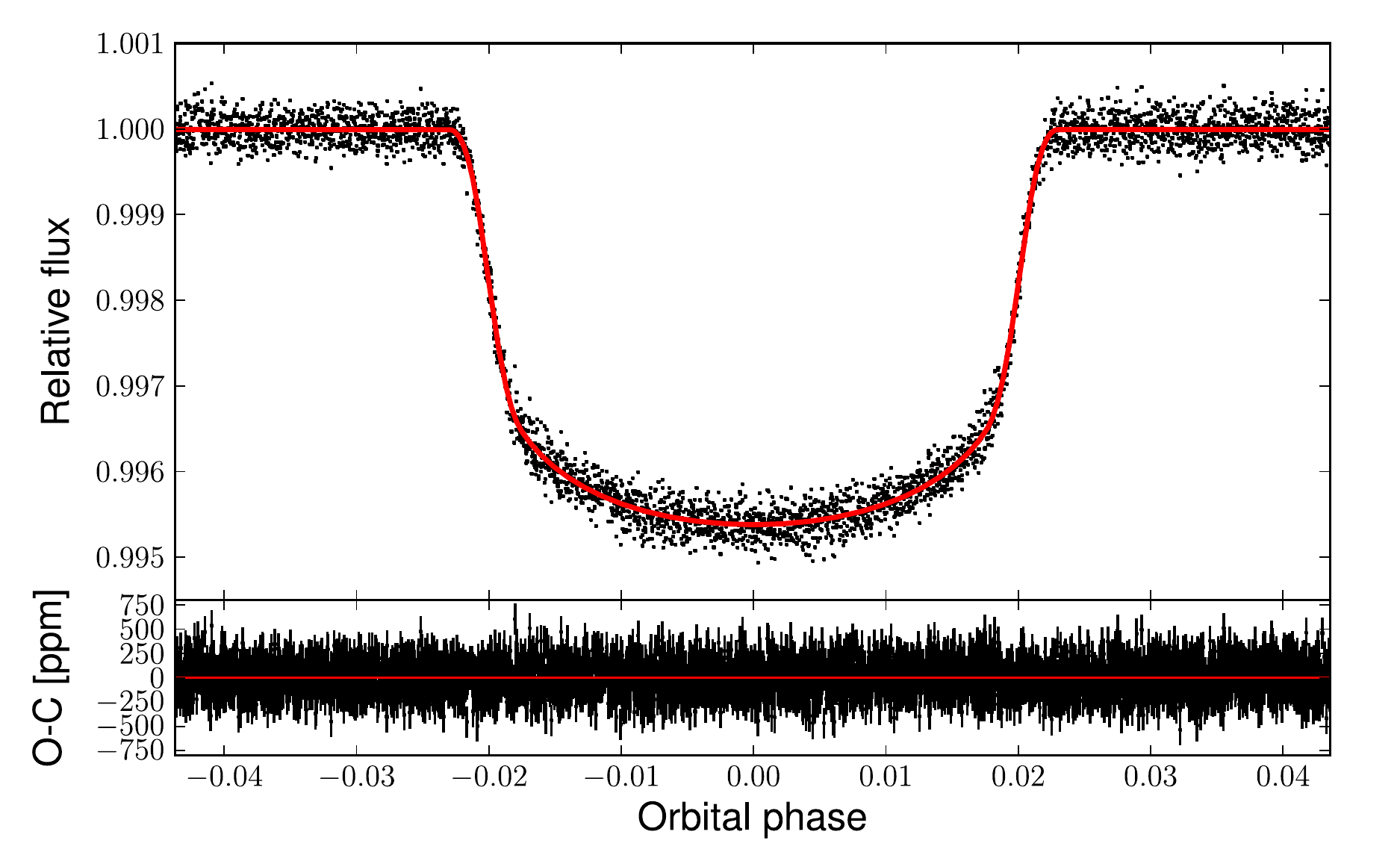}\\
\includegraphics[width=6cm]{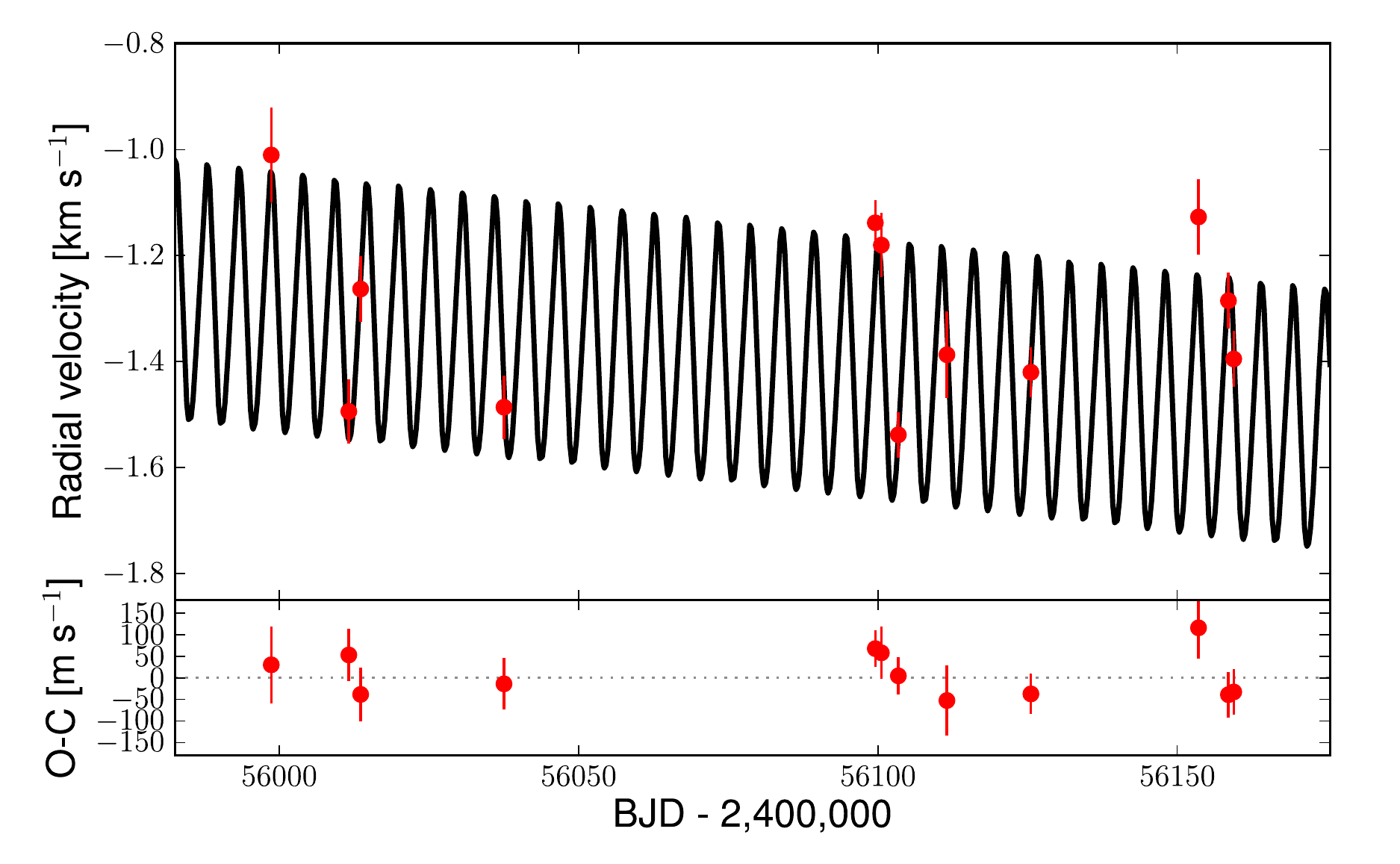}\includegraphics[width=6cm]{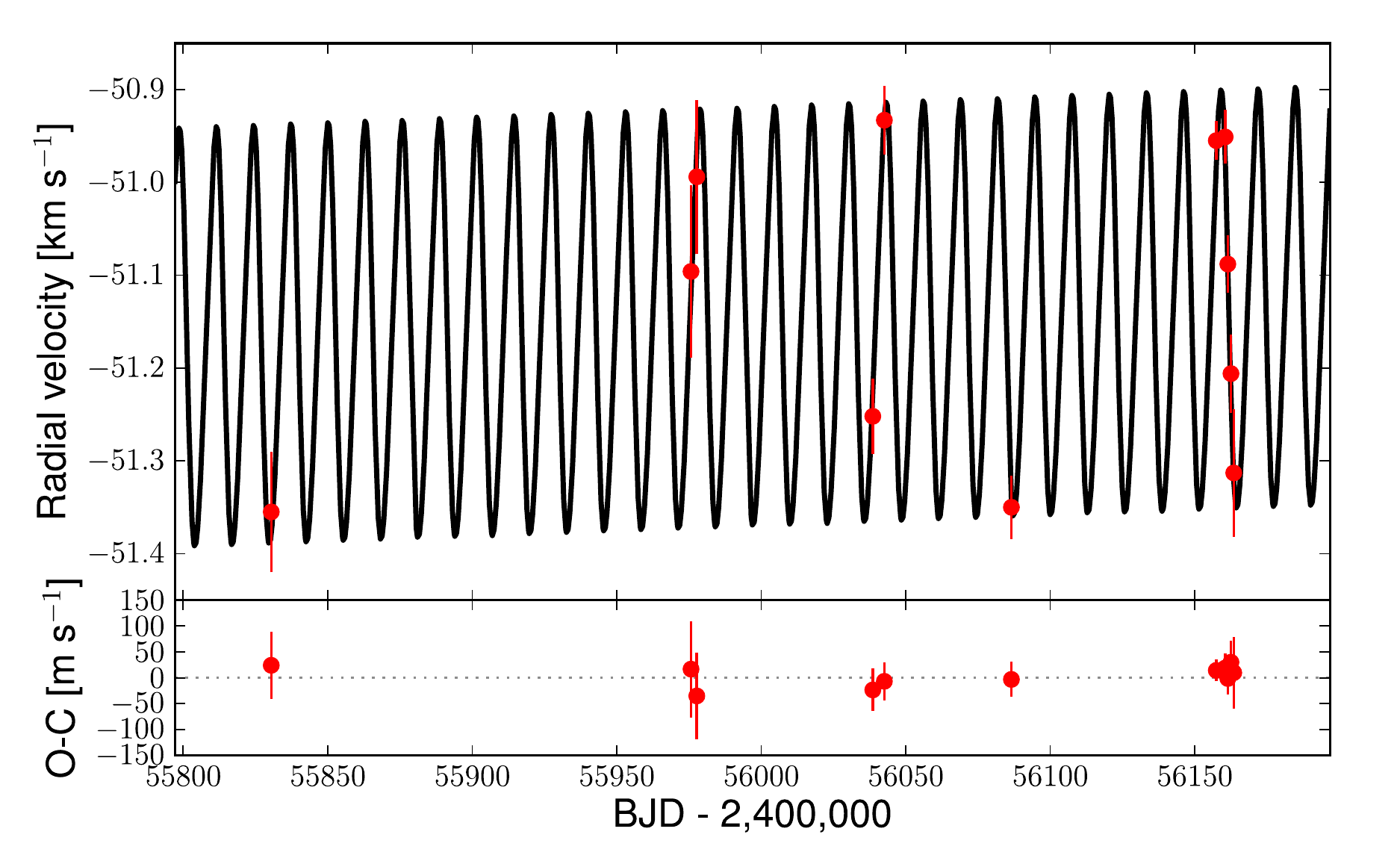}\includegraphics[width=6cm]{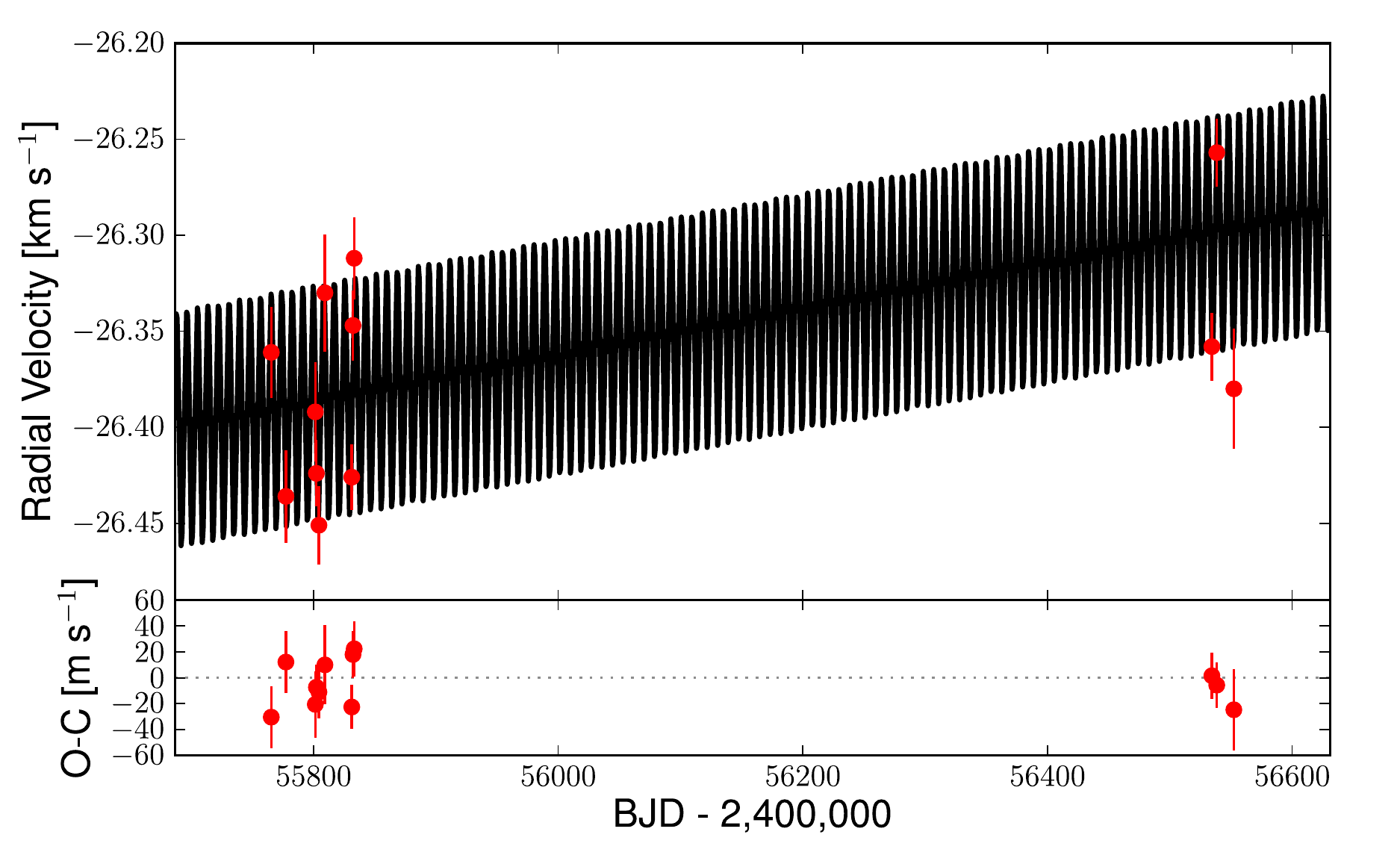}
\includegraphics[width=6cm]{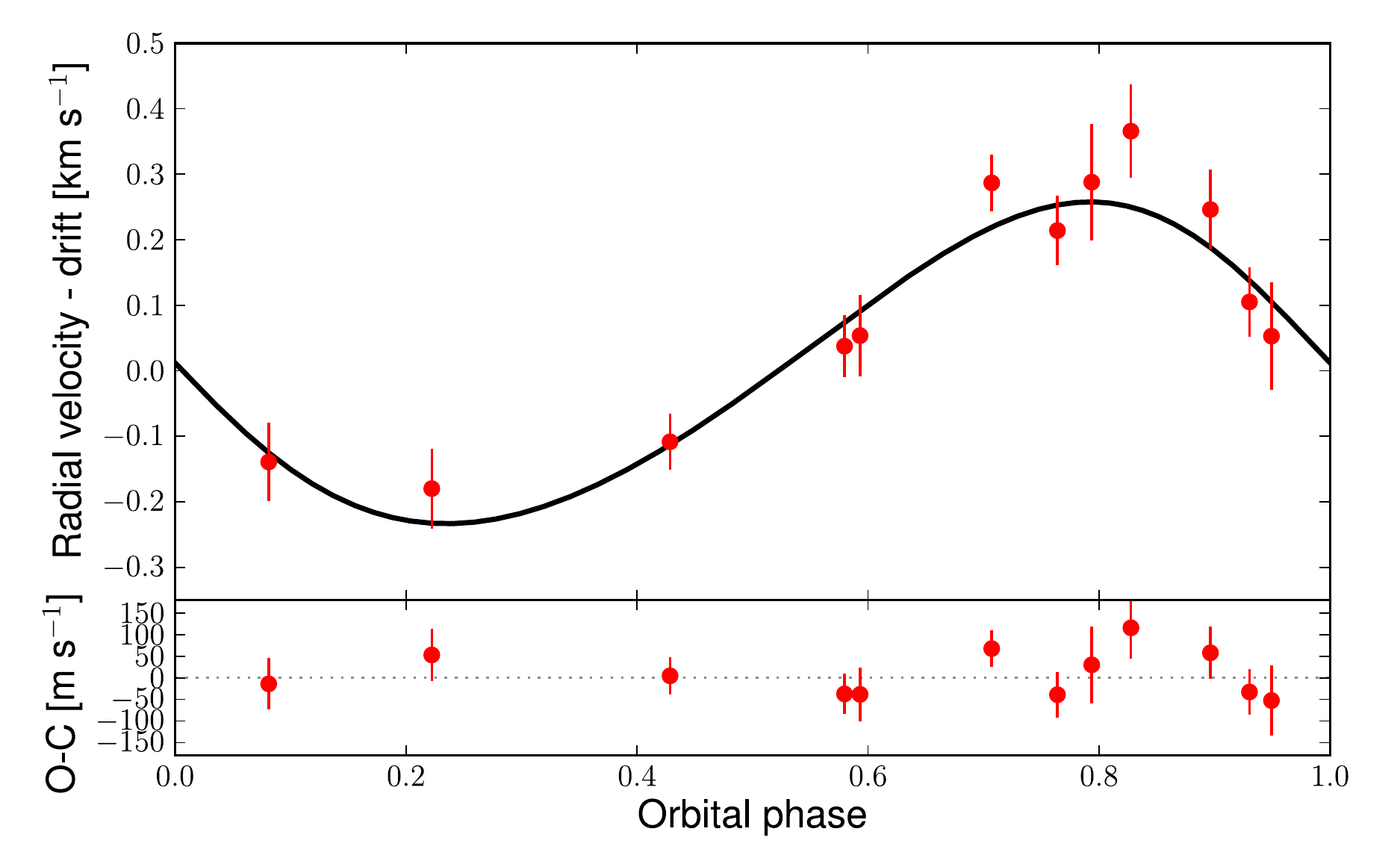}\includegraphics[width=6cm]{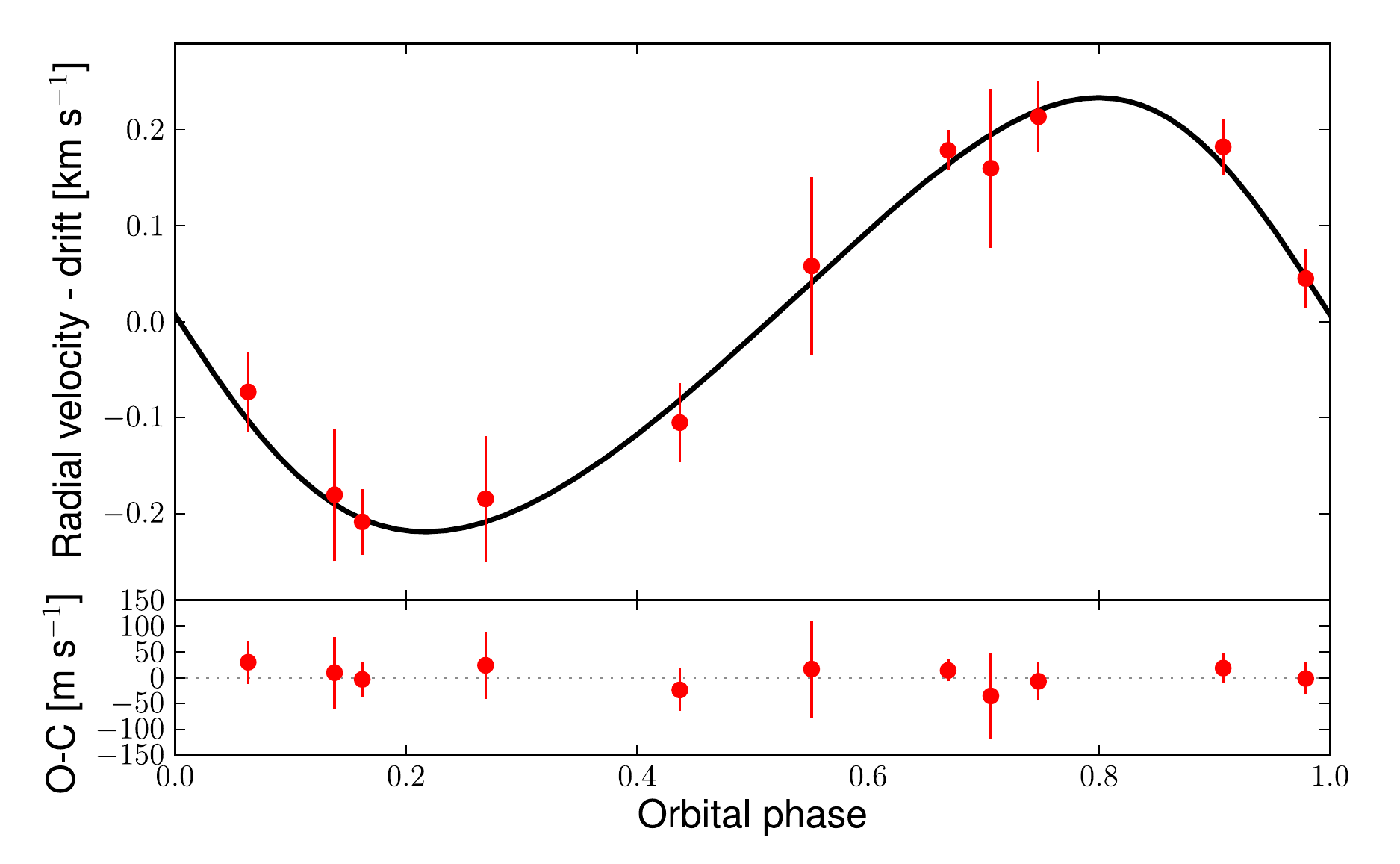}\includegraphics[width=6cm]{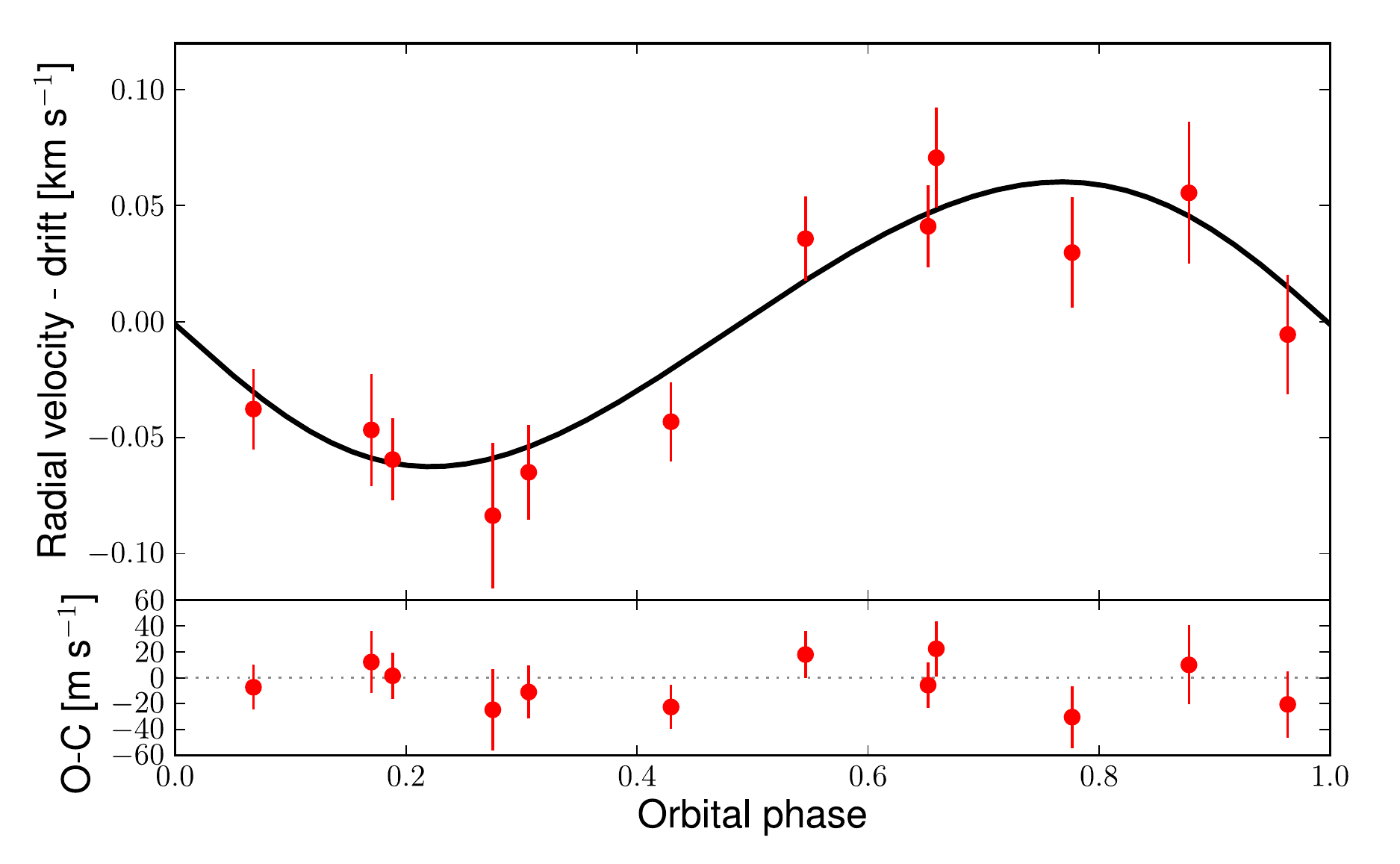}\\
\caption{Model of the maximum likelihood step of the merged chain of \Ksix, \Kfour, and \Kzero\ (from left to right columns) plotted with the data. From top to bottom: SED, light curve, radial velocity in time, and radial velocity in phase. In the bottom panels: residuals after subtracting the model to the observed data. SED plot: the solid line plots the PHOENIX/BT-Settl interpolated synthetic spectrum, in red circles the absolute photometric observations listed in Table~\ref{tableparam}, in blue open circles the result of integrated the synthetic spectrum in the observed bandpasses. Light curve plot: folded transit observed by \Kepler\ in long-the black solid line plots the Keplerian model a for the radial velocities and red circles the SOPHIE observations.}
\label{solplot}
\end{figure*}

\begin{figure*}
\centering
\includegraphics[width=6cm]{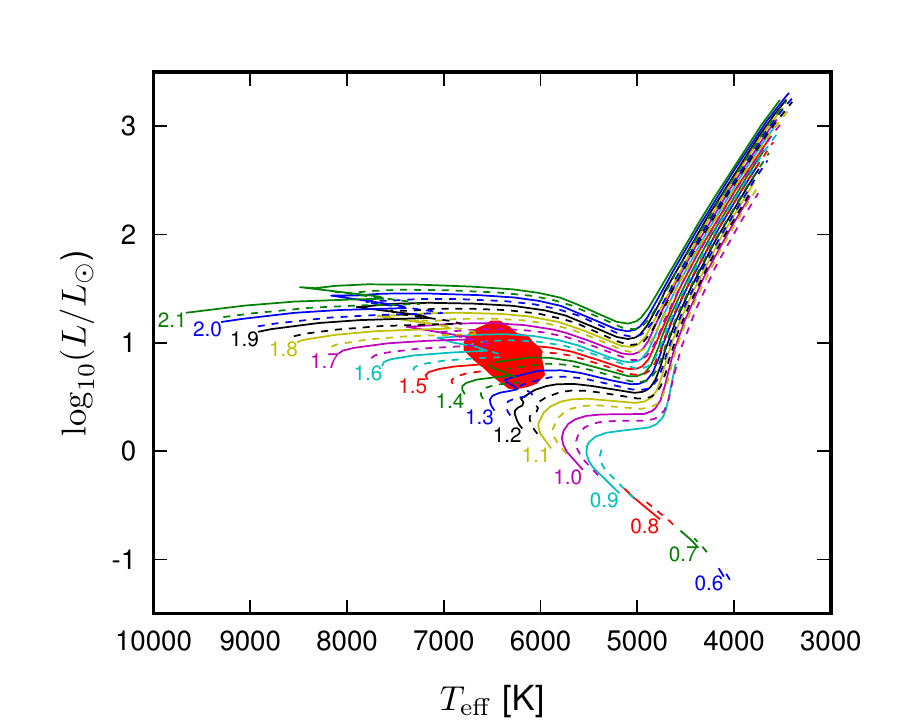}
\includegraphics[width=6cm]{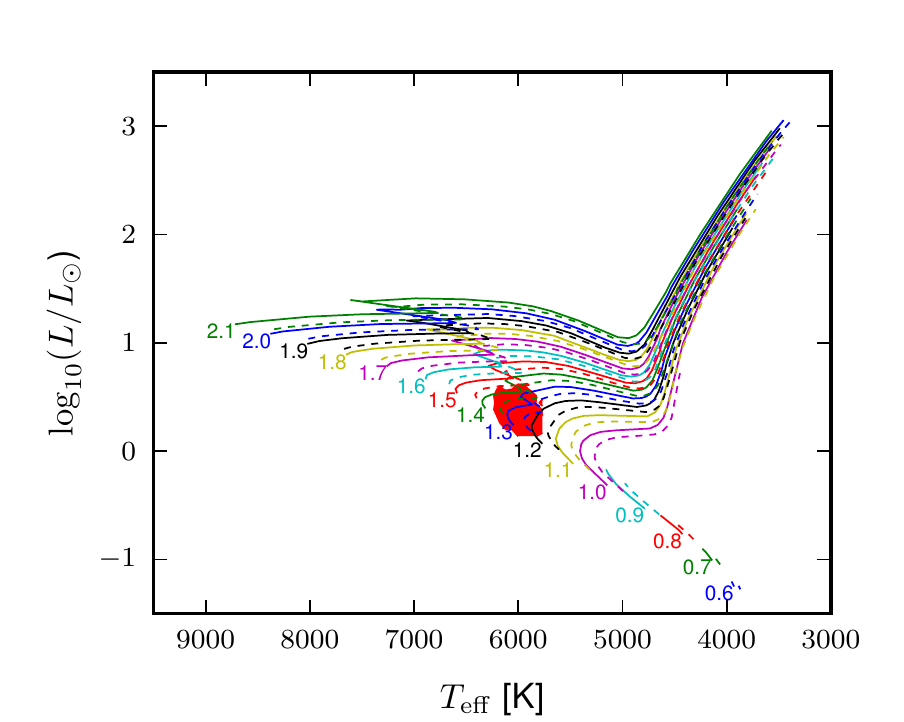}
\includegraphics[width=6cm]{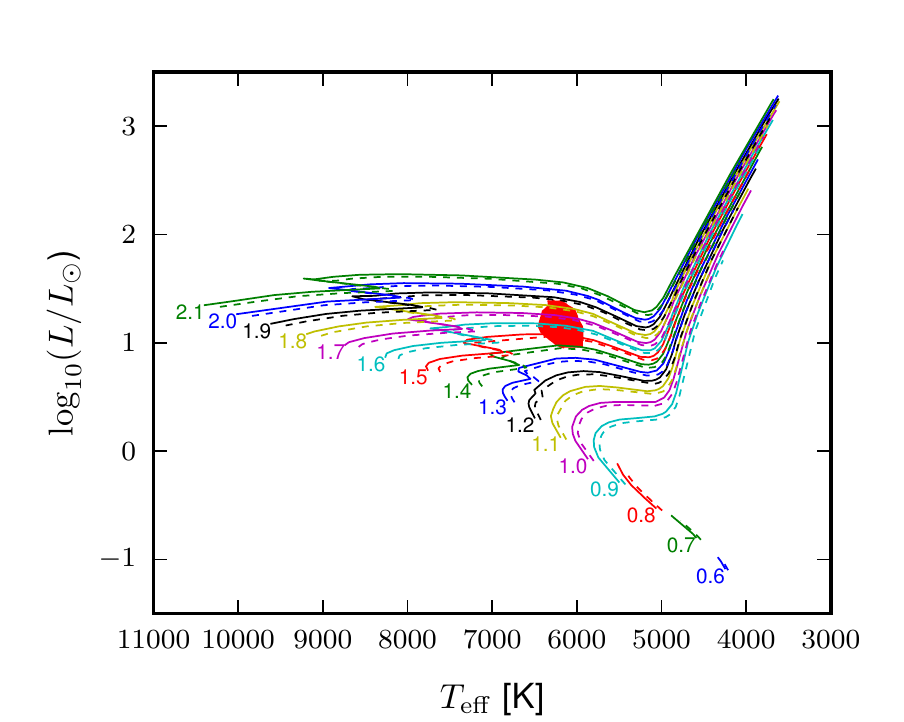}
\caption{Stellar evolution tracks from {\sl STAREVOL}, from the zero-age main-sequence stage to 25~Gyr age, in the luminosity-effective temperature plane for [Fe/H] = 0.00, 0.25, and -0.20 (from left to right, solid line) and [Fe/H] = 0.10, 0.35, and -0.15 (from left to right, dashed line) with the joint posterior distributions of the MCMC (98.9\% joint confidence region is denoted in red) for \Ksix, \Kfour, and \Kzero,\ respectively. The mass in solar masses is annotated at the beginning of the main sequence of each track.}
\label{HR}
\end{figure*}

\begin{figure*}
\hspace{-2cm}\includegraphics[width=22cm]{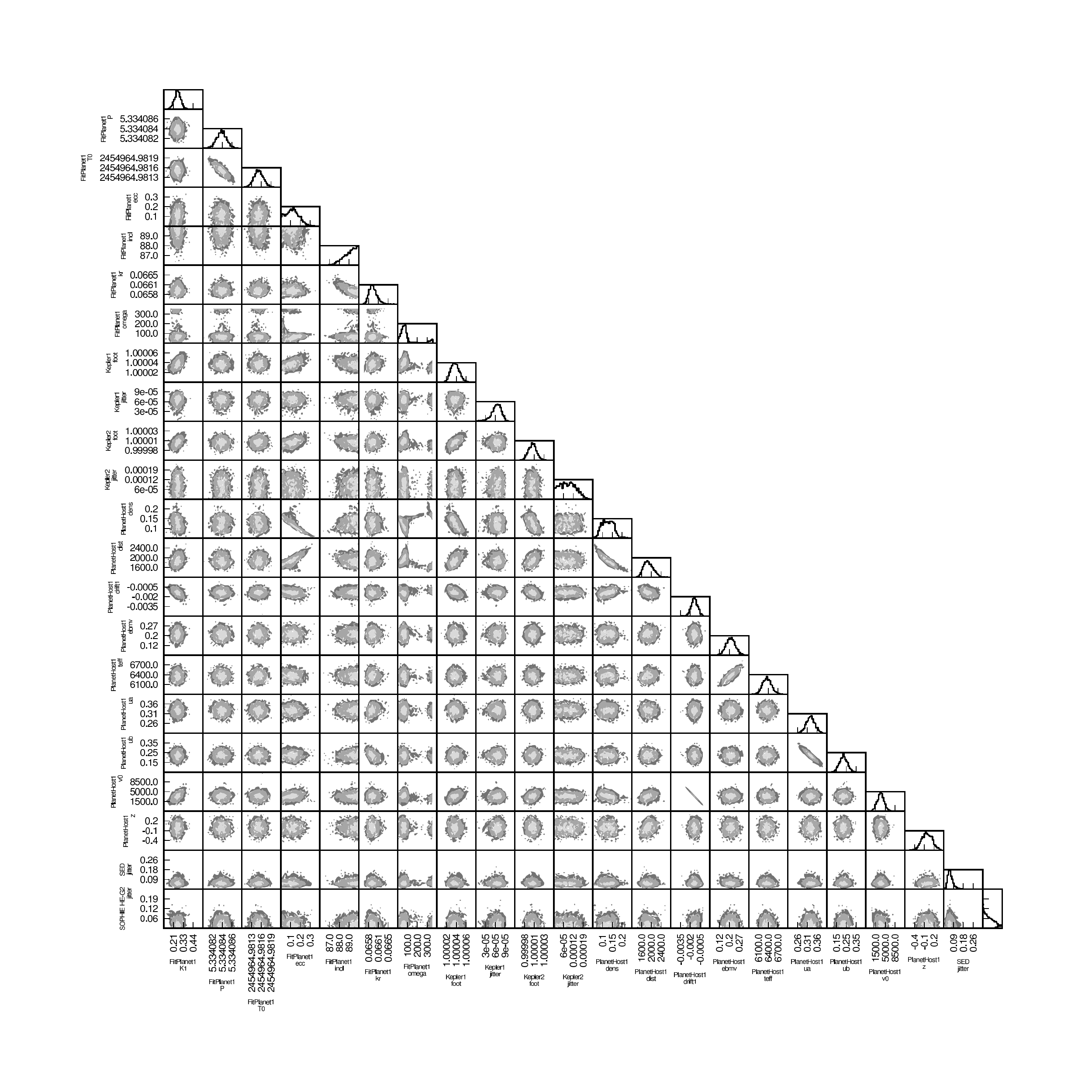}
\caption{Two-parameter joint posterior distributions of the MCMC jump parameters of the fit with stellar models of \Ksix. The 39.3\%, 86.5\%, and 98.9\% joint confidence regions are denoted by three different gray levels. The histogram of each parameter is shown at the top of each column, except for the parameter on the last line that is shown at the end of the line.}
\label{pyram6}
\end{figure*}

\begin{figure*}
\hspace{-2cm}\includegraphics[width=22cm]{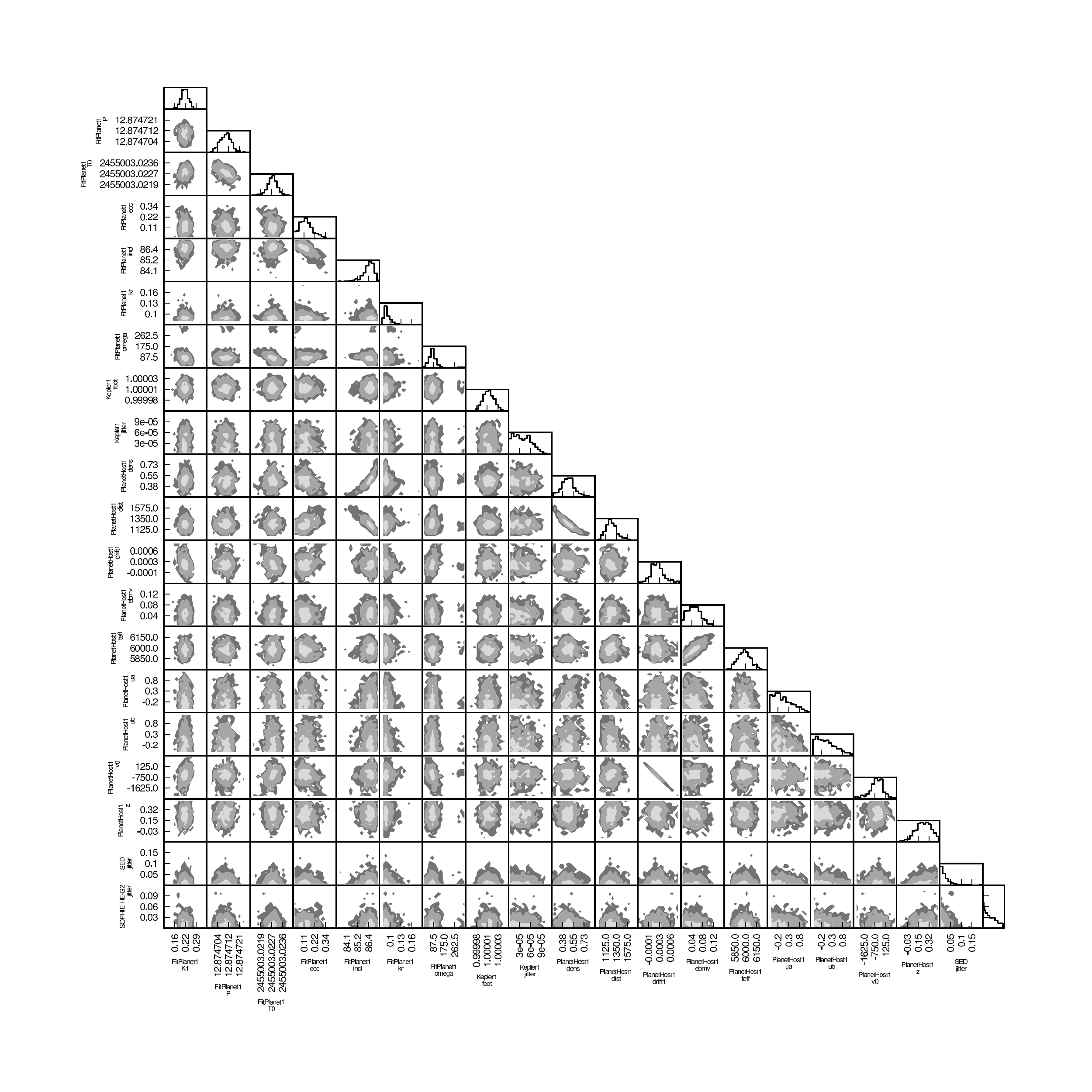}
\caption{Idem Fig.~\ref{pyram6} but for \Kfour.}
\label{pyram4}
\end{figure*}

\begin{figure*}
\hspace{-2cm}\includegraphics[width=22cm]{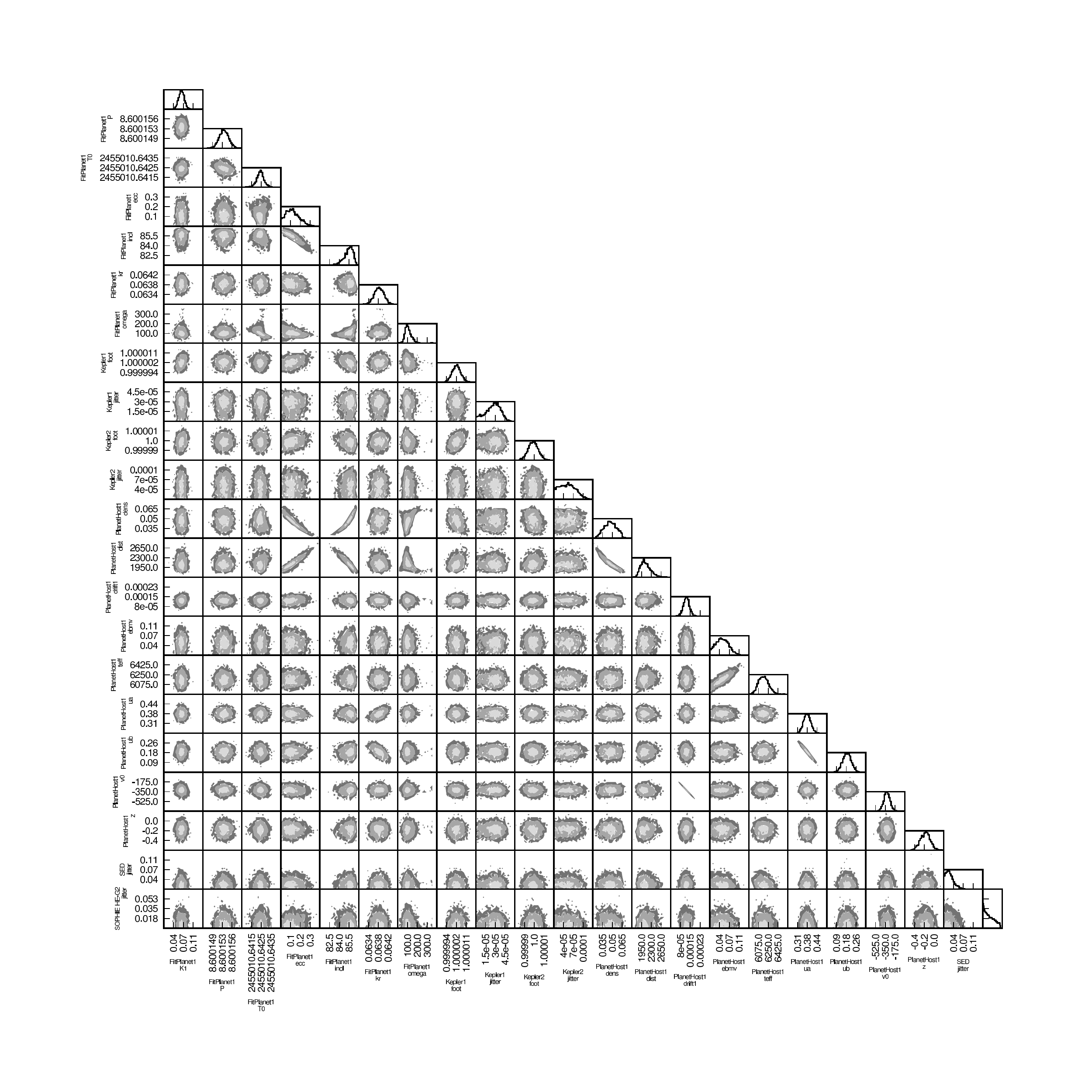}
\caption{Idem Fig.~\ref{pyram6} but for \Kzero.}
\label{pyram0}
\end{figure*}

It is a well known problem that the uncertainty in the stellar parameters obtained from evolutionary models may be underestimated. To account for this, we use different evolutionary models as input for the stellar parameters, and the combined outcome will naturally enlarge the errors in the stellar parameters from the differences between them. For each of the targets, we therefore fit the data three times using a different evolutionary model: {\sl STAREVOL}, Dartmouth, and PARSEC. To ensure a broad exploration of the parameter space, we compute 30 chains of $10^6$~steps, starting at random points drawn from the joint prior. After removing the burn-in interval of each chain, we obtain a merged chain by thinning each chain. The thinning factor is determined by the maximum correlation length found among all parameters in each chain \citep[e.g.][]{2004PhRvD..69j3501T}. This is done to obtain independent points to form the posterior distributions. Then we take the same number of samples of each merged chain from each stellar model, obtaining the combined merged chain. We obtain 3168, 729, and 4564 independent points in the combined merged chain for \Ksix, \Kfour, and \Kzero,\ respectively, which we use to obtain the estimated value and the 68.3\% central confidence interval, both for jump and derived parameters (listed in Table~\ref{tableparam}). The model of maximum likelihood is plotted together with the data in Fig.~\ref{solplot}. The correlation distributions and histograms are shown in Figs.~\ref{pyram6}, \ref{pyram4}, and \ref{pyram0} for the MCMC jump parameters. We found that the distributions of the posteriors obtained from the different stellar models are compatible within 1-$\sigma$. The stellar evolution tracks from {\sl STAREVOL} in a Hertzsprung-Russell diagram are given in Fig.~\ref{HR}, along with the  luminosity-$T_{\rm eff}$ joint posterior distributions to show the evolutionary stage of the host stars.

\section{Planetary evolution models}\label{set}
\begin{figure}[b]
\centering
\hspace{-1.2cm}
\includegraphics[width=9.5cm]{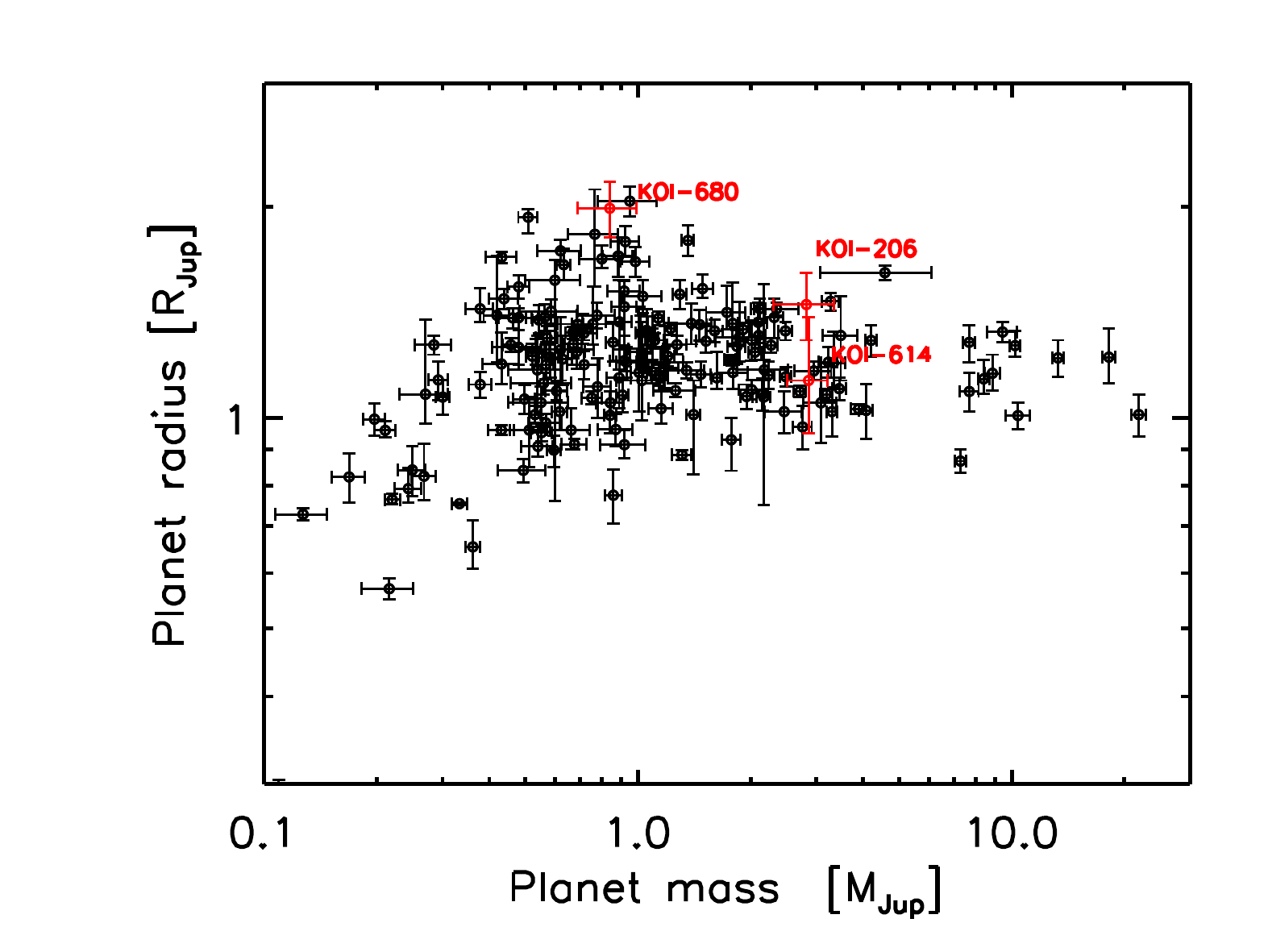}
\caption{Radius versus mass of all known transiting planets as of February 2014 \citep[see http://exoplanets.org/,][]{2011PASP..123..412W}. The three planets presented in this paper are highlighted in red and labeled.}
\label{mpvsrp}
\end{figure}

\Kfourb, \Ksixb,\ and \Kzerob\ are all different from each other: with radii of 1.13, 1.45, and up to 1.99~$\rm R_{Jup}$, and masses of 2.86, 2.82, and down to 0.84~$\rm M_{Jup}$, they occupy distinct regions of the mass-radius diagram as illustrated in Fig.~\ref{mpvsrp}.

To estimate the structure and composition of the three planets, we computed the joint evolution of both planet and host star
for each system, combining the CEPAM \citep{Guillot1995, Guillot2010} and PARSEC \citep{Bressan2012} evolutionary models using SET (see \citealt{GH2011, Havel2011}; see also \citealt{Almenara2013}). Since stellar and planetary absolute parameters are ultimately model dependent, the values we obtained are solely based on observations. Through the use of SET's MCMC algorithm, we thus obtained the posterior probability distributions of the bulk composition of the planets (i.e., their core mass), as well as an independent estimation of planetary and stellar parameters. The latter are entirely consistent with those presented in Section~\ref{pastis}.

\begin{figure*}
\centering
\includegraphics[width=6cm]{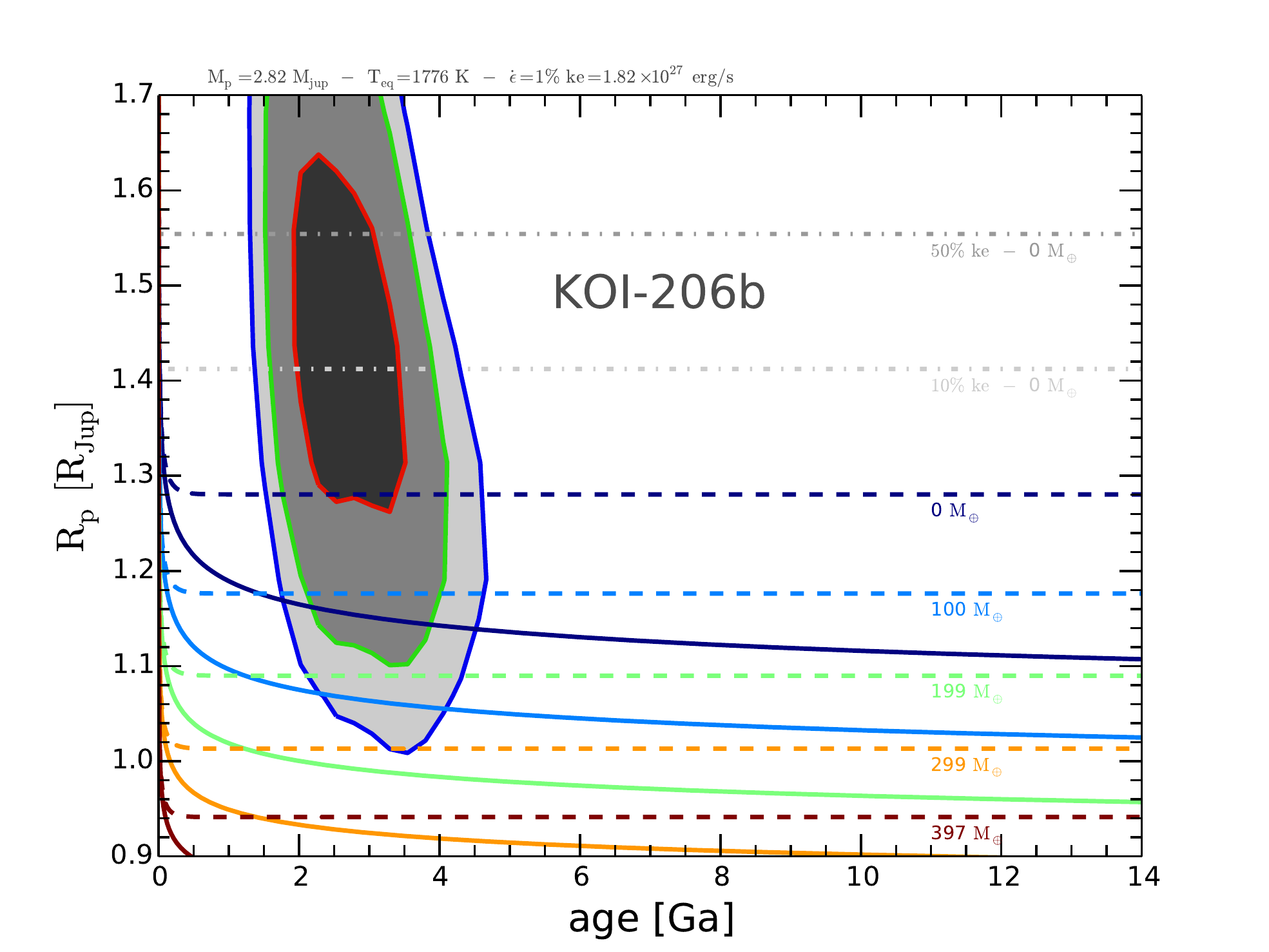}
\includegraphics[width=6cm]{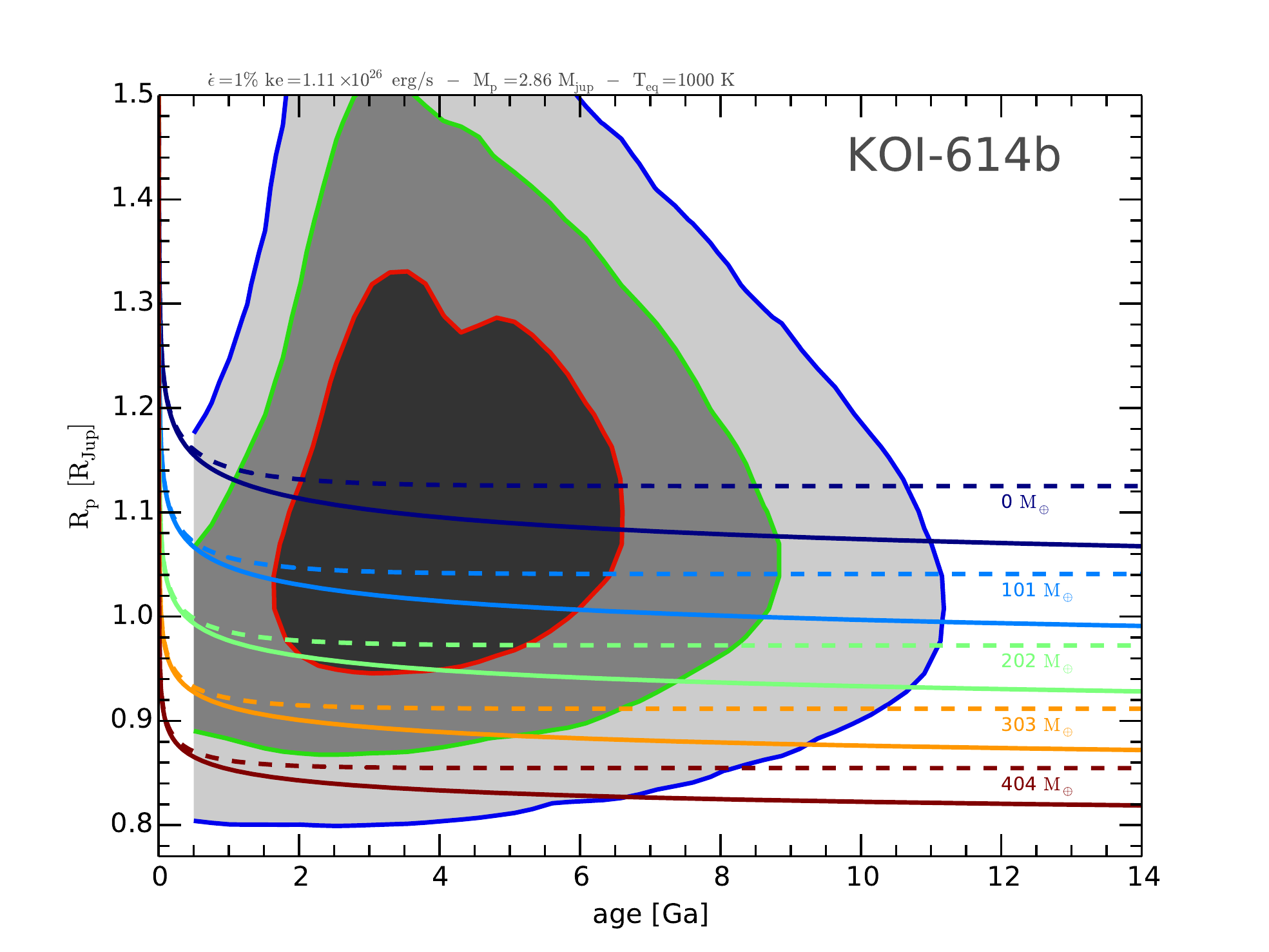}
\includegraphics[width=6cm]{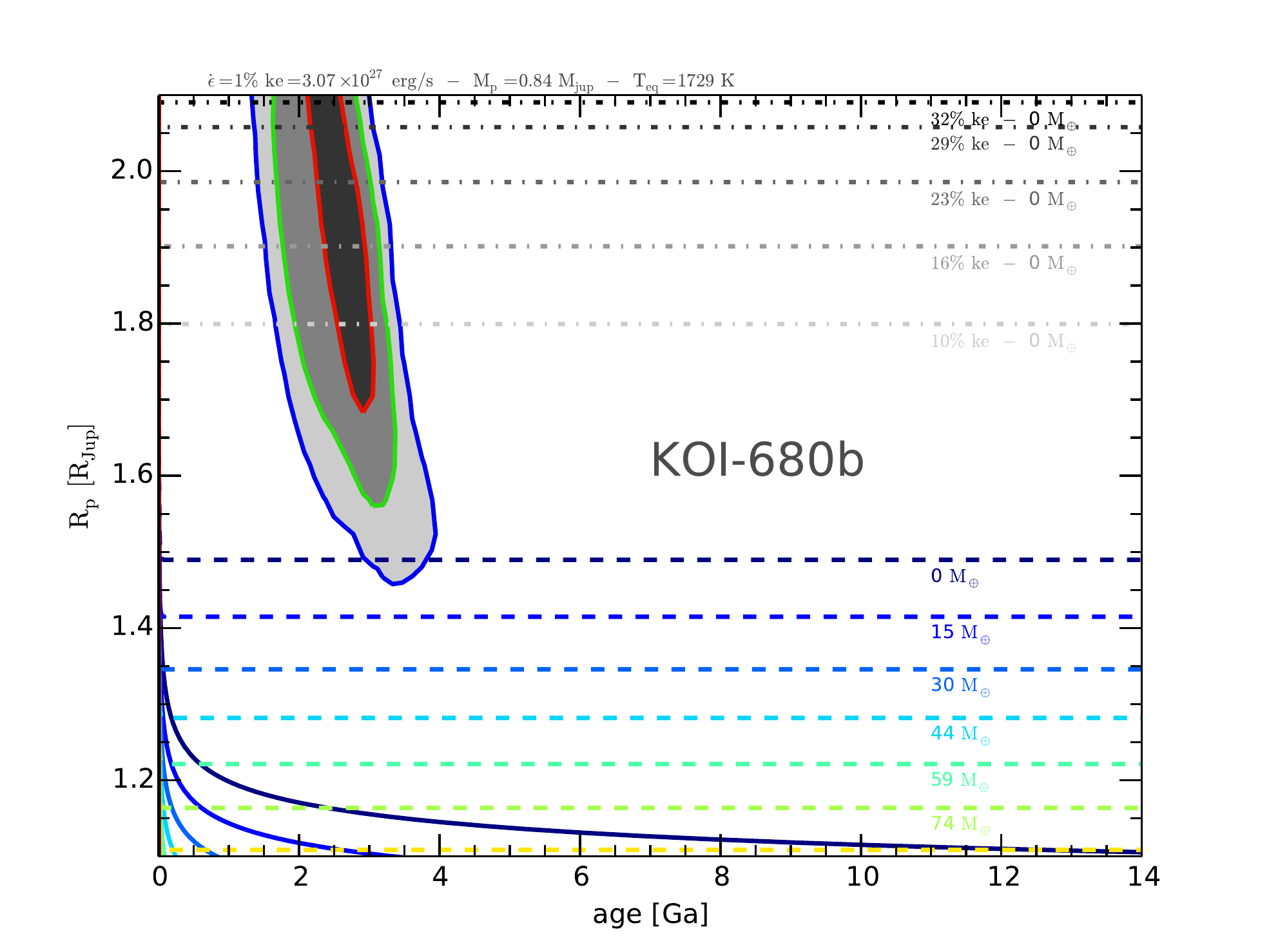}
\caption{Evolution of the planet radius as a function of the age. \Ksixb\ with mass 2.82~M$_{\rm Jup}$ and effective temperature 1776~K is plotted to the left, \Kfourb\ (2.86~M$_{\rm Jup}$ and 1000~K) in the center plot, and \Kzerob\ (0.84~M$_{\rm Jup}$ and 1729~K) in the right plot. The 68.3\%, 95.5\%, and 99.7\% joint confidence regions are denoted by black, dark gray, and light gray areas, respectively. The curves represent the thermal evolution of a respective planet mass and equilibrium temperature. Text labels indicate the amount of heavy elements in the planet (its core mass in Earth masses). Dashed lines represent planetary evolution models for which 0.25\% of the incoming stellar flux ($\dot{\epsilon}$) is dissipated into the core of the planet, whereas plain lines do not account for this dissipation (standard models). Dashed-dotted lines (gray shades) are coreless models with higher amounts of dissipated energy, as labeled.}
\label{figset}
\end{figure*}

The results for \Ksixb, \Kfourb, and \Kzerob\ are presented in terms of planetary radii as a function of age in Fig.~\ref{figset}: the regions show the 1, 2, and 3-$\sigma$ constraints from modeling the star and transit, while the lines show a subset of planetary models for the nominal mass and equilibrium temperature of the planet and different compositions.

All our planetary evolution models assume the following: the planet is made of a central rocky\footnote{38\% SiO${}_2$, 25\% MgO, 25\% FeS, 12\% FeO} core and a solar-composition\footnote{70\% H, 28\% He, 2\% others (Z)} envelope. Of course we do not know whether all heavy elements are in the core or mixed in the envelope or know their exact composition. However, \citet{Baraffe2008} show that having the heavy elements mixed in the envelope generally leads to smaller radii. Therefore our models should provide an upper limit for Z, the fraction of heavy elements in the planet. In addition, we do not expect the envelope to have a solar composition \citep[it is not even the case for Jupiter or Saturn; see, e.g.,][]{Guillot2014}, but the uncertainties in the models themselves (EOS, opacities, atmospheric model) play a larger role than the small adjustments\footnote{as compared to the expected, much higher Z values in planets than in stars \citep{Guillot2006}} in the composition of the envelope due to the different stellar metallicities ($\rm Z_\star \sim 0.013$ for \Kzero\ and $\rm Z_\star \sim 0.042$ for \Kfour).

For consistency with previous modeling of exoplanets, planetary evolution models are calculated in two cases: (i) a ``standard'' case where only the irradiation from the star is accounted for; (ii) a ``dissipated-energy'' model in which, in addition to the standard case, a fraction (0.25\%) of the incoming stellar flux is assumed to be converted into kinetic energy\footnote{Note the use of `ke' labels in figures: 1\% ke = 0.25\% of the incident flux at substellar point or 1\% of the irradiating flux averaged over the planet's area converted into kinetic energy (ke).} and then dissipated at the center of the planet (see, e.g., \citealt{GS2002} and \citealt{Spiegel2013} for detailed discussions).

With a radius of 1.13~$\rm R_{Jup}$ and a mass of 2.86~M$_{\rm Jup}$, both standard and dissipated models provide solutions for the planetary radius of \Kfourb\ that match the available constraints, resulting in a core mass of $81^{+154}_{-63}~\rm M_{\oplus}$. As seen in Fig.~\ref{figset}, models with no core (i.e., pure H/He planet) land in the middle of the 1-$\sigma$ region, but the high metallicity of the star (+0.35) tends to indicate that the planet should probably have a significant number of heavy elements \citep{Guillot2006}.

On the other hand, \Ksixb\ has a similar mass of 2.82~$\rm M_{Jup}$, but a radius of 1.45~$\rm R_{Jup}$, and an inferred density of 1.13~$\rm g\, cm^{-3}$. For such a high radius and nearly three times the mass of Jupiter, \Ksixb\ does not have many similar siblings: the three closest in the mass-radius diagram (Fig.~\ref{mpvsrp}) are CoRoT-2b \citep{Alonso2008}, CoRoT-18b \citep{Hebrard2011}, and Kepler-17b \citep{2011ApJS..197...14D,Bonomo2012}. The former two, for instance, have been modeled and are still challenging the current planetary models \citep[see, e.g.,][]{GH2011,Moutou2013}. As Fig.~\ref{figset} shows, this is also the case of \Ksixb: standard models barely succeed at matching the 2-$\sigma$ age-radius constraint, while dissipated-energy models are just reaching the bottom of the 1-$\sigma$ region. Although an independent MCMC 1-D distribution gives a core mass of $0~\rm M_{\oplus}$ with a 68.3\% confidence interval [0, 74]$~\rm M_{\oplus}$, thus promoting solutions for which the planet has a much lower radius. Extra models with incredibly large amounts of energy dissipated in the planet have to be created to explain the observed radius at 1-$\sigma$. Up to 75\% of the total light reaching the planet would need to be dissipated for these models to work out a solution. There is, to date, no known physical process that would allow such efficiency. However, the recent work of \citet{Spiegel2013} suggests that dissipating the energy in the upper atmosphere instead of the deep interior maintains the planet's ``hot state'' (i.e., large radius) longer for ages below 2-3 Gyr. Last but not least, the use of a newer H/He equation-of-state (EOS) may contribute significantly to solving the problem: while our models are using SCVH \citep[Saumon-Chabrier-Van Horn,][]{Saumon1995}, the EOS from \citet{Militzer2013} may produce planets with bigger radius, up to about 0.2~$\rm R_{Jup}$ larger than ours.

Last of all, with a radius of 1.99~$\rm R_{Jup}$ and a mass of 0.84~$\rm M_{Jup}$, \Kzerob\ is among the largest planets ever known. Only two planets have a higher radius, HAT-P-32b \citep{Hartman2011} and WASP-17 \citep{Anderson2011} (see Fig.~\ref{mpvsrp}), and as few as eleven have a radius larger than 1.7~$\rm R_{Jup}$. WASP-78b \citep{Smalley2012} is the only one of similar mass and radius around an evolved star with low metallicity: unfortunately, it has not been modeled yet. As expected, our standard planetary models cannot account for such a high radius. Even with the dissipation of 1\% of the available irradiation, we hardly reach the 3-$\sigma$ constraint region. Once again, we had to produce some more extreme models with unphysical amounts of energy dissipated into the deep planet to achieve the 2~$\rm R_{Jup}$ limit. Interestingly, we note that dissipating ``just'' 10\%\ to 40\% of the incident stellar flux is enough to match the 1-$\sigma$ region, possibly less than what is required for \Ksixb. Previously suggested improvements over our models for \Ksixb\ should apply for this planet. On the other hand, \Kzerob\ is below a Jupiter mass and should therefore be easier to re-inflate. 

\section{Discussion}\label{discussion}

\begin{figure}
\centering
\includegraphics[width=\hsize]{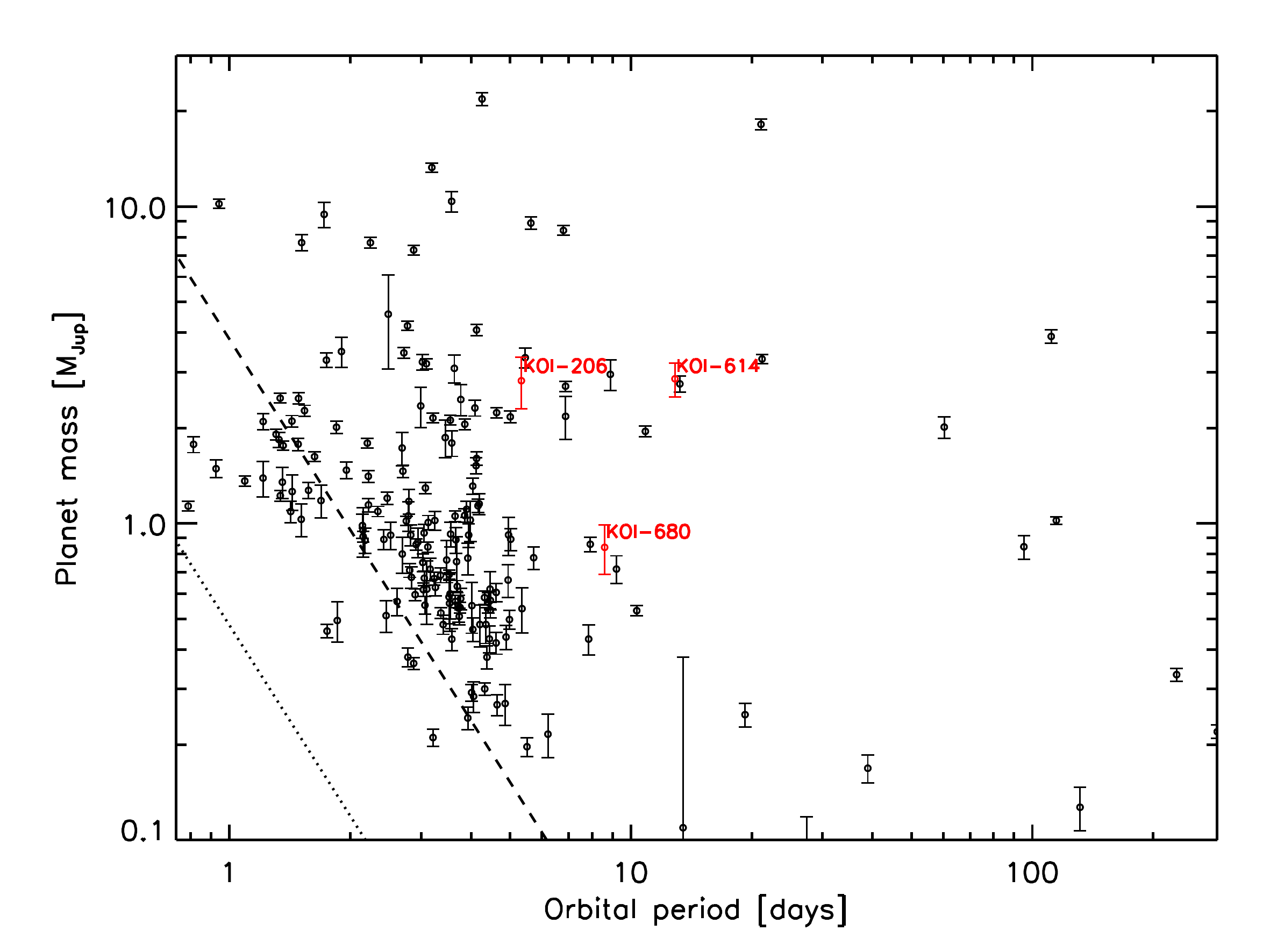} 
\caption{Planetary mass as a function of the orbital period for known transiting giant exoplanets \citep{2011PASP..123..412W}. The three planets presented in this paper are highlighted in red and labeled. The dotted line represents the Roche limit, and the dashed line corresponds to twice this value, which would be the final orbital period of an initial eccentric long-period planet that would have completed its orbit.}
\label{fig:permplan}
\end{figure}

\begin{figure}
\centering
\includegraphics[width=\hsize]{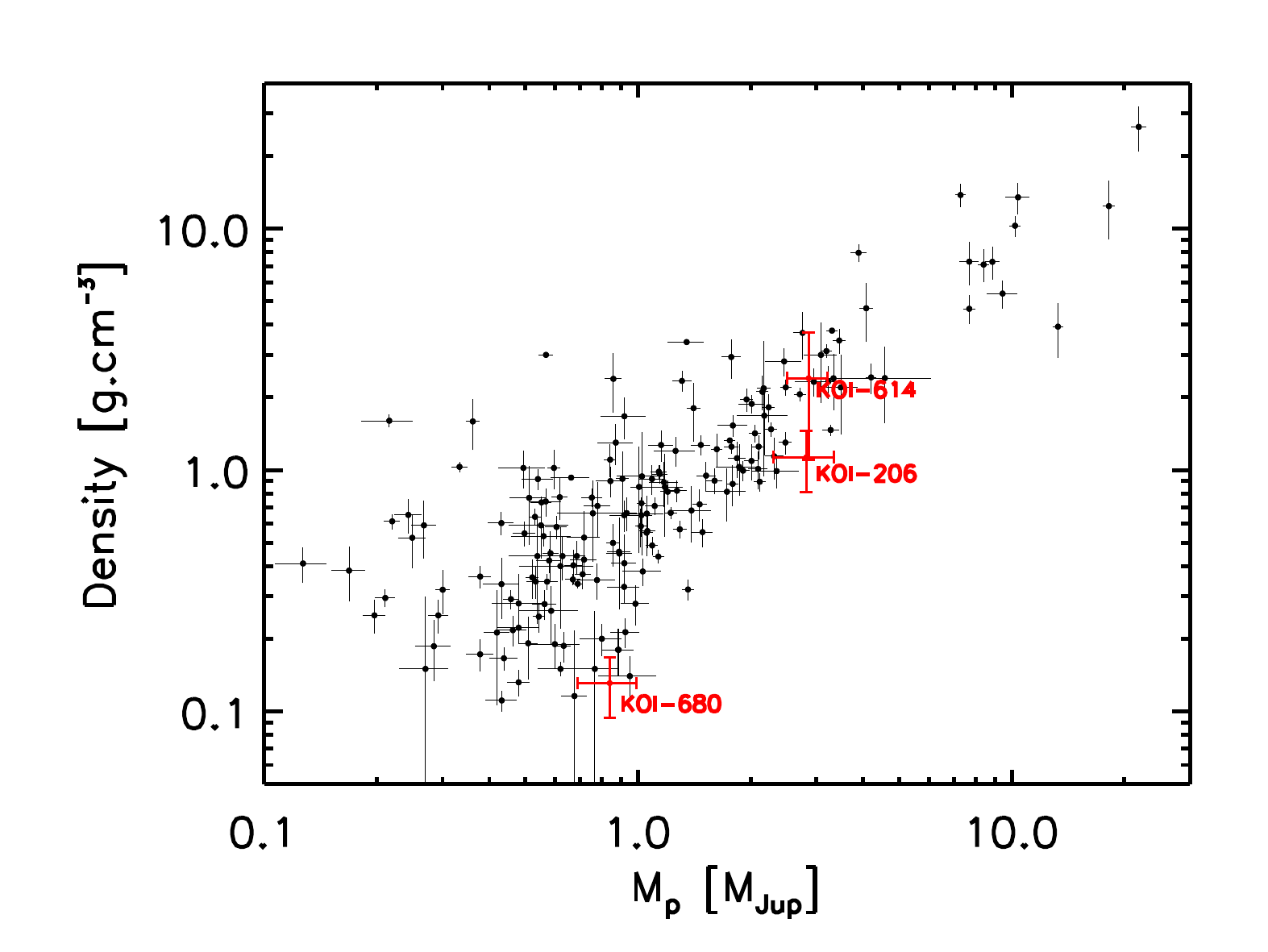} 
\caption{Planetary density as a function of mass for known transiting giant exoplanets \citep{2011PASP..123..412W}. The three planets presented in this paper are highlighted in red and labeled.}
\label{fig:mplandens}
\end{figure}

\subsection{\Kfourb}
\Kfourb\ was a moderate probability candidate, but the radial velocity measurements presented in this paper establish the planetary nature of the companion, its high-impact parameter ($b\approx0.98$) explaining the V-shaped transits. With an equilibrium temperature of $T_{eq}=1000 \pm 45$~K, \Kfourb\ is at the transition between what is usually referred to as \lq\lq{}hot\rq\rq{} and \lq\lq{}warm\rq\rq{} Jupiters and is a good candidate for testing migration theories. \Kfourb\ is one of the few transiting giant planets with an orbital period longer than ten~days, as can be seen in Fig.~\ref {fig:permplan}. Unlike hot Jupiters, this kind of planets has a pericenter distance that is too large to allow efficient tidal dissipation to induce migration through the high-eccentricity migration mechanism \citep{2014ApJ...781L...5D}. In order to access the close pericenter required for migration and circularization by tidal dissipation of the eccentricity, they need to be accompanied by a strong enough perturber to overcome the precession caused by General Relativity. \Kfourb\ has a low eccentricity ($e \lesssim 0.32$), but there is no trace of an additional massive companion in the system. Using the radial velocity residuals of our fit, we can exclude the presence of a companion of more than 2.0~M$_{\rm Jup}$ on orbital periods less than 200 days, at the 3-$\sigma$ level. Given the length of our observation period, we cannot make any conclusion about more distant companions. In addition, \citet{2013ApJS..208...16M} report no significant transit timing variations for this target. The high-eccentricity migration scenario cannot be favored yet.

 Considering the mass of $\sim 3$~M$_{\rm Jup}$ of \Kfourb,\, its formation would have quickly opened a gap in the protoplanetary disk, and another migration scenario would involve type II migration. However, as shown by \citet{2007ApJ...660..845B}, whatever the disk lifetime, type II migration of a single Jovian planet in a truncated disk naturally leads to the creation of period gap between $0.08\lesssim a \lesssim 0.6$~AU in the distribution semi-major axis, which is especially pronounced for systems with stellar masses $M_\star \geq 1.2$~M$_\odot$. As this is precisely the case of this system, the migration of the planet within the disk could be doubted. Nevertheless, while single planets open only narrow gaps, multiple planets can clear wide inner holes \citep{2011ApJ...738..131D},  thus preventing the migration of the Jovian planet down to a short-period orbit. As previously said, our observations exclude the presence of Jovian companions over orbital periods $<100$~days, but we cannot exclude the presence of less massive, terrestrial planets in the system.
In this regard, the composition of the planet can provide additional clues. The estimated radius of $\sim1.13$~R$_{\rm Jup}$ makes \Kfourb\ a high-density giant planet. As can be seen in Fig.\ref{fig:mplandens}, this mass and density are common in the population of the transiting exoplanets discovered so far\footnote{as of Februray 2014 at http://www.exoplanets.org \label{exofoot}}. Within our estimated uncertainties, we have established that the dimensions of this object can encompass internal compositions ranging from pure hydrogen to containing a solid core of heavy elements up to $\sim$ 150~M$_\oplus$. Even if the pure hydrogen scenario cannot be discarded, some authors \citep[see, e.g., ][]{Guillot2006} have noted that there is observational evidence of a correlation between star metallicity and heavy elements in the planet. Considering that \Kfourb\ is orbiting a main-sequence, metallic ([Fe/H] = 0.35~dex), G0V star, it is reasonable to favor a high heavy-element content. In the core-accretion scenario, the \lq\lq{}canonical\rq\rq{} core mass is about 10~M$_\oplus$ \citep{1980PThPh..64..544M}. A greater heavy element content can be obtained with particular disk conditions \citep[the case of \lq\lq{}monarchical growth\rq\rq{},][]{2009A&A...501.1139M,2005SSRv..116...53W} or by the collision with another planet(s) or planetesimals after substantial accretion of the envelope \citep{2006ApJ...650.1150I}.  In the latter case, the disk would have fostered the formation of multiple embryos, which would provide a natural way to stop the migration of the planet at its current semi-major axis ($a\sim0.114$~AU).  

Both migration scenarios require the presence of additional companion(s) in the system. High-eccentricity migration mechanisms require a massive, not very distant perturber to bring the warm Jupiter into a Kozai-Lidov cycle, type II migration requires multiple planets to clear a wide inner hole and prematurely stop the migration of the planet. \Kfourb\ is thus an interesting candidate for future observations with the aim of finding additional low-mass, short-period planetary companions or a more distant and massive perturber.

\subsection{KOI-206}
Like \Kfourb, \Ksixb\ is a relatively massive planet of about $2.8~{\rm M_{Jup}}$, but its radius of $1.45\pm0.16$~R$_{\rm Jup}$ implies a different inner structure with a Jupiter-like density. Although the much tighter orbit of \Ksixb\ ($a\sim0.0679$~AU) puts it in the range of hot Jupiters, and although the higher effective temperature of its host star does imply a stronger irradiation than for \Kfourb, its large radius is particularly challenging for planetary evolution models. 

The radius of a transiting planet is one of the most robust measurements because it only relies on the knowledge of the transit depth and of the stellar radius (with a second-order dependence on limb darkening, and perhaps light curve contamination). Here, the stellar radius is determined using evolutionary models, which are known to be inaccurate, and yet this is partially taken into account by our fitting method by combining several sets of evolutionary tracks in the final result. Concerning the transit depth, it is precisely measured at the 210-$\sigma$ level. Nonetheless, the light curve of \Ksixb\ exhibits rotational modulations due to spot coverage (cf. Sect.~\ref{keplerobs}). 

There is no sign of spot occultation during the transits of the planet, but high-latitude spot could still affect the estimation of the radius. In particular, the presence of such a group of spots could be indistinguishable in the light curve and would induce an overestimation of the radius of the planet. As shown in Fig.~\ref{figset}, typical models of bloated hot Jupiters, where 0.25\% of the incoming stellar flux is dissipated in the core, would agree better with a planetary radius in the bottom of the 1-$\sigma$ confidence region. Still, to reduce the estimated planetary radius to this value, i.e. to about 1.3~R$_{\rm Jup}$, we would need to assume that the value of the flux continuum is actually a factor 1.26 more than the value we are using. It would require a polar spot of angular radius of 25$^{\circ}$, in the very conservative assumption that the contrast of the spot is 0 and the rotational axis of the star is aligned on the line of sight. 

More realistically, if we assume a surface-to-spot contrast of about 0.67 \citep{2003A&A...403.1135L} and the limb darkening coefficients of our best model, this would require a polar spot with a radius of at least 59$^{\circ}$ \citep[estimated using][]{2012MNRAS.427.2487K}. Such a spot could be betrayed by the activity level estimated with the emission cores of the CaII lines. Unfortunately, the signal-to-noise ratio of the SOPHIE spectra in the blue part of the spectrum is too low to estimate the activity with CaII H and K lines. Either way, the dimensions of such a spot is rarely observed \citep{2003A&A...403.1135L}, and it is very unlikely that our planetary radius estimation would be overestimated because of such an effect. 
Besides, the stellar density inferred from the transit (Table.~\ref{tableparam}) is in good agreement with the stellar parameter determined by the spectral analysis, and there is no evidence of a false positive. 

\begin{figure}
\centering
\includegraphics[width=\hsize]{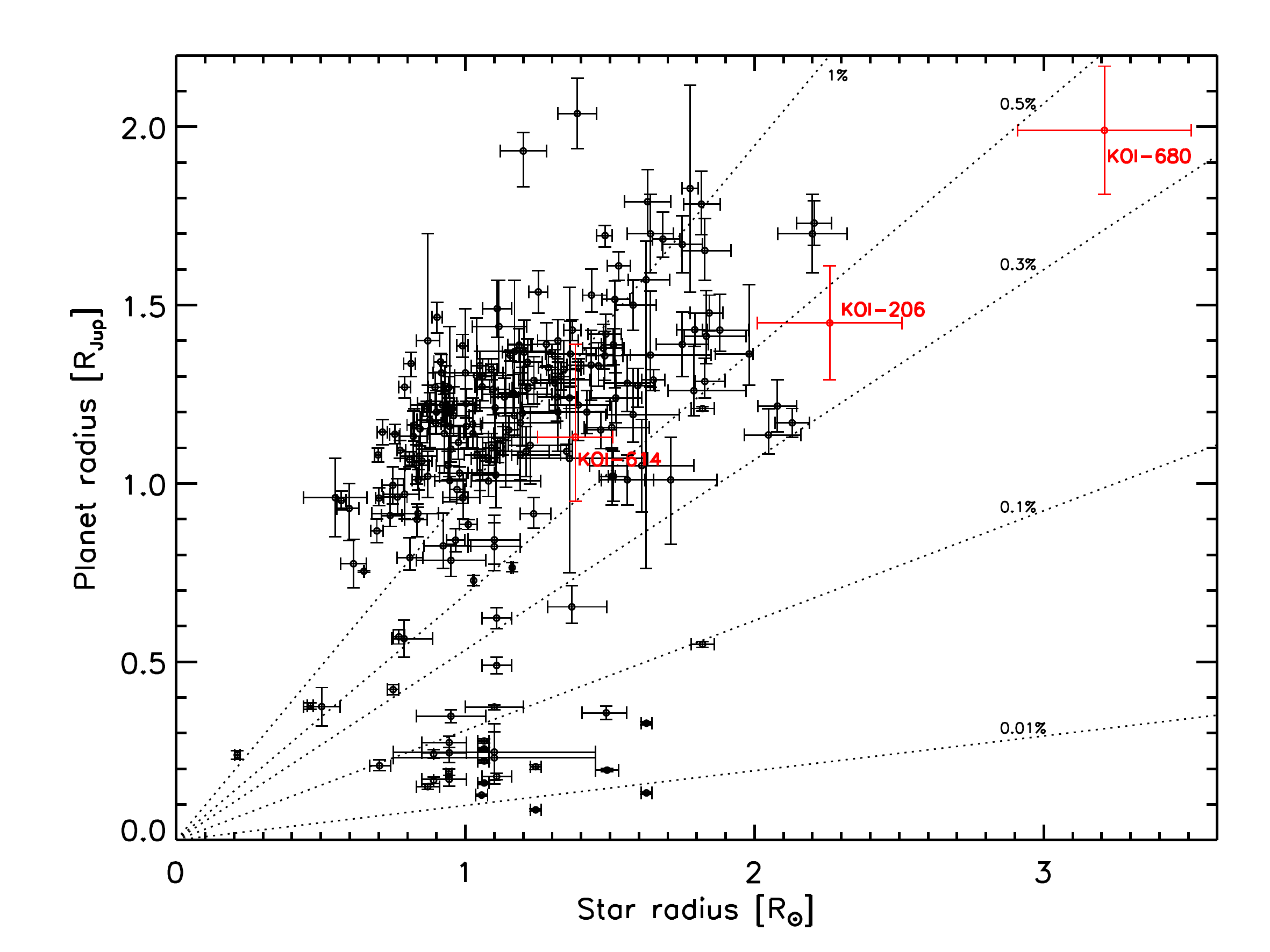} 
\caption{Planetary radius as a function of stellar radius for known transiting exoplanets \citep{2011PASP..123..412W}. The dotted lines give the loci of some typical transit depth given in labels.  The three planets presented in this paper are highlighted in red and labeled.}
\label{fig:rplanrstar}
\end{figure}

At the time of writing, \Ksixb\  is the transiting planet detected around the host with the second largest radius so far\footnote{see note \ref{exofoot}} just after \Kzerob, as shown in Fig.~\ref{fig:rplanrstar}. There are only a few other known bloated planets in this mass range (see Fig.~\ref{mpvsrp}), and \Ksixb\ is the one with the lowest density in this group (see Fig.~\ref{fig:mplandens}). For such a mass, the amount of extra dissipated energy to produce that kind of density is preposterous. \Ksixb\ orbits an evolved F7-type star that has recently left the main sequence (see Fig.~\ref{HR}), and the dynamic aspects related to the rapid and recent evolution of the host might have effects that are unaccounted for. 
As noted by \citet{2003ApJ...594..545B}, there is a difference between the observed transit radius (at $\sim$mbar levels) and the modeled radius, $\Delta$R. Since the evolution of a giant planet is largely controlled by its atmosphere, the planetary evolution code we used, CEPAM, has a dedicated module that computes the exact P-T profile of the atmosphere up to mbar levels (see \citealt{2013arXiv1311.6322P} for more details on the atmospheric model and, e.g., \citealt{2008PhST..130a4023G} for an estimation of $\Delta$R). Therefore, our results for modeling the interior structure of the three giant planets can be directly compared to the observations: $\Delta$R cannot be the explanation for the discrepancy between theoretical models and observed radius.

\subsection{\Kzerob}
\begin{figure}
\centering
\hspace{-1.1cm}
\includegraphics[width=9.5cm]{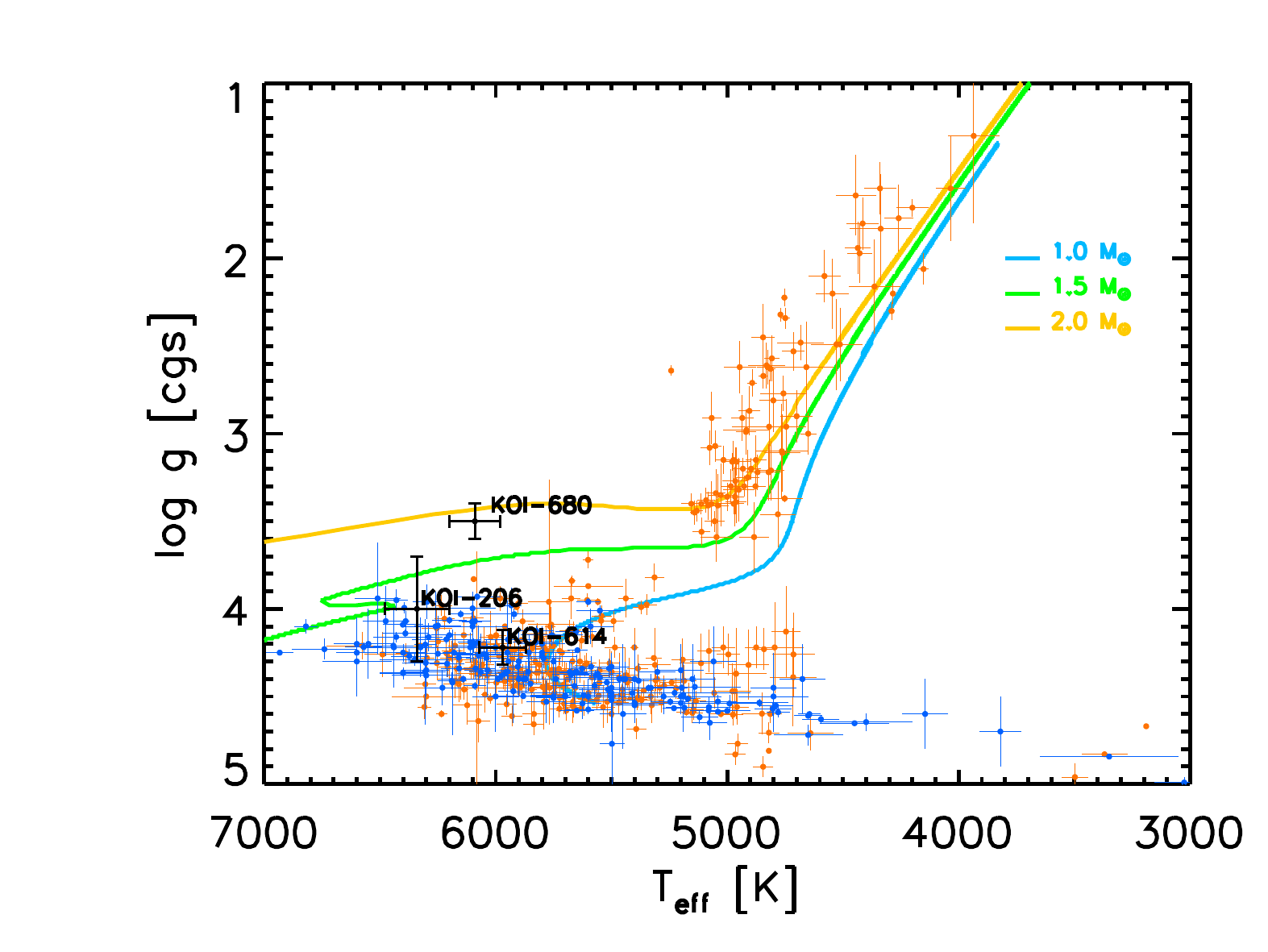} 
\caption{Kiel diagram for the stars of known transiting (in blue) and radial velocity (in orange) exoplanets \citep{2011PASP..123..412W}. \Kfour, \Ksix, and \Kzero\ are drawn in black and labeled. We also plot the {\sl STAREVOL} evolutionary tracks for stars of 1, 1.5, and 2~\Msun\ and solar metallicity.}
\label{HR_diag_exo}
\end{figure}

The host star of \Kzerob\ shares many properties with the one of \Ksixb, but with a mass of only $0.84\pm0.15$~M$_{\rm Jup}$ and a radius of $1.99\pm0.18$~R$_{\rm Jup}$, \Kzerob\ is one of the biggest (Fig.~\ref{mpvsrp}) and least dense (Fig.~\ref{fig:mplandens}) planets detected to date. Such properties exclude a high concentration of heavy elements in its interior  (Fig.~\ref{figset}), which correlates well with the low metallicity of the host star ($\rm{[Fe/H]=-0.17\pm0.10}$~dex). Although standard planetary evolution models fail to reproduce its observed characteristics, the low  mass of \Kzerob\ also means that it is easier to inflate, and improvements in the equation of state might easily solve the problem \citep{Militzer2013}. Not only is the planet one of the biggest ever observed, but it also orbits the bigger star observed with the transit technique (Fig.~\ref{fig:rplanrstar}). Indeed the host star of \Kzerob\ is an F9 subgiant, well on its way toward the beginning of the red giant branch (RGB). Although it is slightly younger than the host of \Ksixb, it is also more massive ($M_\star = 1.538\pm0.088~\rm{M}_\odot$) and thus more evolved. As can be seen in Fig.~\ref{HR_diag_exo}, the host star of \Ksixb\ is observed in a phase of rapid stellar evolution, where no other planets have been detected before. 

With an orbital semi-major axis of $0.0948\pm0.0018$~AU, \Kzerob\ is doomed to be engulfed by the star, even neglecting tidal interaction, the stellar radius will catch up with the current orbit of the planet within the next 260~Myr or so.
We note that we detect a companion of minimum mass $2.9\pm0.6~\rm{M_{Jup}}$ with minimum orbital period of about 790~days (i.e., about 1.9~AU). Further radial velocity follow-up is required to determine the period and the planetary nature of the companion.

\section{Conclusion}\label{conclusion}
We have presented the detection and characterization of \Kfourb, \Ksixb, and \Kzerob, which are\, three new transiting, giant extrasolar planets. They were first detected as promising candidates by the \Kepler\ team from the \Kepler\ light curve. We established their planetary nature with radial velocity follow-up using the spectrograph SOPHIE and characterized the systems through a combined MCMC Bayesian fit of photometric and spectroscopic datasets. We find that \Kfourb\ is a dense, warm Jupiter, with a mass of $2.86\pm0.35~{\rm M_{Jup}}$and a radius of $1.13^{+0.26}_{-0.18}~{\rm R_{Jup}}$, and it is orbiting a G0, metallic ([Fe/H]=$0.35\pm0.15$) dwarf in 12.9 days. \Ksixb\ has a mass of $2.82\pm 0.52~{\rm M_{Jup}}$ and a radius of $1.45\pm0.16~{\rm R_{Jup}}$, and it orbits a slightly evolved F7-type star in a 5.3-day orbit. Orbiting a F9 subgiant in 8.6~days, \Kzerob\ has a much lower mass of $0.84\pm0.5~{\rm M_{Jup}}$ and an uncommonly large radius of $1.99\pm0.18~{\rm R_{Jup}}$. With host stars of masses of $1.46\pm0.17~\Msun$ and $1.54 \pm 0.09~\Msun$ and radii of $2.26\pm0.25~\Rsun$ and $3.21\pm0.30~\Rsun$,  \Ksixb\ and \Kzerob\ are the transiting planets around the two biggest stars observed so far. They are inflated hot Jupiters, which are particularly challenging for planetary models that require unusually large amounts of additional dissipated energy in the planet, especially the massive \Ksixb. For those two targets we also find signs of a distant companion that are compatible with the planetary mass range. Although we did not perform a detailed analysis of the false positive scenarios, we note that there is no evidence of a blend or of unaccounted-for contamination. There is no significant spatial shift at the time of the transits in the \Kepler\ photometric centroid. The transits of \Ksixb\ and \Kzerob\ are well sampled and U-shaped, with a large number of points in the ingress and egress phases. There are no detectable lines of a contaminant in the spectra or in the CCF that shows any mask effect. The bisector span is not correlated to the radial velocities at the 95\% level. For both targets, the projected rotational velocity of the star is compatible with the modulations of the light curves due to spots. The stellar density determined with the transit is in good agreement with the one determined by the spectroscopic analysis. Using three different stellar evolutionary tracks, the inferred stellar masses, radii, and ages are compatible within 1-$\sigma$. Nevertheless, more observations of those systems would be interesting for assessing the nature and orbital period of the companion. That would provide new insight into the survival of planets through the late evolution stages of their stars and help improve the theories of formation and survival of the short-period planets around massive stars. 

\begin{acknowledgements}

This paper includes data collected with SOPHIE and ESPaDOnS and by the \Kepler\ mission. Funding for the \Kepler\ mission is provided by the NASA Science Mission directorate. We thank the technical team at the Observatoire de Haute-Provence for their support with the SOPHIE instrument and the 1.93-m telescope and, in particular, for the essential work of the night assistants. Financial support for the SOPHIE observations from the Programme National de Planétologie (PNP) of CNRS/INSU, France is gratefully acknowledged. We also acknowledge support from the French National Research Agency (ANR-08- JCJC-0102-01). 

Some of the data presented in this paper were obtained from the Mikulski Archive for Space Telescopes (MAST). STScI is operated by the Association of Universities for Research in Astronomy, Inc., under NASA contract NAS5-26555. Support for MAST for non-HST data is provided by the NASA Office of Space Science via grant NNX09AF08G and by other grants and contracts.
This publication makes use of data products from the Two Micron All Sky Survey, which is a joint project of the University of Massachusetts and the Infrared Processing and Analysis Center/California Institute of Technology, funded by the National Aeronautics and Space Administration and the National Science Foundation.

This publication makes use of data products from the Wide-field Infrared Survey Explorer, which is a joint project of the University of California, Los Angeles, and the Jet Propulsion Laboratory/California Institute of Technology, funded by the National Aeronautics and Space Administration.

This research made use of the Exoplanet Orbit Database and the Exoplanet Data Explorer at exoplanets.org.

The team at LAM acknowledges support by CNES grants 98761 (SCCB), 426808 (CD), and 251091 (JMA). AS acknowledge the support from the Euro- pean Research Council/European Community under the FP7 through Starting Grant agreement number 239953. A.S. is supported by the European Union under a Marie Curie Intra-European Fellowship for Career Development with reference FP7-PEOPLE-2013-IEF, number 627202. ASB acknowledges funding from the European Union Seventh Framework Program (FP7/2007-2013) under Grant agreement No. 313014 (ETAEARTH). Part of this research was supported by an appointment to the NASA Postdoctoral Program at the Ames Research Center, administered by Oak Ridge Associated Universities through a contract with NASA. JMA acknowledges funding from the European Research Council under the ERC Grant Agreement n. 337591-ExTrA. 

\end{acknowledgements}

\bibliographystyle{aa}  % A&A bibliography style file (aa.bst)
\bibliography{koi} % your references in file: Yourfile.bib

\begin{thebibliography}{71}
\expandafter\ifx\csname natexlab\endcsname\relax\def\natexlab#1{#1}\fi

\bibitem[{{Allard} {et~al.}(2012){Allard}, {Homeier}, \& {Freytag}}]{Allard}
{Allard}, F., {Homeier}, D., \& {Freytag}, B. 2012, Royal Society of London
  Philosophical Transactions Series A, 370, 2765

\bibitem[{{Almenara} {et~al.}(2013){Almenara}, {Bouchy}, {Gaulme}, {Deleuil},
  {Havel}, {Gandolfi}, {Deeg}, {Wuchterl}, {Guillot}, {Gardes}, {Pasternacki},
  {Aigrain}, {Alonso}, {Auvergne}, {Baglin}, {Bonomo}, {Bord{\'e}}, {Cabrera},
  {Carpano}, {Cochran}, {Csizmadia}, {Damiani}, {Diaz}, {Dvorak}, {Endl},
  {Erikson}, {Ferraz-Mello}, {Fridlund}, {H{\'e}brard}, {Gillon}, {Guenther},
  {Hatzes}, {L{\'e}ger}, {Lammer}, {MacQueen}, {Mazeh}, {Moutou}, {Ollivier},
  {Ofir}, {P{\"a}tzold}, {Parviainen}, {Queloz}, {Rauer}, {Rouan}, {Santerne},
  {Samuel}, {Schneider}, {Tal-Or}, {Tingley}, \& {Weingrill}}]{Almenara2013}
{Almenara}, J.~M., {Bouchy}, F., {Gaulme}, P., {et~al.} 2013, \aap, 555, A118

\bibitem[{{Alonso} {et~al.}(2008){Alonso}, {Auvergne}, {Baglin}, {Ollivier},
  {Moutou}, {Rouan}, {Deeg}, {Aigrain}, {Almenara}, {Barbieri}, {Barge},
  {Benz}, {Bord{\'e}}, {Bouchy}, {de La Reza}, {Deleuil}, {Dvorak}, {Erikson},
  {Fridlund}, {Gillon}, {Gondoin}, {Guillot}, {Hatzes}, {H{\'e}brard},
  {Kabath}, {Jorda}, {Lammer}, {L{\'e}ger}, {Llebaria}, {Loeillet}, {Magain},
  {Mayor}, {Mazeh}, {P{\"a}tzold}, {Pepe}, {Pont}, {Queloz}, {Rauer},
  {Shporer}, {Schneider}, {Stecklum}, {Udry}, \& {Wuchterl}}]{Alonso2008}
{Alonso}, R., {Auvergne}, M., {Baglin}, A., {et~al.} 2008, \aap, 482, L21

\bibitem[{{Anderson} {et~al.}(2011){Anderson}, {Smith}, {Lanotte}, {Barman},
  {Collier Cameron}, {Campo}, {Gillon}, {Harrington}, {Hellier}, {Maxted},
  {Queloz}, {Triaud}, \& {Wheatley}}]{Anderson2011}
{Anderson}, D.~R., {Smith}, A.~M.~S., {Lanotte}, A.~A., {et~al.} 2011, \mnras,
  416, 2108

\bibitem[{{Baraffe} {et~al.}(2008){Baraffe}, {Chabrier}, \&
  {Barman}}]{Baraffe2008}
{Baraffe}, I., {Chabrier}, G., \& {Barman}, T. 2008, \aap, 482, 315

\bibitem[{{Baranne} {et~al.}(1996){Baranne}, {Queloz}, {Mayor}, {Adrianzyk},
  {Knispel}, {Kohler}, {Lacroix}, {Meunier}, {Rimbaud}, \&
  {Vin}}]{1996A&AS..119..373B}
{Baranne}, A., {Queloz}, D., {Mayor}, M., {et~al.} 1996, \aaps, 119, 373

\bibitem[{{Belkacem} {et~al.}(2011){Belkacem}, {Goupil}, {Dupret}, {Samadi},
  {Baudin}, {Noels}, \& {Mosser}}]{2011A&A...530A.142B}
{Belkacem}, K., {Goupil}, M.~J., {Dupret}, M.~A., {et~al.} 2011, \aap, 530,
  A142

\bibitem[{{Boisse} {et~al.}(2010){Boisse}, {Eggenberger}, {Santos}, {Lovis},
  {Bouchy}, {H{\'e}brard}, {Arnold}, {Bonfils}, {Delfosse}, {Desort},
  {D{\'{\i}}az}, {Ehrenreich}, {Forveille}, {Gallenne}, {Lagrange}, {Moutou},
  {Udry}, {Pepe}, {Perrier}, {Perruchot}, {Pont}, {Queloz}, {Santerne},
  {S{\'e}gransan}, \& {Vidal-Madjar}}]{2010A&A...523A..88B}
{Boisse}, I., {Eggenberger}, A., {Santos}, N.~C., {et~al.} 2010, \aap, 523, A88

\bibitem[{{Bonomo} {et~al.}(2012){Bonomo}, {H{\'e}brard}, {Santerne}, {Santos},
  {Deleuil}, {Almenara}, {Bouchy}, {D{\'{\i}}az}, {Moutou}, \&
  {Vanhuysse}}]{Bonomo2012}
{Bonomo}, A.~S., {H{\'e}brard}, G., {Santerne}, A., {et~al.} 2012, \aap, 538,
  A96

\bibitem[{{Borucki} {et~al.}(2010){Borucki}, {Koch}, {Basri}, {Batalha},
  {Brown}, {Caldwell}, {Caldwell}, {Christensen-Dalsgaard}, {Cochran},
  {DeVore}, {Dunham}, {Dupree}, {Gautier}, {Geary}, {Gilliland}, {Gould},
  {Howell}, {Jenkins}, {Kondo}, {Latham}, {Marcy}, {Meibom}, {Kjeldsen},
  {Lissauer}, {Monet}, {Morrison}, {Sasselov}, {Tarter}, {Boss}, {Brownlee},
  {Owen}, {Buzasi}, {Charbonneau}, {Doyle}, {Fortney}, {Ford}, {Holman},
  {Seager}, {Steffen}, {Welsh}, {Rowe}, {Anderson}, {Buchhave}, {Ciardi},
  {Walkowicz}, {Sherry}, {Horch}, {Isaacson}, {Everett}, {Fischer}, {Torres},
  {Johnson}, {Endl}, {MacQueen}, {Bryson}, {Dotson}, {Haas}, {Kolodziejczak},
  {Van Cleve}, {Chandrasekaran}, {Twicken}, {Quintana}, {Clarke}, {Allen},
  {Li}, {Wu}, {Tenenbaum}, {Verner}, {Bruhweiler}, {Barnes}, \&
  {Prsa}}]{2010Sci...327..977B}
{Borucki}, W.~J., {Koch}, D., {Basri}, G., {et~al.} 2010, Science, 327, 977

\bibitem[{{Borucki} {et~al.}(2011){Borucki}, {Koch}, {Basri}, {Batalha},
  {Brown}, {Bryson}, {Caldwell}, {Christensen-Dalsgaard}, {Cochran}, {DeVore},
  {Dunham}, {Gautier}, {Geary}, {Gilliland}, {Gould}, {Howell}, {Jenkins},
  {Latham}, {Lissauer}, {Marcy}, {Rowe}, {Sasselov}, {Boss}, {Charbonneau},
  {Ciardi}, {Doyle}, {Dupree}, {Ford}, {Fortney}, {Holman}, {Seager},
  {Steffen}, {Tarter}, {Welsh}, {Allen}, {Buchhave}, {Christiansen}, {Clarke},
  {Das}, {D{\'e}sert}, {Endl}, {Fabrycky}, {Fressin}, {Haas}, {Horch},
  {Howard}, {Isaacson}, {Kjeldsen}, {Kolodziejczak}, {Kulesa}, {Li}, {Lucas},
  {Machalek}, {McCarthy}, {MacQueen}, {Meibom}, {Miquel}, {Prsa}, {Quinn},
  {Quintana}, {Ragozzine}, {Sherry}, {Shporer}, {Tenenbaum}, {Torres},
  {Twicken}, {Van Cleve}, {Walkowicz}, {Witteborn}, \&
  {Still}}]{2011ApJ...736...19B}
{Borucki}, W.~J., {Koch}, D.~G., {Basri}, G., {et~al.} 2011, \apj, 736, 19

\bibitem[{{Bouchy} {et~al.}(2013){Bouchy}, {D{\'{\i}}az}, {H{\'e}brard},
  {Arnold}, {Boisse}, {Delfosse}, {Perruchot}, \&
  {Santerne}}]{2013A&A...549A..49B}
{Bouchy}, F., {D{\'{\i}}az}, R.~F., {H{\'e}brard}, G., {et~al.} 2013, \aap,
  549, A49

\bibitem[{{Bouchy} {et~al.}(2009){Bouchy}, {H{\'e}brard}, {Udry}, {Delfosse},
  {Boisse}, {Desort}, {Bonfils}, {Eggenberger}, {Ehrenreich}, {Forveille},
  {Lagrange}, {Le Coroller}, {Lovis}, {Moutou}, {Pepe}, {Perrier}, {Pont},
  {Queloz}, {Santos}, {S{\'e}gransan}, \& {Vidal-Madjar}}]{2009A&A...505..853B}
{Bouchy}, F., {H{\'e}brard}, G., {Udry}, S., {et~al.} 2009, \aap, 505, 853

\bibitem[{{Bressan} {et~al.}(2012{\natexlab{a}}){Bressan}, {Marigo}, {Girardi},
  {Salasnich}, {Dal Cero}, {Rubele}, \& {Nanni}}]{2012MNRAS.427..127B}
{Bressan}, A., {Marigo}, P., {Girardi}, L., {et~al.} 2012{\natexlab{a}},
  \mnras, 427, 127

\bibitem[{{Bressan} {et~al.}(2012{\natexlab{b}}){Bressan}, {Marigo}, {Girardi},
  {Salasnich}, {Dal Cero}, {Rubele}, \& {Nanni}}]{Bressan2012}
{Bressan}, A., {Marigo}, P., {Girardi}, L., {et~al.} 2012{\natexlab{b}},
  \mnras, 427, 127

\bibitem[{{Bruntt} {et~al.}(2010{\natexlab{a}}){Bruntt}, {Bedding}, {Quirion},
  {Lo Curto}, {Carrier}, {Smalley}, {Dall}, {Arentoft}, {Bazot}, \&
  {Butler}}]{2010MNRAS.405.1907B}
{Bruntt}, H., {Bedding}, T.~R., {Quirion}, P.-O., {et~al.} 2010{\natexlab{a}},
  \mnras, 405, 1907

\bibitem[{{Bruntt} {et~al.}(2004){Bruntt}, {Bikmaev}, {Catala}, {Solano},
  {Gillon}, {Magain}, {Van't Veer-Menneret}, {St{\"u}tz}, {Weiss}, {Ballereau},
  {Bouret}, {Charpinet}, {Hua}, {Katz}, {Ligni{\`e}res}, \&
  {Lueftinger}}]{2004A&A...425..683B}
{Bruntt}, H., {Bikmaev}, I.~F., {Catala}, C., {et~al.} 2004, \aap, 425, 683

\bibitem[{{Bruntt} {et~al.}(2008){Bruntt}, {De Cat}, \&
  {Aerts}}]{2008A&A...478..487B}
{Bruntt}, H., {De Cat}, P., \& {Aerts}, C. 2008, \aap, 478, 487

\bibitem[{{Bruntt} {et~al.}(2010{\natexlab{b}}){Bruntt}, {Deleuil}, {Fridlund},
  {Alonso}, {Bouchy}, {Hatzes}, {Mayor}, {Moutou}, \&
  {Queloz}}]{2010A&A...519A..51B}
{Bruntt}, H., {Deleuil}, M., {Fridlund}, M., {et~al.} 2010{\natexlab{b}}, \aap,
  519, A51

\bibitem[{{Burkert} \& {Ida}(2007)}]{2007ApJ...660..845B}
{Burkert}, A. \& {Ida}, S. 2007, \apj, 660, 845

\bibitem[{{Burrows} {et~al.}(2003){Burrows}, {Sudarsky}, \&
  {Hubbard}}]{2003ApJ...594..545B}
{Burrows}, A., {Sudarsky}, D., \& {Hubbard}, W.~B. 2003, \apj, 594, 545

\bibitem[{{D{\'e}sert} {et~al.}(2011){D{\'e}sert}, {Charbonneau}, {Demory},
  {Ballard}, {Carter}, {Fortney}, {Cochran}, {Endl}, {Quinn}, {Isaacson},
  {Fressin}, {Buchhave}, {Latham}, {Knutson}, {Bryson}, {Torres}, {Rowe},
  {Batalha}, {Borucki}, {Brown}, {Caldwell}, {Christiansen}, {Deming},
  {Fabrycky}, {Ford}, {Gilliland}, {Gillon}, {Haas}, {Jenkins}, {Kinemuchi},
  {Koch}, {Lissauer}, {Lucas}, {Mullally}, {MacQueen}, {Marcy}, {Sasselov},
  {Seager}, {Still}, {Tenenbaum}, {Uddin}, \& {Winn}}]{2011ApJS..197...14D}
{D{\'e}sert}, J.-M., {Charbonneau}, D., {Demory}, B.-O., {et~al.} 2011, \apjs,
  197, 14

\bibitem[{{D{\'{\i}}az} {et~al.}(2014){D{\'{\i}}az}, {Almenara}, {Santerne},
  {Moutou}, {Lethuillier}, \& {Deleuil}}]{2014MNRAS.441..983D}
{D{\'{\i}}az}, R.~F., {Almenara}, J.~M., {Santerne}, A., {et~al.} 2014, \mnras,
  441, 983

\bibitem[{{Dodson-Robinson} \& {Salyk}(2011)}]{2011ApJ...738..131D}
{Dodson-Robinson}, S.~E. \& {Salyk}, C. 2011, \apj, 738, 131

\bibitem[{{Dong} {et~al.}(2014){Dong}, {Katz}, \&
  {Socrates}}]{2014ApJ...781L...5D}
{Dong}, S., {Katz}, B., \& {Socrates}, A. 2014, \apjl, 781, L5

\bibitem[{{Dotter} {et~al.}(2008){Dotter}, {Chaboyer}, {Jevremovi{\'c}},
  {Kostov}, {Baron}, \& {Ferguson}}]{2008ApJS..178...89D}
{Dotter}, A., {Chaboyer}, B., {Jevremovi{\'c}}, D., {et~al.} 2008, \apjs, 178,
  89

\bibitem[{{Etzel}(1981)}]{1981psbs.conf..111E}
{Etzel}, P.~B. 1981, in Photometric and Spectroscopic Binary Systems, ed. E.~B.
  {Carling} \& Z.~{Kopal}, 111

\bibitem[{{Gray}(2005)}]{2005oasp.book.....G}
{Gray}, D.~F. 2005, {The Observation and Analysis of Stellar Photospheres}

\bibitem[{{Guillot}(2008)}]{2008PhST..130a4023G}
{Guillot}, T. 2008, Physica Scripta Volume T, 130, 014023

\bibitem[{{Guillot}(2010)}]{Guillot2010}
{Guillot}, T. 2010, \aap, 520, A27

\bibitem[{{Guillot} \& {Gautier}(2014)}]{Guillot2014}
{Guillot}, T. \& {Gautier}, D. 2014, ArXiv:1405.3752

\bibitem[{{Guillot} \& {Havel}(2011)}]{GH2011}
{Guillot}, T. \& {Havel}, M. 2011, \aap, 527,

\bibitem[{{Guillot} \& {Morel}(1995)}]{Guillot1995}
{Guillot}, T. \& {Morel}, P. 1995, \aaps, 109, 109

\bibitem[{{Guillot} {et~al.}(2006){Guillot}, {Santos}, {Pont}, {Iro}, {Melo},
  \& {Ribas}}]{Guillot2006}
{Guillot}, T., {Santos}, N.~C., {Pont}, F., {et~al.} 2006, \aap, 453, L21

\bibitem[{{Guillot} \& {Showman}(2002)}]{GS2002}
{Guillot}, T. \& {Showman}, A.~P. 2002, \aap, 385, 156

\bibitem[{{Gustafsson} {et~al.}(2008){Gustafsson}, {Edvardsson}, {Eriksson},
  {J{\o}rgensen}, {Nordlund}, \& {Plez}}]{2008A&A...486..951G}
{Gustafsson}, B., {Edvardsson}, B., {Eriksson}, K., {et~al.} 2008, \aap, 486,
  951

\bibitem[{{Hartman} {et~al.}(2011){Hartman}, {Bakos}, {Torres}, {Latham},
  {Kov{\'a}cs}, {B{\'e}ky}, {Quinn}, {Mazeh}, {Shporer}, {Marcy}, {Howard},
  {Fischer}, {Johnson}, {Esquerdo}, {Noyes}, {Sasselov}, {Stefanik},
  {Fernandez}, {Szklen{\'a}r}, {L{\'a}z{\'a}r}, {Papp}, \&
  {S{\'a}ri}}]{Hartman2011}
{Hartman}, J.~D., {Bakos}, G.~{\'A}., {Torres}, G., {et~al.} 2011, \apj, 742,
  59

\bibitem[{{Havel} {et~al.}(2011){Havel}, {Guillot}, {Valencia}, \&
  {Crida}}]{Havel2011}
{Havel}, M., {Guillot}, T., {Valencia}, D., \& {Crida}, A. 2011, \aap, 531,

\bibitem[{{H{\'e}brard} {et~al.}(2008){H{\'e}brard}, {Bouchy}, {Pont},
  {Loeillet}, {Rabus}, {Bonfils}, {Moutou}, {Boisse}, {Delfosse}, {Desort},
  {Eggenberger}, {Ehrenreich}, {Forveille}, {Lagrange}, {Lovis}, {Mayor},
  {Pepe}, {Perrier}, {Queloz}, {Santos}, {S{\'e}gransan}, {Udry}, \&
  {Vidal-Madjar}}]{2008A&A...488..763H}
{H{\'e}brard}, G., {Bouchy}, F., {Pont}, F., {et~al.} 2008, \aap, 488, 763

\bibitem[{{H{\'e}brard} {et~al.}(2011){H{\'e}brard}, {Evans}, {Alonso},
  {Fridlund}, {Ofir}, {Aigrain}, {Guillot}, {Almenara}, {Auvergne}, {Baglin},
  {Barge}, {Bonomo}, {Bord{\'e}}, {Bouchy}, {Cabrera}, {Carone}, {Carpano},
  {Cavarroc}, {Csizmadia}, {Deeg}, {Deleuil}, {D{\'{\i}}az}, {Dvorak},
  {Erikson}, {Ferraz-Mello}, {Gandolfi}, {Gibson}, {Gillon}, {Guenther},
  {Hatzes}, {Havel}, {Jorda}, {Lammer}, {L{\'e}ger}, {Llebaria}, {Mazeh},
  {Moutou}, {Ollivier}, {Parviainen}, {P{\"a}tzold}, {Queloz}, {Rauer},
  {Rouan}, {Santerne}, {Schneider}, {Tingley}, \& {Wuchterl}}]{Hebrard2011}
{H{\'e}brard}, G., {Evans}, T.~M., {Alonso}, R., {et~al.} 2011, \aap, 533, A130

\bibitem[{{Howard} {et~al.}(2010){Howard}, {Marcy}, {Johnson}, {Fischer},
  {Wright}, {Isaacson}, {Valenti}, {Anderson}, {Lin}, \&
  {Ida}}]{2010Sci...330..653H}
{Howard}, A.~W., {Marcy}, G.~W., {Johnson}, J.~A., {et~al.} 2010, Science, 330,
  653

\bibitem[{{Ikoma} {et~al.}(2006){Ikoma}, {Guillot}, {Genda}, {Tanigawa}, \&
  {Ida}}]{2006ApJ...650.1150I}
{Ikoma}, M., {Guillot}, T., {Genda}, H., {Tanigawa}, T., \& {Ida}, S. 2006,
  \apj, 650, 1150

\bibitem[{{Jenkins} {et~al.}(2010){Jenkins}, {Caldwell}, {Chandrasekaran},
  {Twicken}, {Bryson}, {Quintana}, {Clarke}, {Li}, {Allen}, {Tenenbaum}, {Wu},
  {Klaus}, {Van Cleve}, {Dotson}, {Haas}, {Gilliland}, {Koch}, \&
  {Borucki}}]{2010ApJ...713L.120J}
{Jenkins}, J.~M., {Caldwell}, D.~A., {Chandrasekaran}, H., {et~al.} 2010,
  \apjl, 713, L120

\bibitem[{{Kipping}(2012)}]{2012MNRAS.427.2487K}
{Kipping}, D.~M. 2012, \mnras, 427, 2487

\bibitem[{{Lanza} {et~al.}(2003){Lanza}, {Rodon{\`o}}, {Pagano}, {Barge}, \&
  {Llebaria}}]{2003A&A...403.1135L}
{Lanza}, A.~F., {Rodon{\`o}}, M., {Pagano}, I., {Barge}, P., \& {Llebaria}, A.
  2003, \aap, 403, 1135

\bibitem[{{Mazeh} {et~al.}(2013){Mazeh}, {Nachmani}, {Holczer}, {Fabrycky},
  {Ford}, {Sanchis-Ojeda}, {Sokol}, {Rowe}, {Zucker}, {Agol}, {Carter},
  {Lissauer}, {Quintana}, {Ragozzine}, {Steffen}, \&
  {Welsh}}]{2013ApJS..208...16M}
{Mazeh}, T., {Nachmani}, G., {Holczer}, T., {et~al.} 2013, \apjs, 208, 16

\bibitem[{{Militzer} \& {Hubbard}(2013)}]{Militzer2013}
{Militzer}, B. \& {Hubbard}, W.~B. 2013, \apj, 774, 148

\bibitem[{{Mizuno}(1980)}]{1980PThPh..64..544M}
{Mizuno}, H. 1980, Progress of Theoretical Physics, 64, 544

\bibitem[{{Mordasini} {et~al.}(2009){Mordasini}, {Alibert}, \&
  {Benz}}]{2009A&A...501.1139M}
{Mordasini}, C., {Alibert}, Y., \& {Benz}, W. 2009, \aap, 501, 1139

\bibitem[{{Moutou} {et~al.}(2013){Moutou}, {Deleuil}, {Guillot}, {Baglin},
  {Bord{\'e}}, {Bouchy}, {Cabrera}, {Csizmadia}, \& {Deeg}}]{Moutou2013}
{Moutou}, C., {Deleuil}, M., {Guillot}, T., {et~al.} 2013, \icarus, 226, 1625

\bibitem[{{Nelson} \& {Davis}(1972)}]{1972ApJ...174..617N}
{Nelson}, B. \& {Davis}, W.~D. 1972, \apj, 174, 617

\bibitem[{{Parmentier} {et~al.}(2013){Parmentier}, {Guillot}, {Fortney}, \&
  {Marley}}]{2013arXiv1311.6322P}
{Parmentier}, V., {Guillot}, T., {Fortney}, J.~J., \& {Marley}, M.~S. 2013,
  ArXiv:1311.6322

\bibitem[{{Pepe} {et~al.}(2002){Pepe}, {Mayor}, {Galland}, {Naef}, {Queloz},
  {Santos}, {Udry}, \& {Burnet}}]{2002A&A...388..632P}
{Pepe}, F., {Mayor}, M., {Galland}, F., {et~al.} 2002, \aap, 388, 632

\bibitem[{{Perruchot} {et~al.}(2008){Perruchot}, {Kohler}, {Bouchy}, {Richaud},
  {Richaud}, {Moreaux}, {Merzougui}, {Sottile}, {Hill}, {Knispel}, {Regal},
  {Meunier}, {Ilovaisky}, {Le Coroller}, {Gillet}, {Schmitt}, {Pepe}, {Fleury},
  {Sosnowska}, {Vors}, {M{\'e}gevand}, {Blanc}, {Carol}, {Point}, {Laloge}, \&
  {Brunel}}]{2008SPIE.7014E..17P}
{Perruchot}, S., {Kohler}, D., {Bouchy}, F., {et~al.} 2008, in Society of
  Photo-Optical Instrumentation Engineers (SPIE) Conference Series, Vol. 7014,
  Society of Photo-Optical Instrumentation Engineers (SPIE) Conference Series

\bibitem[{{Pollacco} {et~al.}(2008){Pollacco}, {Skillen}, {Collier Cameron},
  {Loeillet}, {Stempels}, {Bouchy}, {Gibson}, {Hebb}, {H{\'e}brard}, {Joshi},
  {McDonald}, {Smalley}, {Smith}, {Street}, {Udry}, {West}, {Wilson},
  {Wheatley}, {Aigrain}, {Alsubai}, {Benn}, {Bruce}, {Christian}, {Clarkson},
  {Enoch}, {Evans}, {Fitzsimmons}, {Haswell}, {Hellier}, {Hickey}, {Hodgkin},
  {Horne}, {Hrudkov{\'a}}, {Irwin}, {Kane}, {Keenan}, {Lister}, {Maxted},
  {Mayor}, {Moutou}, {Norton}, {Osborne}, {Parley}, {Pont}, {Queloz}, {Ryans},
  \& {Simpson}}]{2008MNRAS.385.1576P}
{Pollacco}, D., {Skillen}, I., {Collier Cameron}, A., {et~al.} 2008, \mnras,
  385, 1576

\bibitem[{{Popper} \& {Etzel}(1981)}]{1981AJ.....86..102P}
{Popper}, D.~M. \& {Etzel}, P.~B. 1981, \aj, 86, 102

\bibitem[{{Press} \& {Rybicki}(1989)}]{1989ApJ...338..277P}
{Press}, W.~H. \& {Rybicki}, G.~B. 1989, \apj, 338, 277

\bibitem[{{Raghavan} {et~al.}(2010){Raghavan}, {McAlister}, {Henry}, {Latham},
  {Marcy}, {Mason}, {Gies}, {White}, \& {ten Brummelaar}}]{2010ApJS..190....1R}
{Raghavan}, D., {McAlister}, H.~A., {Henry}, T.~J., {et~al.} 2010, \apjs, 190,
  1

\bibitem[{{Robin} {et~al.}(2003){Robin}, {Reyl{\'e}}, {Derri{\`e}re}, \&
  {Picaud}}]{2003A&A...409..523R}
{Robin}, A.~C., {Reyl{\'e}}, C., {Derri{\`e}re}, S., \& {Picaud}, S. 2003,
  \aap, 409, 523

\bibitem[{{Santerne} {et~al.}(2012){Santerne}, {D{\'{\i}}az}, {Moutou},
  {Bouchy}, {H{\'e}brard}, {Almenara}, {Bonomo}, {Deleuil}, \&
  {Santos}}]{2012A&A...545A..76S}
{Santerne}, A., {D{\'{\i}}az}, R.~F., {Moutou}, C., {et~al.} 2012, \aap, 545,
  A76

\bibitem[{{Santerne} {et~al.}(2014){Santerne}, {H{\'e}brard}, {Deleuil},
  {Havel}, {Correia}, {Almenara}, {Alonso}, {Arnold}, {Barros}, {Behrend},
  {Bernasconi}, {Boisse}, {Bonomo}, {Bouchy}, {Bruno}, {Damiani},
  {D{\'{\i}}az}, {Gravallon}, {Guillot}, {Labrevoir}, {Montagnier}, {Moutou},
  {Rinner}, {Santos}, {Abe}, {Audejean}, {Bendjoya}, {Gillier}, {Gregorio},
  {Martinez}, {Michelet}, {Montaigut}, {Poncy}, {Rivet}, {Rousseau}, {Roy},
  {Suarez}, {Vanhuysse}, \& {Verilhac}}]{2014A&A...571A..37S}
{Santerne}, A., {H{\'e}brard}, G., {Deleuil}, M., {et~al.} 2014, \aap, 571, A37

\bibitem[{{Saumon} {et~al.}(1995){Saumon}, {Chabrier}, \& {van
  Horn}}]{Saumon1995}
{Saumon}, D., {Chabrier}, G., \& {van Horn}, H.~M. 1995, \apjs, 99, 713

\bibitem[{{Smalley} {et~al.}(2012){Smalley}, {Anderson}, \&
  {Collier-Cameron}}]{Smalley2012}
{Smalley}, B., {Anderson}, D.~R., \& {Collier-Cameron}. 2012, \aap, 547, A61

\bibitem[{{Southworth}(2011)}]{2011MNRAS.417.2166S}
{Southworth}, J. 2011, \mnras, 417, 2166

\bibitem[{{Spiegel} \& {Burrows}(2013)}]{Spiegel2013}
{Spiegel}, D.~S. \& {Burrows}, A. 2013, \apj, 772, 76

\bibitem[{{Tegmark} {et~al.}(2004){Tegmark}, {Strauss}, {Blanton}, {Abazajian},
  {Dodelson}, {Sandvik}, {Wang}, {Weinberg}, {Zehavi}, {Bahcall}, {Hoyle},
  {Schlegel}, {Scoccimarro}, {Vogeley}, {Berlind}, {Budavari}, {Connolly},
  {Eisenstein}, {Finkbeiner}, {Frieman}, {Gunn}, {Hui}, {Jain}, {Johnston},
  {Kent}, {Lin}, {Nakajima}, {Nichol}, {Ostriker}, {Pope}, {Scranton},
  {Seljak}, {Sheth}, {Stebbins}, {Szalay}, {Szapudi}, {Xu}, {Annis},
  {Brinkmann}, {Burles}, {Castander}, {Csabai}, {Loveday}, {Doi}, {Fukugita},
  {Gillespie}, {Hennessy}, {Hogg}, {Ivezi{\'c}}, {Knapp}, {Lamb}, {Lee},
  {Lupton}, {McKay}, {Kunszt}, {Munn}, {O'Connell}, {Peoples}, {Pier},
  {Richmond}, {Rockosi}, {Schneider}, {Stoughton}, {Tucker}, {vanden Berk},
  {Yanny}, \& {York}}]{2004PhRvD..69j3501T}
{Tegmark}, M., {Strauss}, M.~A., {Blanton}, M.~R., {et~al.} 2004, \prd, 69,
  103501

\bibitem[{{Valenti} \& {Fischer}(2005)}]{2005ApJS..159..141V}
{Valenti}, J.~A. \& {Fischer}, D.~A. 2005, \apjs, 159, 141

\bibitem[{{Valenti} \& {Piskunov}(1996)}]{1996A&AS..118..595V}
{Valenti}, J.~A. \& {Piskunov}, N. 1996, \aaps, 118, 595

\bibitem[{{Weidenschilling}(2005)}]{2005SSRv..116...53W}
{Weidenschilling}, S.~J. 2005, Space Science Reviews, 116, 53

\bibitem[{{Wright} {et~al.}(2011){Wright}, {Fakhouri}, {Marcy}, {Han}, {Feng},
  {Johnson}, {Howard}, {Fischer}, {Valenti}, {Anderson}, \&
  {Piskunov}}]{2011PASP..123..412W}
{Wright}, J.~T., {Fakhouri}, O., {Marcy}, G.~W., {et~al.} 2011, \pasp, 123, 412

\bibitem[{{Wright} \& {Gaudi}(2013)}]{2013pss3.book..489W}
{Wright}, J.~T. \& {Gaudi}, B.~S. 2013, {Exoplanet Detection Methods}, ed.
  T.~D. {Oswalt}, L.~M. {French}, \& P.~{Kalas} (Springer), 489

\end{thebibliography}

\Online

\end{document}